\documentclass{article}
\usepackage{amsmath,amsfonts,amssymb,graphics,graphicx,epsfig,color,bbm}

\newcommand{\be}{\begin{eqnarray}}
\newcommand{\ee}{\end{eqnarray}}
\newcommand{\bea}{\begin{eqnarray}}
\newcommand{\eea}{\end{eqnarray}}
\newcommand{\bma}{\begin{subequations}}
\newcommand{\ema}{\end{subequations}}

\newcommand{\MPS}{{{\rm MPS}}}
\newcommand{\NRG}{{{\rm NRG}}}
\newcommand{\openone}{I}

\newcommand{\ket}[1]{| \, #1 \, \rangle}
\newcommand{\bra}[1]{\langle \, #1 \, |}

\newcommand{\scal}[2]{\bra{#1} \, #2 \, \rangle}
\newcommand{\expect}[1]{\langle #1 \rangle}

\newcommand{\adj}[1]{#1^{\dagger}}

\newcommand{\sx}[1]{\sigma_x^{(#1)}}
\newcommand{\sy}[1]{\sigma_y^{(#1)}}
\newcommand{\sz}[1]{\sigma_z^{(#1)}}

\newcommand{\Omicron}[1]{\textrm{O} \left[ #1 \right] }

\DeclareMathOperator{\X}{X} \DeclareMathOperator{\XX}{\mathbb{X}} 
 \DeclareMathOperator{\XXt}{\tilde{\mathbb{X}}} 
\DeclareMathOperator{\tr}{Tr} 

\def\tr{ \mbox{tr}}
\def\qed{\leavevmode\unskip\penalty9999 \hbox{}\nobreak\hfill
     \quad\hbox{\leavevmode  \hbox to.77778em{%
               \hfil\vrule   \vbox to.675em%
               {\hrule width.6em\vfil\hrule}\vrule\hfil}}
     \par\vskip3pt}

\newtheorem{lemma}{Lemma}

\def\qed{\leavevmode\unskip\penalty9999 \hbox{}\nobreak\hfill
     \quad\hbox{\leavevmode  \hbox to.77778em{%
               \hfil\vrule   \vbox to.675em%
               {\hrule width.6em\vfil\hrule}\vrule\hfil}}
     \par\vskip3pt}

\begin{document}

\markboth{MPS, PEPS and variational renormalization group methods for quantum spin systems}{MPS, PEPS and variational renormalization group methods for quantum spin systems}

\title{Matrix Product States, Projected Entangled Pair States, and variational renormalization group methods for quantum spin systems}

\author{F. Verstraete$^a$, J.I. Cirac$^b$ and V. Murg$^{a,b}$\\
%\vspace{.5cm}
\\
$^a$ Faculty of Physics, University of Vienna, Austria\\
%\vspace{.5cm}
%\\
$^b$ Max-Planck-Institut f\"ur Quantenoptik, Garching, D-85748, Germany}
\maketitle

\maketitle

\begin{abstract}
This article reviews recent developments in the theoretical understanding and the
numerical implementation of variational renormalization group methods using
matrix product states and projected entangled pair states.

\end{abstract}

%%%%%%%%%%%%%%%%%%%%%%%%%%%%%%%%%%%%%%%%%%%%%%%%%%%%%%%%%%%%%%%%%%%%%%

% \newpage
\tableofcontents

\newpage

One of the biggest challenges in physics is to develop efficient methods for simulating Hamiltonians: the basic laws of quantum mechanics are very well understood, but the crucial understanding of the collective behavior of interacting elementary particles and of emerging phenomena relies on our ability to approximate or simulate the corresponding
equations. Simulation is particularly relevant in the context of strongly interacting quantum systems, where conventional perturbation theory fails.
Popular methods to simulate such systems are the numerical renormalization group algorithms by K. Wilson~\cite{wilson75} and S. White~\cite{white92,white92b,schollwoeck04,peschel} and quantum Monte Carlo methods. Both of these have been used with great success, but they also have severe limitations: the numerical renormalization group and DMRG only work in cases where the original Hamiltonian can be mapped to a local Hamiltonian defined on a 1-dimensional chain, while Monte Carlo methods suffer from the so--called sign problem~\cite{troyer04} which makes them inappropriate for the description of fermionic and frustrated quantum systems. Very recently however, new insights coming from the field of quantum information and entanglement theory
have shown how those numerical RG-methods can be generalized to higher dimensions, to finite temperature, to random systems etc. \cite{vidal03,zwolak04,vidal07,verstraeteporras04,verstraeteripoll04,verstraetecirac04,murgverstraete07,murgverstraete05,paredesverstraete05,andersplenio06} These developments bear the promise to revolutionize the way quantum many-body systems can be simulated and understood, which will prove essential in the light of the ongoing miniaturization of electronic devices and of fundamental problems such as the study of the role of the Hubbard model in high $T_c$ superconductivity.

In this review, we will focus on those new developments. In particular, we will focus on the simulation of quantum lattice systems of spins and/or fermions. The Hamiltonians arising in this context can be thought of as effective Hamiltonians of more complicated systems that capture the physics for the relevant low-energy degrees of freedom. Studying quantum spin systems has a long and very interesting history, starting with Dirac and Heisenberg in the 1920's where they proposed the so-called Heisenberg model~\cite{heisenberg26,dirac26} as being illustrative for the basic mechanism giving rise to magnetism. The simulation of quantum spin systems and fermionic systems on a lattice has however turned out to be extremely complicated. Consider for example the Hubbard model in 2 dimensions that is representing the simplest possible model of fermions hopping on a lattice and exhibiting on-site interactions: despite considerable efforts because of the connection with high $T_c$ superconductivity, it is absolutely unclear how the low-energy wavefunctions look like in the relevant parameter range. This is the most notable illustration of the biggest bottleneck in condensed matter theory: numerical simulation of effective simple Hamiltonians. The new tools described in this review bear the promise of opening up that bottleneck.

This review is organized as follows. In a first, rather technical, section, we will review basic properties of quantum spin systems and provide the intuitive justification for the numerical tools used later. This section can be skipped by those who are more interested in the variational numerical renormalization methods and less in the underlying theory; however one of the major advantages of the methods is precisely that there is a solid theoretical justification for using them. In section~\ref{sec:nrg}, we review the numerical renormalization group method as introduced by Wilson, and show how this gives rise to the concept of matrix product states. Section 3 discusses the features of matrix product states, and goes on to applications on spin chains such as a reformulation of the density matrix renormalization group (DMRG) and generalizations that allow to treat excited states. Section~\ref{sec:timeevol} is devoted to the issue of real and imaginary time-evolution of 1-dimensional quantum spin systems, and section~\ref{sec:mpo} to the concept of matrix product operators leading to extensions of DMRG to finite temperature systems, to random spin chains and to the simulation of 2-D classical partition functions. In section~\ref{sec:peps}, we will show how all these ideas can be generalized to the 2-dimensional case by introducing the class of Projected Entangled Pair States (PEPS), and we will discuss the applicability and limitations of these methods. Finally, we discuss the convex set of reduced density operators of translational invariant states in appendix~\ref{sec:reddensityop}, the relation between block entropy and the accuracy of a matrix product state approximation in appendix~\ref{sec:mpsgs}, and we include some explicit matlab programs of matrix product state algorithms in appendix~\ref{sec:matlab}.

At this point, we would like to warn the reader that this is not a review of DMRG methods and possible extensions. In fact, there exist very good reviews of that topic (see, for example, \cite{schollwoeck04,peschel}), where the important progress experienced by DMRG-like methods is extensively discussed. Here, we offer a review of the new methods that have been introduced during the last few years, and that have evolved from ideas in the context of Quantum Information Theory. To make the review self-contained and uniform, we will mostly focus on the methods introduced by ourselves and collaborators, and will also touch upon the approach advocated by Vidal. The main reason is that they can be viewed as (generalized) variational techniques over sets of entangled states, and thus we can present in a unified form the vast variety of algorithms whose description is thusfar scattered in several publications. Furthermore, we will also discuss some issues which have not be published so far.

\newpage
\section{Spin systems: general features} \label{sec:spinsystems}

One of the main characteristics of quantum mechanics is that the underlying Hilbert space is endowed with a tensor product structure:  the Hilbert space of two interacting systems is given by the tensor product space of the two individual ones. This structure of the Hilbert space is a direct consequence of the superposition principle, and opens up the possibility of entanglement and new phases of matter. This simple tensor product structure, however, makes it very clear what the main obstacle is in developing a general theory of many-body quantum systems: the size of the Hilbert space grows exponentially in the number of basis constituents, and hence we would, in principle, need exponentially many variables to specify the wavefunction of a $N$-particle system. However, it is a priori not clear whether Nature can fully exploit and explore these vast territories of  Hilbert space, because another main characteristic of quantum mechanics is that interactions always seem to happen locally and only between a few bodies.

As it turns out, all physical states live on a tiny submanifold of Hilbert space
\footnote{Let us for example consider a quantum system of $N$ spin $1/2$'s on a cubic lattice and all spins polarized in the $z$-direction, and let us pick a random state in the $2^N$ dimensional Hilbert space according to the unitarily invariant Haar measure. Let us furthermore ask the question to find a lower bound on the time it would take to evolve the polarized state into one that has an overlap with the random one that is not exponentially small, and this with local interactions. Using tools developed in the context of quantum information theory, one can show that this lower bound scales exponentially in the number of spins \cite{verstraetewinter07}: any evolution over a time that scales only polynomially in the number of spins would only allow to get an exponentially small overlap with the chosen random state. If we would for example consider a system of a few hundred of spins, it would already take much longer than the lifetime of the universe to come close to any random point in its Hilbert space. This shows that almost all points in the Hilbert space of a many-body quantum system are unphysical as they are inaccessible: all physical states, i.e. states that can ever be created, live on a tiny submanifold of measure zero.}. This opens up a very interesting perspective in the context of the description of quantum many-body systems, as there might exist an efficient parametrization of such a submanifold that would provide the natural language for describing those systems. In the context of many-body quantum physics, one is furthermore mainly interested in describing the low energy sector of local Hamiltonians, and as we will discuss later, this puts many extra constraints on the allowed wavefunctions.

As a trivial example of such a submanifold, let us consider a system of $N\rightarrow\infty$ spins that all interact with each other via a permutation invariant 2-body Hamiltonian such as the Heisenberg interaction~\cite{heisenberg26,dirac26}. Note that there is always a ground state that exhibits the same symmetry as the Hamiltonian. As a consequence of the quantum de-Finnetti theorem \cite{stormer69} one can show that the manifold of all density operators with permutation symmetry is exactly the set of all separable states $\rho=\sum_ip_i\rho_i^{\otimes N}$ when $N\rightarrow\infty$. Ground states correspond to the extreme points of this set, which are exactly the product states that have no entanglement. All ground states of permutational invariant systems therefore lie on the submanifold of separable states which indeed have an efficient representation (only $N$ vectors, one for each individual spin, have to be specified); this is the equivalent statement as saying that mean-field theory becomes exact in the thermodynamic limit. This can also be understood in the context of the monogamy property of entanglement \cite{coffman00,OsborneVerstraete}: a spin has only a finite entanglement susceptibility, and if it has to share the entanglement with infinitely many other spins then the amount of entanglement between 2 spins will be go to zero \footnote{In the case of a translational and rotational invariant system on a regular lattice, the finite versions of the de-Finnetti theorem \cite{koenig05,chistandl06} allow one to find lower bounds on the distance of the reduced 2-body density operators to the separable ones; a scaling of the form $1/c$ is obtained where $c$ is the coordination number. See also appendix A.}.

In the more general situation when no permutation symmetry is present but only a smaller symmetry group such as the group of translations, the characterization of the relevant manifold is much harder. In contrast to the case of permutation symmetry, where every pure state of $N$ spin $s$ systems can be written as a linear combination of at most $N^{2s}/(2s)!$ \emph{Dicke} states \footnote{Dicke states are well studied in the context of quantum optics, and are obtained by taking the linear superposition of all permutations of a pure separable state in the standard basis; in the case of spin 1/2, you can label them by their expectation value of $\sum_iS^z_i$. See e.g. \cite{stockton-2003-67} for an extensive discussion of their entanglement properties.}, the size of a complete basis of translational invariant states is exponentially increasing as $(2s+1)^N/N$, and hence the manifold of interest has exponentially many parameters. But ground states of local Hamiltonians have many more nongeneric properties, most notably the fact that they have extremal local properties such as to minimize the energy: as the energy is only dependent on the local properties, and the ground state is determined by the condition that its energy is extremal, ground states have extremal local properties and the global properties only emerge to allow for these local properties to be extremal. As an example, let us consider a spin $1/2$ antiferromagnetic chain with associated Hamiltonian
\begin{displaymath}
\mathcal{H}_{\rm{Heis}}=\sum_{\langle i,j\rangle}\vec{S}_i\vec{S}_j
\end{displaymath}
where the notation $\langle i,j\rangle$ denotes the sum over nearest neighbours. Each individual term in the Hamiltonian corresponds to an exchange interaction and would be minimized if spins $i$ and $j$ are in the singlet state
\begin{displaymath}
|\psi\rangle=\frac{1}{\sqrt{2}}\left(|01\rangle-|10\rangle\right),
\end{displaymath}
but due to the monogamy or frustration properties of entanglement, a spin $1/2$ cannot be in a singlet state with more than one neighbour. As a result, the ground state becomes a complicated superposition of all possible singlet coverings, and as an interesting by-product quasi long-range order may arise. The important point, however, is that this wavefunction arises from the condition that its local properties are extremal: finding ground states of local translational invariant 2-body Hamiltonians is equivalent to characterizing the convex set of 2-body density operators compatible with the fact that they originate from a state with the right symmetry (see appendix~\ref{sec:reddensityop}).

Clearly, we would like to parameterize the manifold of states $\{|\psi_{ex}\rangle\}$ with extremal local properties. In practice, it is enough to parameterize a manifold of states $\{|\psi_{appr}\rangle\}$ such that there always exist a large overlap with the exact states $\{|\psi_{ex}\rangle\}$:  $\forall |\psi_{ex}\rangle, \exists |\psi_{appr}\rangle : \||\psi_{ex}\rangle-|\psi_{appr}\rangle\|\leq \epsilon$. Let us consider any local  Hamiltonian of $N$ spins that exhibits the property that there is a unique ground state $|\psi_{ex}\rangle$ and that the  gap is $\Delta(N)$. Let us furthermore consider the case when $\Delta(N)$ decays not faster than an inverse polynomial in $N$ (this condition is satisfied for all gapped systems and for all known critical translational invariant systems in 1D). Then let us assume that there exists a state $|\psi_{appr}\rangle$ that reproduces well the local properties of all nearest neighbour reduced density operators: $\|\rho_{appr}-\rho_{ex}\|\leq \delta$. Then it follows that the global overlap is bounded by
\begin{displaymath}
\||\psi_{ex}\rangle-|\psi_{appr}\rangle\|^2\leq \frac{N\delta}{\Delta(N)}.
\end{displaymath}
This is remarkable as it shows that it is enough to reproduce the local properties well to guarantee that also the global properties are reproduced accurately: for a constant global accuracy $\epsilon$, it is enough to reproduce the local properties well to an accuracy $\delta$ that scales as an inverse polynomial (as opposed to exponential) in the number of spins. This is very relevant in the context of variational simulation methods: if the energy is well reproduced and if the computational effort to get a better accuracy in the energy only scales polynomially in the number of spins, then a scalable numerical method can be constructed that reproduces all global properties well (here scalable means essentially a polynomial method).

The central question is thus: is it possible to find an efficient parameterization of a manifold of states whose local properties approximate well all possible local properties? A very interesting new development, mainly originating in the field of quantum information and entanglement theory, has shown that this is indeed possible. The main idea is to come up with a class of variational wave functions that captures the physics of the low-energy sector of local quantum spin Hamiltonians.

So what other properties do ground states of local quantum Hamiltonians exhibit besides the fact that all global properties follow from their local ones? The key concept to understand their structure is to look at the amount of entanglement present in those states~\cite{vidallatorre03}: entanglement is the crucial ingredient that forces quantum systems to behave differently than classical ones, and it is precisely the existence of entanglement that is responsible for such exotic phenomena like quantum phase transitions and topological quantum order~\cite{wen,levinwen06}. It is also the central resource that gives rise to the power of quantum computing~\cite{nielsenchuang}, and it is known that a lot of entanglement is needed between the different qubits as otherwise the quantum computation can be simulated on a classical computer \cite{jozsa02,vidal03}. This is because the amount of entanglement effectively quantifies the relevant number of degrees of freedom that have to be taken into account, and if this is small then the quantum computation could be efficiently simulated on a classical computer. In the case of ground states of strongly correlated quantum many-body systems, there is also lot's of entanglement (in the case of pure states, the connected correlation functions can be nonzero iff there is entanglement), but the key question is obviously to ask how much entanglement is present there: maybe the amount of entanglement is not too big such that those systems can still be simulated classically?

\begin{figure}[t]
\centering
\includegraphics[width=\linewidth]{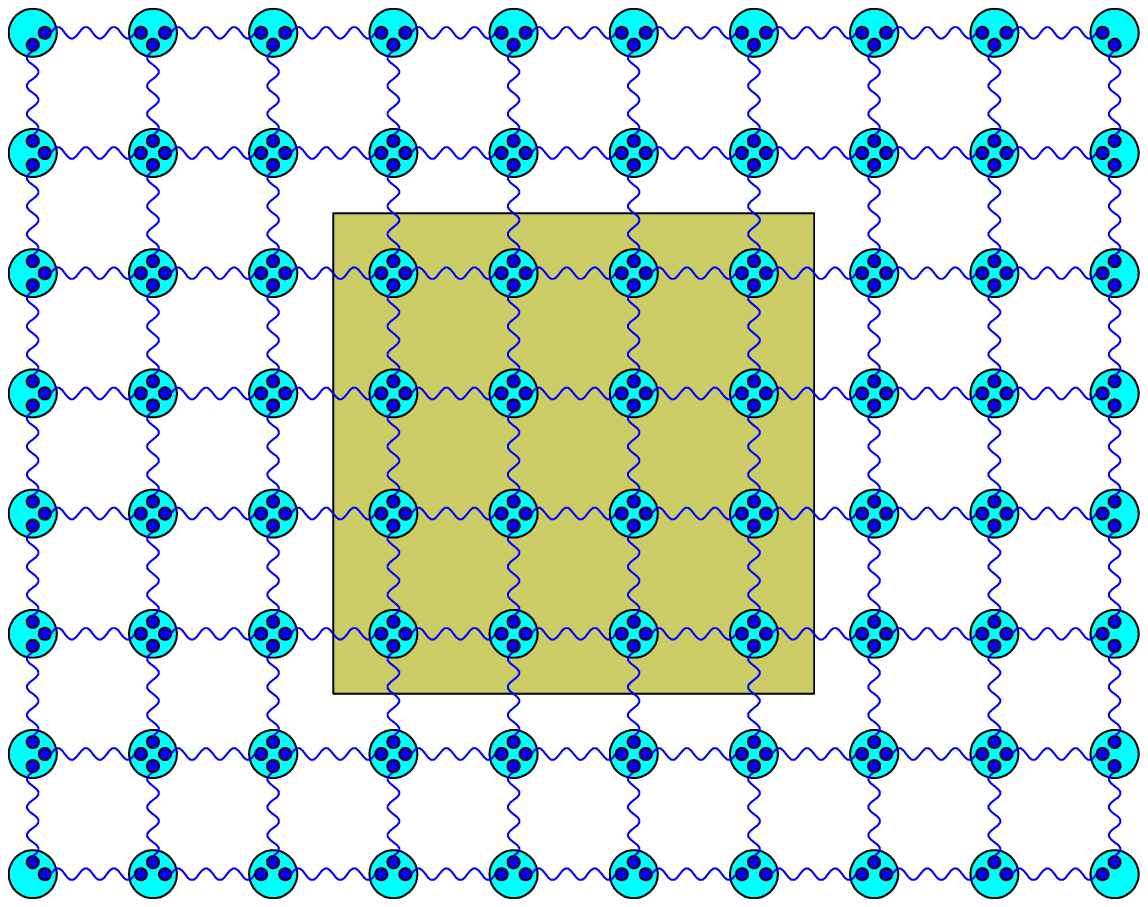}
\caption{The entropy of a block of spins scales like the perimeter of the block.}
\label{figarealaw}
\end{figure}

Let us for example consider a quantum spin system on a n-dimensional infinite lattice, and look at the reduced density operator $\rho_L$ of a block of spins in an $L\times L\times ...L$ hypercube (Figure \ref{figarealaw}). The von-Neumann entropy of $\rho_L$ is a coarse-grained measure that quantifies the number of modes in that block that are entangled with the outside~\cite{nielsenchuang}, and the relevant quantity is to study how this entropy scales with the size of the cube. This question was first studied in the context of black-hole entropy~\cite{bombelli86,sredniicki93,hooft85,callan94,fiola-1994-50} and has recently attracted a lot of attention~\cite{vidallatorre03,plenio-2005-94,calabrese04}. Ground states of local Hamiltonians of spins or bosons seem to have the property that the entropy is not an extensive property but that the leading term in the entropy only scales as the boundary of the block (hence the name area-law):
\bea
S(\rho_L)\simeq c L^{n-1}.
\label{arealaw}
\eea
This has a very interesting physical meaning: it shows that most of the entanglement must be concentrated around the boundary, and therefore there is much less entanglement than would be present in a random quantum state (where the entropy would be extensive and scale like $L^n$). This is very encouraging, as it indicates that the wavefunctions involved exhibit some form of locality, and we might be able to exploit that to come up with efficient local parameterizations of those ground states.

The area law (\ref{arealaw}) is mildly violated in the case of 1-D critical spin systems where the entropy of a block of spins scales like~\cite{vidallatorre03,calabrese04}
\begin{displaymath}
S(\rho_L)\simeq \frac{c+\bar{c}}{6}\log L,
\end{displaymath}
but even in that case the amount of entanglement is still exponentially smaller then the amount present in a random state. It is at present not clear to what extent such a logarithmic correction will occur in the case of higher dimensional systems: the block-entropy of a critical 2-D system of free fermions scales like $L\log L$ \cite{wolf06,gioev06,barthel-2006-74,cramer-2007-98}, while critical 2-D spin systems were reported where no such logarithmic correction are present~\cite{verstraetewolf06}, but in any case the amount of entanglement will be much smaller than for a random state. It is interesting to note that this violation of an area law is a pure quantum phenomenon as it occurs solely at zero temperature: in a recent paper \cite{wolfverstrate07}, it has been shown that the block entropy, as measured by the mutual information (which is the natural measure of correlations for mixed states), obeys an exact area law for all local classical and quantum Hamiltonians. The logarithmic corrections therefore solely arise due to the zero-temperature quantum fluctuations.  From the practical point of view, that might indicate that thermal states at low temperature are simpler to simulate than exact ground states.

\begin{figure}[t]
  %\centering
  \includegraphics[width=\linewidth]{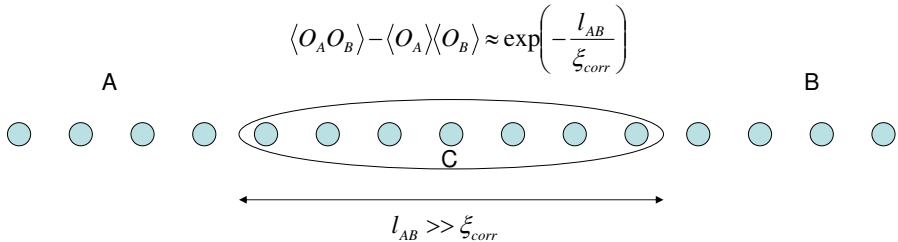}
  \caption{A one dimensional spin chain with finite correlation length $\xi_{corr}$; $l_{AB}$ denotes the distance between the block $A$ (left) and $B$ (right). Because $\l_{AB}$ is much larger than the correlation length $\xi_{corr}$, the state $\rho_{AB}$ is essentially a product state.}
  \label{figurecorr}
\end{figure}

The existence of an area law for the scaling of entropy is intimately connected to the fact that typical quantum spin systems exhibit a finite correlation length. In fact, M. Hastings has recently proven that all connected correlation functions between two blocks in a gapped system have to decay exponentially as a function of the distance of the blocks~\cite{hastings-2004-93}. Let us therefore consider a 1-D gapped quantum spin system with correlation length $\xi_{corr}$. Due to the finite correlation length, the reduced density operator $\rho_{AB}$ obtained when tracing out a block $C$ of length $l_{AB}\gg \xi_{corr}$ (see figure \ref{figurecorr}) will be equal to
\begin{equation}
\rho_{AB}\simeq \rho_A\otimes\rho_B\label{hasg}
\end{equation}
up to exponentially small corrections
\footnote{From a purely mathematical point of view, this is not correct as, surprisingly, there exist states for which all connected correlations functions are negligable while they are very far from being tensor product states \cite{hayden03,hastings07}: examples exist with negligable correlation functions but with the mutual information in the order of the number of qubits in every block. To remedy this, we pointed out in a recent paper that a more sensible way of defining a correlation length is by using the concept of mutual information, and when there is an exponential decay of mutual information, then the state $\rho_{AB}$ is guaranteed to be close to a tensor product state \cite{wolfverstrate07}.}.
The original ground state
$|\psi_{ABC}\rangle$ is a purification of this mixed state, but it is of course also possible to find another purification of the form $|\psi_{AC_l}\rangle\otimes|\psi_{BC_r}\rangle$ (up to exponentially small corrections) with no correlations whatsoever between $A$ and $B$; here $C_l$ and $C_r$ together span the original block $C$. It is however well known that all possible purifications of a mixed state are equivalent to each other up to local unitaries on the ancillary Hilbert space. This automatically implies that there exists a unitary operation $U_C$ on the block $C$ (see figure \ref{figurecorr}) that completely disentangles the left from the right part:
\begin{displaymath}
I_A\otimes U_{C}\otimes I_B|\psi_{ABC}\rangle\simeq|\psi_{AC_l}\rangle\otimes|\psi_{BC_r}\rangle.
\end{displaymath}
This implies that there exists a tensor $A_{\alpha,\beta}^i$ with indices $1\leq \alpha,\beta,i\leq D$ (where $D$ is the dimension of the Hilbert space of $C$) and states $|\psi^A_\alpha\rangle,|\psi_i^C\rangle,|\psi^B_\beta\rangle$ defined on the Hilbert spaces belonging to $A,B,C$ such that
\begin{displaymath}
|\psi_{ABC}\rangle\simeq \sum_{\alpha,\beta,i}A^i_{\alpha,\beta}|\psi^A_\alpha\rangle|\psi^C_i\rangle|\psi^B_\beta\rangle.\label{firstMPS}
\end{displaymath}
Applying this argument recursively leads to a matrix product state (MPS) description of the state (we will define those MPS later) and gives a strong hint that ground states of gapped Hamiltonians are well represented by MPS. It turns out that this is even true for critical systems \cite{verstraetecirac05}; a proof is presented in appendix B.

A remarkable feature of any gapped spin chain is thus that one could imagine dividing the whole chain in segments of size $l\gg\xi_{corr}$, and then apply a disentangling operation on all blocks in parallel. This would then lead to a product state of many parts, and as such gives a procedure of how such a ground state could be prepared using a quantum circuit with a logical depth that is only dependent on the correlation length (and independent of the number of spins!).

In the case of critical systems, we do not expect that this disentangling procedure works as correlations on all length scales appear. However, G. Vidal showed how this can be remedied by introducing some more advanced disentangling scheme that acts on many different length scales \cite{vidal05}. Basically, the idea is follow up a disentangling step by a coarse-graining step, and do this recursively until there is only one spin left. The procedure for doing so is called the Multiscale Entanglement Renormalization Ansatz (MERA), and may lead to alternative methods for simulating quantum spin systems.

It is interesting to note that the situation can again be very different in two dimensions: in that case, gapped quantum spin systems can exhibit topological quantum order, and in \cite{bravyi06} it was proven that the depth of any quantum circuit preparing such a topological state has to scale linearly in the size of the system. In other words, it is impossible to device a scheme/pattern by which one could disentangle such states in a parallel way as can be done in 1-D. However, also in this 2-D case it is possible to come up with a variational class of states, the so-called projected entangled pair states \cite{verstraetecirac04,murgverstraete07}, that captures the essential physics for describing those systems. By a recent argument of M. Hastings \cite{hastings06,hastings07}, it can again be proven that basically every ground state of a local Hamiltonian can be well represented by a state within this class.

\newpage
\section{Wilson's numerical renormalization group method} \label{sec:nrg}

The exact description of strongly correlated condensed matter systems poses formidable difficulties due to the exponentially large dimension of the associated Hilbert space. However, K. Wilson was the first one to understand that the locality of the interactions between particles or modes enforces ground states to be of a very specific form and could be exploited to simulate them. This insight led to the development of numerical renormalization group algorithms (NRG) \cite{wilson75}. The reason for the remarkable accuracy of NRG if applied to quantum impurity problems (e.g. the Kondo and Anderson Hamiltonians \cite{kondo64,krishnamurthy80}) can be traced back to the ability of mapping the related Hamiltonians to momentum space such that they become inhomogeneous 1-D quantum lattice Hamiltonians with nearest neighbor interactions\footnote{The basic reason why a 1-D model is obtained even though the original model is concerned with a magnetic impurity in a 3-D system is that the Kondo Hamiltonian only affects the s-wave part of the wavefunction; the s-wave modes are therefore dominant at low energies.}. NRG is then a recursive method for finding the low-energy spectrum of such Hamiltonians, and yields very accurate results when there is a clear separation of energies, reflected by e.g. an exponential decay of the couplings within the 1-D hopping Hamiltonian. The NRG method then recursively diagonalizes the Hamiltonian from large to small energies: at each iteration, a tensor product of the larger energy modes with lower energy modes is made and then projected on a subspace of the lower energy modes of the combined system. Thereafter the Hamiltonian is rescaled. Hence the basic assumption is that the low energy modes are affected by their high energy counterparts, but not vice-versa.

Let us illustrate Wilson's method on the hand of the NRG treatment of the  impurity Anderson model (SIAM). This model can be mapped to a hopping Hamiltonian after a logarithmic discretization of the conduction band \cite{krishnamurthy80}:
\begin{eqnarray}
 \mathcal{H}&=&\sum_{n=0}^{N}\xi_n
 \left(f^\dagger_{n\mu}f_{(n+1)\mu}+\rm{h.c.}\right)+\frac{1}{D}\left(\epsilon_d+\frac{U}{2}\right)c^\dagger_{d\mu}c_{d\mu}\nonumber\\
&&+\sqrt{\frac{2\Gamma}{\pi
D}}(f^\dagger_{0\mu}c_{d\mu}+\rm{h.c.})+\frac{U}{2D}\left(c^\dagger_{d\mu}c_{d\mu}-1\right)^2\label{SIAM}\\
\xi_n&=&\frac{\Gamma^{-n/2}}{2}\frac{(1+\Gamma^{-1})(1-\Gamma^{-n-1})}{\sqrt{(1-\Gamma^{-2n-1})(1-\Gamma^{-2n-3})}}\label{xi}
\end{eqnarray}
Here $\mu$ can take the values $\downarrow,\uparrow$, $c_{d\mu}$ denotes the annihilation operator of the impurity and $f_{n\mu}$  of the $n$'th fermion with spin $\mu$, summation over $\mu$ has been assumed, and $N\rightarrow\infty$. As the dimension of the associated Hilbert space is $2^{2N}$, an exact diagonalization is impossible, and approximations must be made. The hopping terms are decaying exponentially in $n$, and the basic idea of NRG is to treat the largest $n_0$ terms first which involves the diagonalization of a $2^{2n_0}$ matrix. Next we specify a control parameter $D\leq
2^{2n_0}$, retain only the eigenvectors
\begin{displaymath}
\{|\psi^{n_0}_\alpha\rangle\}_{\alpha=1..D}
\end{displaymath}
corresponding to the $D$ lowest eigenvalues, and project the first $n_0$ terms of the Hamiltonian onto that subspace using the projector
\begin{displaymath}
P^{[n_0]}=\sum_\alpha|\alpha\rangle\langle \psi^{n_0}_\alpha|
\end{displaymath}
yielding the $D\times D$ matrix $\mathcal{H}^{n_0}$. In the first step of the iteration, the extra term involving the coupling between the $n_0$'th and the $(n_0+1)'$th fermions is considered, and an exact diagonalization in the corresponding $4D$-dimensional Hilbert space is performed yielding the eigenvectors. To avoid that the dimensions of the effective Hamiltonian blow up, we project the Hamiltonian onto the new $D$-dimensional eigenspace corresponding to the lowest eigenvalues
\begin{displaymath}
P^{[n_0+1]}=\sum_\alpha|\alpha\rangle\langle\psi^{n_0+1}_{\alpha}|
\end{displaymath}
yielding $\mathcal{H}^{n_0+1}$. Note that $P^{[n_0+1]}$ is a $D\times 4D$ matrix, and for later reference  we write its coefficients in tensor form as $P^{i[n_0+1]}_{\alpha,\beta}$, $1\leq \alpha,\beta\leq D,1\leq i\leq 4$.  Now we iterate this procedure $N-n_0$ times, and NRG is typically said to have converged when $\Gamma\mathcal{H}^{N-1}=\mathcal{H}^{N}+\rm{cst}$ up to a unitary transformation. Note that the computational complexity of the NRG procedure scales as $ND^3$.

Let us now consider more closely the subspace on which we projected the original Hamiltonian. The $D$ states $\{|\psi^{N}_{\alpha_{N}}\rangle\}$ at the end of the iterations can be written as
\begin{equation} |\psi^{N}_{\alpha_{N}}\rangle=\sum_{\alpha_{n_0}...\alpha_{n_{N-1}}}\sum_{i_{n_0}i_{n_0+1}\ldots i_N}P^{i_{n_0}[n_0]}_{\alpha_{n_0}}P^{i_{n_0+1}[n_0+1]}_{\alpha_{n_0}\alpha_{n_0+1}}\ldots P^{i_{N}[N]}_{\alpha_{N-1}\alpha_N}|i_{n_0}\rangle|i_{n_0+1}\rangle
\ldots|i_N\rangle.
\label{MMMMPS}
\end{equation}
These states are exactly of the form as the ones mentioned in the introduction, and due to the feature that they are defined as a product of matrices, these are called matrix product states (MPS) \cite{oestlund95,rommer97,perezgarcia06,kluemper93,kluemper94} and were originally introduced in the mathematical physics community under the name of finitely correlated states \cite{fannes92,FCS89} (a precursor of it appeared in the context of quantum Markov chains \cite{Accardi-1981}). The energies calculated using NRG are thus energies of an effective Hamiltonian which is the original one projected onto a subspace of MPS. In practice, the NRG method is highly successful for problems where the different terms in the Hamiltonian act on a different scale of energy, and in that case results up to essentially machine precision can be obtained; this can only be true if the class of the matrix product states indeed capture all the physics needed to describe the low-energy physics of these Hamiltonians.

However, by looking at NRG by means of matrix product states, it is already clear that it can in principle be formulated as a variational method within the set of MPS. It turns out that this is exactly what S. White did by introducing the density matrix renormalization group (DMRG) \cite{white92,white92b}, but the way of looking at both methods from that point of view of MPS was only discovered much later~\cite{verstraeteweichselbaum05}.

\newpage
\section{Matrix product states and ground states of spin chains}

\subsection{Construction and calculus of MPS}

\subsubsection{The AKLT-model}

The notion of matrix product states (MPS) already appeared naturally in the section describing spin chains with finite correlation length and in the context of NRG. They were first studied in the work of Affleck, Kennedy, Lieb and Tasaki (AKLT)~\cite{affleck88}, where it was proven that the exact ground state of the spin-1 spin chain with Hamiltonian
\begin{displaymath}
\mathcal{H}_{AKLT}=\sum_{\langle
i,j\rangle}\underbrace{\left(\vec{S}_i.\vec{S}_j+\frac{1}{3}\left(\vec{S}_i.\vec{S}_j\right)^2+\frac{2}{3}\right)}_{=P_{ij}}
\end{displaymath}
can be parameterized exactly as a matrix product state. To see this, they observed that the terms $P_{ij}$ are projectors $\left(P_{ij}\right)^2=P_{ij}$ onto the 5-dimensional spin 2 subspace of 2 spin 1's, and proceeded by constructing the unique ground state $|\psi_{AKLT}\rangle$ which is annihilated by all projectors $P_{ij}$ acting on nearest neighbours. This state $|\psi_{AKLT}\rangle$ can be constructed as follows:
\begin{itemize}
\item Imagine that the 3-dimensional Hilbert space of the a spin 1 particle is effectively the low-energy subspace of the Hilbert space spanned by 2 spin 1/2's, i.e. the 3-D Hilbert space is the symmetric subspace of 2 spin 1/2 particles.

\item To assure that the global state defined on the spin chain has spin zero, let us imagine that each one of the spin 1/2's is in a singlet state with a spin 1/2 of its neighbours (see figure \ref{AKLT1}).

\begin{figure}[t]
  %\centering
  \includegraphics[width=\linewidth]{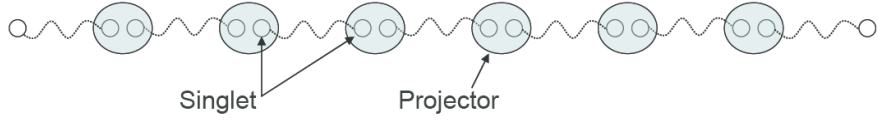}
  \caption{Building up the AKLT state by partial projections on bipartite singlets}
  \label{AKLT1}
\end{figure}

\item The AKLT state can now be represented by locally  projecting the pair of spin 1/2's in the symmetric subspace onto the spin-1 basis $\{|1\rangle,|0\rangle,|-1\rangle\}$:
\begin{eqnarray}
P&=&|-1\rangle\left(\frac{\langle 00|-\langle 11|}{\sqrt{2}}\right)+|0\rangle\left(\frac{\langle 01|+\langle
10|}{\sqrt{2}}\right)+|1\rangle\left(\frac{\langle 00|+\langle 11|}{\sqrt{2}}\right)\nonumber\\
&\equiv&\sum_{\alpha=x,y,z}|\alpha\rangle\left(\frac{\langle 00|+\langle 11|}{\sqrt{2}}\right)\tau_\alpha\otimes
\tau_y \nonumber
\end{eqnarray}
where $\tau_x=\sigma_x,\tau_y=i\sigma_y,\tau_z=\sigma_z$ with $\{\sigma_\alpha\}$ the Pauli matrices
and where we identified $|-1\rangle= |x\rangle, |0\rangle= |z\rangle, |1\rangle= |y\rangle$.
\end{itemize}

Historically, the AKLT state was very important as it shed new lights into the conjecture due to Haldane~\cite{haldane83,haldane83b} that integer spin Heisenberg chains give rise to a gap in the spectrum. That is a feature shared by all generic matrix product states: they are always ground states of local gapped quantum Hamiltonians.

Let us first try to rewrite $|\psi_{AKLT}\rangle$ in the MPS representation. Let us assume that we have an AKLT system of N spins with periodic boundary conditions; projecting the wavefunction in the computational basis leads to the following identity:
\begin{displaymath}
\langle \alpha_1,\alpha_2, ...\alpha_N|\psi_{AKLT}\rangle=\rm{Tr}\left(\tau_{\alpha_1}.\tau_y.\tau_{\alpha_2}.\tau_y...\tau_{\alpha_N}\tau_y.\right).
\end{displaymath}
The different weights can therefore be calculated as a trace of a product of matrices. The complete AKLT state can therefore be represented as
\begin{displaymath}
|\psi_{AKLT}\rangle=\sum_{\alpha_1,\alpha_2,...\alpha_N}\rm{Tr}\left(\tau_{\alpha_1}.\tau_y.\tau_{\alpha_2}.\tau_y...\tau_{\alpha_N}\tau_y.\right)|\alpha_1\rangle|\alpha_2\rangle...|\alpha_N\rangle
\end{displaymath}
which is almost exactly of the same form as the matrix product states introduced in \ref{MMMMPS}. The only difference between them is the occurence of the matrices $\tau_y$ between the different products. This is however only a consequence of the fact that we connected the different nodes with singlets, and we could as well have used maximally entangled states of the form
\begin{displaymath}
|I\rangle=\sum_{i=1}^2|ii\rangle
\end{displaymath}
and absorbing $\tau_y$ into the projector; this way we recover the standard notation for MPS.

So what did we learn from this AKLT-example? Basically, that there is a way of parameterizing the exact ground state of a particular strongly correlated quantum spin chain using a construction involving virtual bipartite entanglement and projections. From the point of view of quantum information theory, this parametrization is very appealing, as it gives an explicit way of constructing a highly entangled multipartite quantum state out of bipartite building blocks. More importantly, it is immediately clear how this picture can be generalized~\cite{verstraete-2004-92,verstraeteporras04}: instead of taking spin $D=2$ bonds corresponding to spin $1/2$'s, we can take much larger $D$ for the virtual spins, and furthermore it is obvious that the projectors can be replaced with any linear map. What is really exciting in doing so is the fact that the states arising from this are translational invariant by construction~\cite{fannes92}; this is highly relevant  as there does not seem to be another simple way of parameterizing translational invariant states. Recalling the discussion in section 1, it is now obvious that this class of translational invariant MPS , originally introduced in the literature under the name of finitely correlated states~\cite{fannes92}, would form a very good ansatz for parameterizing ground states. Indeed, ground states of local Hamiltonians are characterized by extremal local properties that are still compatible with the global translational symmetry, and as the MPS are build up by projecting the underlying maximally entangled states with extremal local correlations, it is very plausible that the MPS are perfectly suited for this\footnote{Besides the AKLT-model, many variations of that Hamiltonian have been studied with exact MPS-states as ground states \cite{Kol3,Kol2,Kol1}; this is particularly interesting because analytical solutions of spin chains are very rare.}.

Note that the number of parameters that we have to our disposition in these translational invariant MPS scales as $dD^2$ with $d$ the physical dimension of the spins (indeed, we have to parameterize $d$ matrices $A_\alpha$ of dimension $D \times D$), and hence the natural question to be asked now is how the convex set of all local reduced density operators so obtained compares to the exact convex set of all possible translational invariant systems. This problem is considered in Appendix \ref{sec:reddensityop} and \ref{sec:mpsgs}, and it is found that the number of parameters needed has to scale as a  constant or at most polynomially in the number of spins.

\subsubsection{Matrix Product States} \label{sec:mps}

\begin{figure}[t]
  %\centering
  \includegraphics[width=\linewidth]{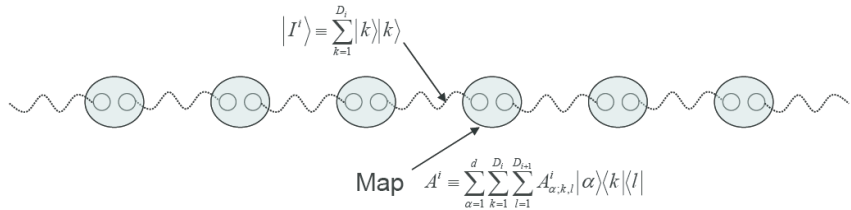}
  \caption{A general Matrix Product State}
  \label{AKLT2}
\end{figure}

As already explained in the previous section, the obvious generalization of the AKLT-states is obtained by making the dimensions of the virtual spins $D$ larger and to consider general linear maps $A_\alpha$ instead of projectors, but we can make a further generalization by making $D$ (the dimension of the virtual subsystems) and those maps site-dependent and write them as $D_i$ and $A^i_\alpha$. The most general form of a MPS on $N$ spins of dimension $d$ is then given by
\bea |\psi\rangle=\sum_{\alpha_1,\alpha_2,...\alpha_N}^d{\rm{Tr}}\left(A^1_{\alpha_1}A^2_{\alpha_2}...A^N_{\alpha_N}\right)|\alpha_1\rangle|\alpha_2\rangle...|\alpha_N\rangle \label{MPS}
\eea
The matrices $A^i$ have dimension $D_i\times D_{i+1}$ (here we take the convention that $D_{N+1}=D_1$),   and a system with open boundary conditions is obtained by choosing $D_1=1$ (i.e. no "singlet" between the endpoints). A pictorial representation is given in figure~\ref{AKLT2}. Before continuing, let us remark that every state of $N$ spins has an exact representation as a MPS if we let $D$ to grow exponentially in the number of spins; this can easily be shown by making use of the tool of quantum teleportation as shown in \cite{verstraeteporras04}. However, the whole point of MPS is that ground states can typically be represented by MPS where the dimension $D$ is small and scales at most polynomially in the number of spins; this is the basic reason why renormalization group methods are exponentially more efficient than exact diagonalization.

\subsubsection{Calculus of MPS}\label{calculus}

\begin{figure}[t]
  %\centering
  \includegraphics[width=\linewidth]{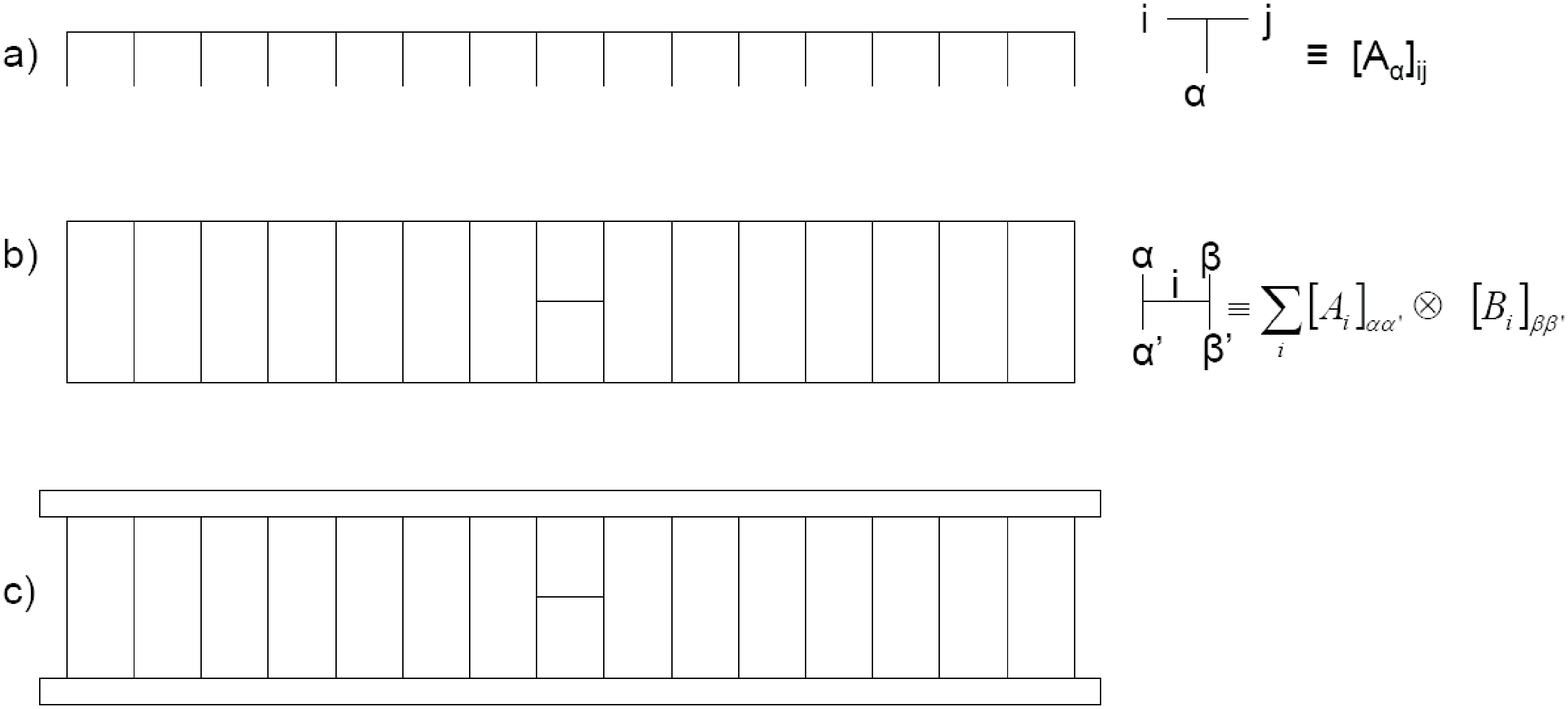}
  \caption{a) An MPS with open boundary conditions as a tensor network. The open bonds correspond to the uncontracted physical indices and the closed bonds to contracted indices arising from taking the products; b) calculating the expectation value of an nearest neighbour operator over a MPS with open boundary conditions by contracting the tensor network completely; c) an extra bond has to be contracted in the case of periodic boundary conditions.}
  \label{contraction}
\end{figure}

Let us next explain the calculus of those MPS. A crucial aspect of those MPS is indeed the fact that the expectation value of a large class of observables can be calculated efficiently. More specifically, this holds for all observables that are a tensor product of local observables, such as $\sigma_x^i\otimes \sigma_x^j$. The easiest way to understand this is by considering a so-called tensor network, in which connected bonds correspond to contracted indices and open bonds to uncontracted ones [see figure \ref{contraction}]. More specifically, the tensor network in figure \ref{contraction}a corresponds to the quantity
\begin{displaymath}
A^1_{\alpha_1}.A^2_{\alpha_2}...A^N_{\alpha_N}
\end{displaymath}
as the "virtual" indices are all contracted (corresponding to taking products of matrices) and the physical ones are left open. To calculate expectation values of observables, it is clear that we also have to contract those physical indices, after sandwiching in the observables, and hence expectation values can be represented by a completely contracted network. The relevant thing of course is now to quantify the computational complexity of actually doing those contractions. Naively, one expects that the calculation of the expectation value of a quantum state consisting of let us say $N$ spin $1/2$'s would involve a number of operations in the order of the size of the Hilbert space and hence exponential in $N$. This strategy would correspond to contracting the tensor network from top to bottom. However, we can be smarter and contract the tensor network from left to right such that we never have to store more than $D_i\times D_i$ variables during the computation (i.e. $D_i$ from the upper row and $D_i$ from then lower row).  Furthermore, the contraction of tensors from left to right can be done in a smart way. Let us e.g. assume that all virtual bonds have dimension $D$ and physical bonds dimension $d$;  the leftmost contraction corresponds to contracting the physical bond, leaving a tensor with two indices $i,j$ of dimension $D$ and $D$. Next we contract one of those with a tensor $A_i$, and thereafter with $\bar{A}_i$. The total number of multiplications that had to be done in that procedure is given by  $d^2D^3$, and this computational cost of contracting the whole  network is therefore $Nd^2D^3$ as we have to repeat the previous step $N$ times. If $D$ is not too big (i.e. $D<1000$), this can still be done efficiently on present day computers \footnote{Calculations up to $D=10000$ have been done in cases where very high precision was needed, but such cases are only tractable when explicitely making use of quantum numbers \cite{schollwoeck04}}.  Note that it is also obvious how to calculate the expectation value of any observable that is defined as a tensor product over local observables \footnote{Note also that any observable can be written as a sum of tensor products; this can easily be seen by using e.g. the singular value decomposition} ; this is useful to calculate e.g. the decay of correlations.

In the case of periodic boundary conditions, the network to contract is very similar (figure \ref{contraction}c) but one has to keep track of more variables (namely the ones going to the left and to the right), and the total computational cost amounts to $Nd^2D^5$; the extra factor of $D^2$ is responsible for the fact that simulations with periodic boundary conditions are considerably slower than with open boundary conditions~\cite{verstraeteporras04,white92,white92b}, but one has to note that the scaling is still only polynomial and that a $D$ in the order of $100$ is still achievable in principle, which is more than enough to get a good accuracy in actual calculations.

The following part of this subsection is pretty technical and can be skipped for a reader who is not really interested in technical details that are relevant for actually implementing numerical renormalization group methods. Let us first show how to efficiently calculate expectation values of generic Hamiltonians with only terms that act on nearest neighbour spins:
\begin{displaymath}
\mathcal{H}=\sum_{k=1}^{N}O^k+\sum_{k=1}^{N-1}\sum_\gamma B^k_\gamma\otimes B^{k+1}_\gamma.
\end{displaymath}
Instead of calculating the expectation value of each term in the sum separately, one can be a bit smarter and store the information of the result of the contraction from left to right and arrange everything such that the total computational cost is basically still $Nd^2D^3$, which in practice means a huge gain in computing time. See Appendix \ref{sec:matlab} for an actual implementation of this in matlab.

Another technical tool that is of great practical value is the fact that a matrix product state does not have a unique representation in terms of tensors $A^i$ but that so-called gauge transformation on the virtual level leave the physical state invariant. This follows from the fact that a multiplication of two matrices is left invariant when a nontrivial resolution of the identity is inserted between them:
\begin{eqnarray}
A B&=&A X X^{-1} B=A' B'\nonumber\\
A'&=&A X\nonumber\\
B'&=&X^{-1} B\nonumber
\end{eqnarray}
It happens that those gauge degrees of freedom can be exploited to make the forthcoming numerical optimization methods better conditioned. In the particular case of a MPS with open boundary conditions, we would like them to have some orthonormality properties. Consider e.g. again the numerical renormalization group of Wilson that was sketched in Section~\ref{sec:nrg}. There, one obtained collections of MPS $\{|\psi_{\alpha_k}^k\rangle\}$ at each step of the recursion such that they were all orthonormal to each other.  It is easy to see that a gauge transformation can always be found such that this is fulfilled in the case of MPS with open boundary conditions. Consider therefore  a MPS with tensors $A^1,A^2,...A^N$, and let us look at the two collections of $D_k$ MPS defined on the left and right parts of the qubits respectively  by opening the $k'$th contracted bonds
\begin{eqnarray}
|\psi^k_n\rangle&=&\sum_{\alpha_1\alpha_2...\alpha_k}A_{\alpha_1}A_{\alpha_2}...A_{\alpha_k}.e_n|\alpha_1\rangle|\alpha_2\rangle...|\alpha_k\rangle\\
|\chi^k_n\rangle&=&\sum_{\alpha_{k+1}e_n^T\alpha_{k+2}...\alpha_N}A_{\alpha_{k+1}}A_{\alpha_{k+2}}...A_{\alpha_N}|\alpha_{k+1}\rangle|\alpha_{k+2}\rangle...|\alpha_N\rangle
\end{eqnarray}
where $e_n$ denote the different unit vectors in a $D_k$ dimensional vector space. Let us furthermore consider the matrices
\begin{eqnarray}
A_{nn'}&=&\langle\psi^{k}_{n'}|\psi^k_n\rangle\\
B_{nn'}&=&\langle\chi^{k}_{n'}|\chi^k_{n}\rangle
\end{eqnarray}
and take their respective square roots $A=XX^\dagger$ and $B=YY^\dagger$. It is clear that if we would now do the gauge transformation $A^k_i\rightarrow A^k_i X^{-1}$ and repeat the procedure described above, we would see that the set $\{|\psi^{'k}_n\rangle\}$ would form an orthonormal set as $A'$ would be equal to the identity. Similarly, we could make the MPS at the right side orthonormal.

It is interesting to note that the Schmidt coefficients obtained by considering the bipartite cut over the $k$'th bond are precisely given by the singular values of the matrix $X Y^T$; indeed, the matrices $X^{-1}$ and $Y^{-T}$ were used to make the left and right hand side orthonormal, and the original MPS can of course be written as
\begin{displaymath}
|\psi\rangle=\sum_{n=1}^{D_k}|\psi^k_n\rangle|\chi^k_n\rangle.
\end{displaymath}
Writing out the singular value decomposition of $XY^T=U\Sigma V^T$ explicitely, a natural gauge transformation to represent the MPS with open boundary conditions would then be obtained by implementing the gauge-transformation
\begin{eqnarray*}
A^k_i&\rightarrow A^k_i X^{-1} U\\
A^{k+1}_i&\rightarrow V^T Y^{-T} A^{k+1}_i
\end{eqnarray*}
for all $k=1...N$ and by writing out the matrix product state in such a form that the Schmidt coefficients are appearing explicitely:
\begin{displaymath}
|\psi\rangle=\sum_{\alpha_1\alpha_2...\alpha_N}A^1_{\alpha_1}\Sigma_1A^2_{\alpha_2}\Sigma_2...A^{N}_{\alpha_N}|\alpha_1\rangle...|\alpha_N\rangle
\end{displaymath}
Here the $\Sigma_i$ are the diagonal matrices containing the Schmidt coefficients. This parametrization of MPS with open boundary conditions coincides exactly with the definition of the family of states introduced by G. Vidal in the context of the simulation of real time evolution of quantum many-body systems \cite{vidal04}, and we have just shown that every MPS with open boundary conditions can written like that.

It would be very appealing to have a similar parametrization in the case of MPS with periodic boundary conditions. However, it is clear that no notion of singular values or orthonormality can exist in that case as the left side is always connected to the right side. However, in a quantum spin system defined on a ring with a correlation length much shorter than the length of the ring, the boundary effects are expected not to be too pronounced, and an approximate orthonormalization can still be carried out~\cite{verstraeteporras04}.  This will be discussed in a later section. Related to this issue, let us suppose that we have a MPS with periodic boundary conditions for which we know that there exists a MPS description with all tensors $A^i$ equal to each other (the state is hence obviously translational invariant). But let us assume that we have a MPS description of that state with all tensors $A^i$ different from each other (i.e. the symmetry has been spoiled by site-dependent gauge transformations), e.g. as the result of a variational optimization. Is there a way to recover the symmetric description?
% -------------- vm --------------
For this, we have to find the gauge transformation $\tilde{A}^i_{\alpha} = X_i A^i_{\alpha} X^{-1}_{i+1}$, $i=1 \ldots N$, such that
\begin{displaymath}
\tilde{A}^1_{\alpha} = \tilde{A}^2_{\alpha} = \cdots = \tilde{A}^N_{\alpha}
\equiv \tilde{A}_{\alpha}.
\end{displaymath}
Without loss of generality, we can assume that $X_1$ is equal to the identity (indeed, we still have the freedom of a global gauge transformation). But then we can find $X_2$ and $X_N$ as the equations  $A_{\alpha}^1X_2^{-1}=X_NA_{\alpha}^N$, $\alpha=1..d$ has at least as many equations as unknowns (in practice, this should be done using a least squares algorithm). This can then be iterated until all gauge transformation $X_i$ have been determined.
Alternatively, the new matrices $\tilde{A}_{\alpha}$ can be found by considering the gauge-invariant normal form of the MPS: in terms of $A^i_{\alpha}$, the norm is expressed as ${\rm Tr}(E_1 \cdots E_N)$, with $E_i = \sum_{\alpha=1}^d A^i_{\alpha} \otimes \bar{A}^i_{\alpha}$. This expression is equal to the norm expressed in terms of $\tilde{A}_{\alpha}$, namely $\tilde{E}^N$, with $\tilde{E} = \sum_{\alpha=1}^d \tilde{A}_{\alpha} \otimes \tilde{A}_{\alpha}$. Therefore, the $\tilde{A}_{\alpha}$'s are obtained by taking the $N$.th root of $E_1 \cdots E_N$ and performing a Schmidt-decomposition of the result.

% --------------------------------

% ------------- old --------------
%For this, we have to find the gauge transformations $X_i$, $i=1...N$ such that
%\begin{displaymath}
%X_1A_{\alpha}^1X_2^{-1}=X_2A_{\alpha}^2X_3^{-1}=...=X_NA_{\alpha}^NX_1^{-1}.
%\end{displaymath}
%Without loss of generality, we can assume that $X_1$ is equal to the identity (indeed, we still have the freedom of a global gauge transformation). But then we can find $X_2$ and $X_N$ as the equations  $A_{\alpha}^1X_2^{-1}=X_NA_{\alpha}^N$, $\alpha=1..d$ has at least as many equations as unknowns (in practice, this should be done using a least squares algorithm). This can then be iterated until all gauge transformation $X_i$ have been determined.
% --------------------------------

To summarize, we have shown that MPS are very appealing from the computational point of view as all correlations functions and the expectation value of local Hamiltonians can be calculated efficiently. Moreover, we have shown how gauge transformations can be used to bring a MPS with open boundary conditions into a normal form.

A relevant observation is that we can also efficiently calculate the overlap between two different MPS. This makes the MPS-approach a very nice tool to detect quantum phase transitions, because small changes in the Hamiltonian lead to big changes in the ground state around those transition points, and those changes can be detected by calculating the overlap between the different MPS-approximations of the respective ground states \cite{cozzini-2006}.

A final remark is that it is also simple to calculate expectation values of Hamiltonians with long-range terms. This allows to easily generalize the methods introduced in later sections to situations of long-range Hamiltonians.

\subsubsection{Generalization of MPS}

The valence bond state representation can readily be generalized to trees or higher dimensions. The generalization to higher dimensions will be discussed in Section~\ref{sec:peps}, and there it will become clear that the calculation of expectation values is much more involved than in the 1-D case and can only be done approximately. In contrast to this, it is easy to see that the generalization of MPS to tree-networks without loops leads to a family of states from whom all expectation values can be calculated efficiently without having to make approximations \cite{FannesCayley}.

\begin{figure}[t]
  \centering
  \includegraphics[width=\linewidth]{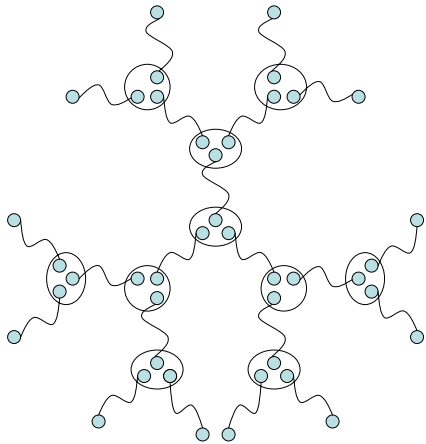}
  \caption{A MPS in the form of a Cayley tree with coordination number 3: just as in the case of spin chains, the big open circles represent projections of three virtual spins to one physical one.}
  \label{Figtree}
\end{figure}

Let us illustrate this for the particular case of a matrix product state on a Cayley tree with coordination number 3 (see figure \ref{Figtree}). To calculate expectation values of such a state $\langle\psi|O|\psi\rangle$, we start contracting the indices from the boundary inwards. Because there are no loops, it is easy to see that the number of variables to keep track off does not explode exponentially but remains bounded just like in the case of matrix product states. Let us for simplicity assume that all projectors $P$ are all equal to each other. At any point of the contraction when going from the outside to the inside, we have two incoming vertices and one outgoing one; as the earlier contractions of the incoming vertices did never entangle with each other, their joint state is in a product $\rho_k\otimes\rho_k$. Analogously as in the MPS case, the role of the projector $P$ is to apply a completely positive map on this product state: if we define the set of Kraus operators $A^i_{\alpha;\beta\gamma}=P^i_{\alpha\beta\gamma}$, then

\[\rho_{k+1}=\sum_{i=1}^d A_i\rho_k\otimes\rho_k A_i^\dagger\]

and yields the input for the next level of contractions. As the map from $\rho_k\rightarrow\rho_{k+1}$ is clearly a nonlinear completely positive map, its structure is much richer than in the case of matrix product states and associated completely positive linear maps \cite{inpreparation}: it happens that such nonlinear maps have multiple fixed points with different basins of attractions, and hence the boundary terms can affect the expectation values in the bulk of the system (note that this seems to be a consequence of the fact that we work with a Cayley tree, where the number of vertices increases exponentially as a function of the distance to the center of the tree).

Anyway, the important point to make is that expectation values of any product observable can be readily calculated efficiently by multiplying matrices with each other: in the case of a tree with coordination number $c$, the computational cost is $D^{c+1}$ with $D$ the dimension of the bonds (note that this yields $D^3$ for $c=2$ (i.e. MPS) and $D^4$ for the Cayley tree with $c=3$). Note also that it is straightforward to obtain a canonical normal form for the tree in the same way as we obtained one for the MPS by making use of the singular value decomposition \cite{ShiVidal2006}. Most of the simulation techniques described in the following sections can therefore also immediately be generalized to this setting of trees.

It can be seen that it is also possible to contract the tensor network exactly and efficiently if there are loops, but not too many of them. A clear demonstration of this is of course the one-dimensional case of MPS with periodic boundary conditions: numbering the d-dimensional spins from 1 to $2N$ in the case of a ring with an even number of sites, one can define $N$ new spins by merging spins $i$ and $2N-i+1$ into composite pairs. The corresponding MPS with open boundary conditions has spin dimension $d^2$ and bond dimension $D^2$, which can still be contracted efficiently. Similarly, we will be able to contract a generic tree with loops efficiently if it can be mapped onto a different tree without loops by merging collections of spins together; this procedure will still lead to a polynomial computational complexity iff the local dimension of the new spins is bounded by a polynomial in the size of the system. This condition is equivalent to the requirement that the maximal amount of spins that has been merged into a new superspin scales  at most logarithmic in the size of the system\footnote{It turns out that this notion of transforming an arbitrary graph into a tree by merging vertices together is a well studied problem in graph theory, and that the tree width of a graph follows from the optimal way of doing this such as to minimize the maximal number of vertices that has to be merged \cite{treewidth} over all possible mappings from a graph to a tree: the tree width is then the maximal number of vertices that is merged in this optimal solution.  Although calculating the treewidth of a general graph seems to be NP-hard, efficient (i.e. P) approximation algorithms  exist that approximate  the tree width to within a constant. The connection between tree width and MPS was pointed out in \cite{ShiMarkov}.} (which guarantees a complexity $poly(D,N)$).

\begin{figure}[t]
  %\centering
  \includegraphics[width=\linewidth]{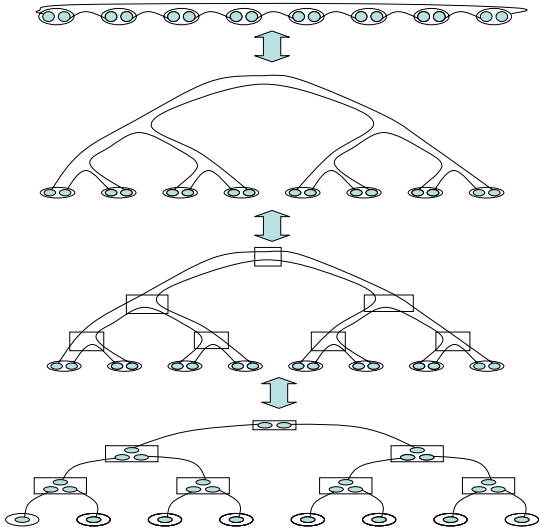}
  \caption{Every MPS can be represented in the form of a tree network in which the vertices are projectors on a 1-dimensional subspace (depicted as square boxes as opposed to ellipses for the case of projectors to physical spins); in the last step, it is understood that the original matrices $A^i_{\alpha\beta}$ are absorbed into the square blocks on the lowest level which are tensors with 6 D-dimensional indices.}
  \label{FigTree3}
\end{figure}

One can also formulate a variant of the tree network case, in which the nodes of the tree do not carry any physical spins but only the end points of the branches do; in other words, we start from a collection of virtual singlets, order them in the form of  a tree, and put a projector on all the vertices that maps the $D^c$ dimensional Hilbert space to a 1-dimensional one. Contraction of such a tensor network can again be done in an efficient way by starting the contraction at the leaves and working oneself inwards. An interesting observation is the fact that every MPS with a given $D$ can be expressed as such a Cayley tree with coordination number $c=3$ and bond dimension $D^2$ (this is true both for the case of open and periodic boundary conditions \footnote{Note that the opposite situation is different: if we want to represent a generic state on the Cayley tree of bond dimension $D$ as a MPS, the MPS will have a site-dependent bond dimension bounded above by $D^{log_2(N)}=N^{log_2(D)}$ (which is still polynomial).}).  This can easily be seen from the construction depicted in figure \ref{FigTree3}. Such a description was used to develop a formalism in which one can do successive renormalization  group transformations on quantum states as opposed on the usual level of Hamiltonians \cite{VerstraeteLatorre05}.

In some further extension, Vidal observed that more general types of tensor networks can be contracted efficiently: if all tensors appearing in the tree (with only physical spins at the bottom) are isometries, then one can additionally put a collection of so--called disentangling unitaries between the different branches. The reason why expectation values of local observables (or more generally observables that act nontrivially only on a constant number of spins) can still be calculated exactly is that in the expression $\langle\psi|O|\psi\rangle$ most of those unitaries disappear cancel each other and hence play no role: one only has to keep track of the unitaries within the \emph{lightcone} of the local observable, and it can be shown that the cost of this is still polynomial (albeit with a very large power, especially if one considers the 2-dimensional variants of this construction). This approach is called the {\em multi-scale entanglement renormalization ansatz} (MERA)~\cite{vidal05,vidal06}, and has a particularly  nice interpretation as  successive rescaling transformations in the sense of the renormalization group. Due to the presence of the tree network, the Schmidt number in the 1-D setting can grow logarithmically with the system size, and the method is therefore very promising for describing critical systems. As opposed to the class of MPS however, there does not seem to be a straightforward well conditioned way of doing a variational optimization over the class of MERA to find the MERA with minimal energy. Several promising approaches have been described in \cite{vidal05,vidal06,vidalMERA07}; the simplest approach is to parameterize the unitaries and isometries as exponentials of anti-Hermitean operators, and implement the steepest descent
optimization algorithm on those parameters \footnote{Such optimization techniques over the manifold of unitaries have been studied in great detail in the context of control theory \cite{Brockett} and are called flow equations;  they were also used in the context of entanglement theory \cite{Audenaert00} and first implemented on the level of MERA in \cite{flowOsborne}.}. Further extensions are possible, and there is currently an effort in trying to identify the broadest class of quantum states for which all local properties can be calculated efficiently (i.e. polynomial complexity). As an example, one can make use of the concept of matrix product operators to show that an efficient contraction is still possible when one acts with a quantum circuit of arbitrary depth but  consisting only of commuting gates (e.g. $exp(i\alpha \sigma^i_z\otimes\sigma^j_z)$) between any two spins  (even ones that are very far apart from each other) of an arbitrary MPS; such states can violate the area law in an extreme sense because they might lead to volume laws, and such states might be relevant in simulations of systems very far from equilibrium.

% --------------------------------

\subsection{Reformulating numerical renormalization group methods as variational
methods in the class of MPS}

Let us now look back at the numerical renormalization group, and reformulate it as a variational method within the class of matrix product states. To start with, let us aim to find the \emph{best
possible} description of the ground state of e.g. the Heisenberg model within the space of all MPSs of the form (\ref{MPS}), using the matrix elements of the matrices $\{ A^{n} \}$ as \emph{variational parameters} to minimize the energy (note that we do not impose the condition that the $A^n$ are projectors anymore). Using a Lagrange multiplier to ensure normalization, this leads to the following optimization problem:
\begin{displaymath}
\min_{|\psi^N\rangle\in\{ {\rm MPS_D} \}} \left[ \langle \psi^N |
\mathcal{H}^N | \psi^N \rangle -\lambda\langle\psi^N|\psi^N
\rangle \right].
\end{displaymath}
This cost function is multiquadratic in the $d N$ matrices $\{A_k\}$ with a multiquadratic constraint; indeed, every matrix appears only once in the bra and once in the ket, and so this problem is basically equivalent to a cost function of the form
\begin{displaymath}
\min_{x^1,x^2,...}\sum_{k_1,k_2,...l_1,l_2,...}x^1_{k_1} \bar{x}^1_{l_1}x^2_{k_2} \bar{x}^2_{l_2}\cdots Q_{k_1,k_2,...,l_1,l_2,...}
\end{displaymath}
where the $x^k$ are vectors with $d\times D_k\times D_{k+1}$ elements and the big tensor $Q$ contains the information of the Hamiltonian. The Lagrange constraint is of a similar form, and it that case $Q_{k_1,k_2,...}=\delta_{k_1l_2}\delta_{k_2l_2}...$. There is a standard way of solving such an optimization problem\footnote{Multiquadratic optimization problems can in principle be NP-hard to solve \cite{bental98}, and so there is no guarantee that the alternating least squares method will converge to the global optimum. However, in practice this does not seem to occur. See also \cite{eisert06}.} which is called alternating least squares (ALS). This ALS is an iterative method that works as follows: after making an initial guess of the vectors $x^k$, we keep $x^2,...,x^N$ fixed and optimize over $x^1$. That subproblem is of the form
\begin{displaymath}
\min_{x^1}  x^{1\dagger} H_{eff} x^1-\lambda x^{1\dagger} N_{eff} x^1
\end{displaymath}
(note that $H_{eff}$ and $N_{eff}$ depend on all the other $x^k$) and is exactly solvable as it is a quadratic problem with a quadratic constraints. The solution is that we have to choose $x$ equal to the smallest eigenvalue of the generalized eigenvalue problem $H_{eff}x=\lambda N_{eff}x$. A crucial point in all this is that there is an efficient way of calculating $H_{eff}$ and $N_{eff}$ given an MPS and the Hamiltonian: these are obtained by contracting tensor networks of the type discussed in \ref{calculus}, and an important point in the actual implementation of this is to store the relevant matrices for later use. In the next step, we will fix $x^1,x^3,x^4,...$ and repeat the same procedure, until we reach $N$ and repeat the procedure again by going from $N$ to $1$ and so on until convergence. Note that at each step of this procedure, the energy is going down and we are hence guaranteed to  converge to some value that is hopefully close to the  minimal energy of the Hamiltonian~\cite{verstraeteporras04}. We will call this procedure the variational matrix product state method (VMPS).

And what about excitations? Within the framework discussed until now, this can easily be done variationally~\cite{porrasverstraete06}. Suppose we found the MPS $|\psi_0\rangle$ with lowest energy, the variational problem is then to find the MPS $|\psi_1\rangle$ with the lowest energy that is orthogonal to the $|\psi_0\rangle$. This amounts to taking another Lagrange constraint in the optimization problem:
\begin{displaymath}
\min_{|\psi_1\rangle\in\{ {\rm MPS_D} \}} \left[ \langle \psi_1 |
\mathcal{H}^N | \psi_1 \rangle -\lambda\langle\psi_1|\psi_1\rangle-\mu\langle\psi_1|\psi_0\rangle \right]
\end{displaymath}
and this can still be solved iteratively in a very similar way. Basically, at every step we have a subproblem of the form
\begin{displaymath}
\min_{x}  x^\dagger H_{eff} x-\lambda x^\dagger N_{eff} x-\mu y^\dagger x
\end{displaymath}
whose solution is given by the solution to the generalized eigenvalue problem $PH_{eff}Px=\lambda PN_{eff}Px$ with $P$ the projector on the subspace orthogonal to the vector $y$. Doing this iteratively, this will again converge to a state $|\psi_1\rangle$ that is the best possible approximation of the first excited state using a MPS. Of course, a similar procedure can now be used to construct more excited states.

Let us now look back at the numerical renormalization group method of Wilson. In essence, the goal is to find the low-energy spectrum of the Hamiltonian. NRG can be understood as a different way of doing the optimization discussed above, but the NRG method is suboptimal and is only applicable in the particular case where there is a clear separation of energies (such as for a Kondo impurity or SIAM). Basically, the steps done in the case of NRG are equivalent to the steps one would do using the variational method discussed above during the first sweep from left to right. However, the NRG method is stopped after that first sweep, which effectively means that one never takes into account the influence of the low energy modes on the larger energy ones \footnote{This feedback may be small in practice, but it is not strictly zero, and its importance increases as the logarithmic
discretization is refined by taking $\Lambda \to 1$.}. This is clearly suboptimal, and a better result can be obtained by sweeping back and forth a few more times until complete convergence is obtained~\cite{verstraeteweichselbaum05}. The computational complexity of this is no greater than the one of the original NRG method.  The method obtained like that is then basically equivalent to the density matrix renormalization group (DMRG) introduced by S. White \cite{white92,white92b}, and it is indeed well known that DMRG has a much wider range of applicability than NRG. Note however that NRG and DMRG were always considered to be rather different, and it is only by reformulating everything in terms of MPS that the many similarities become apparent.

\begin{figure}[t]
\includegraphics[width=0.95\linewidth]{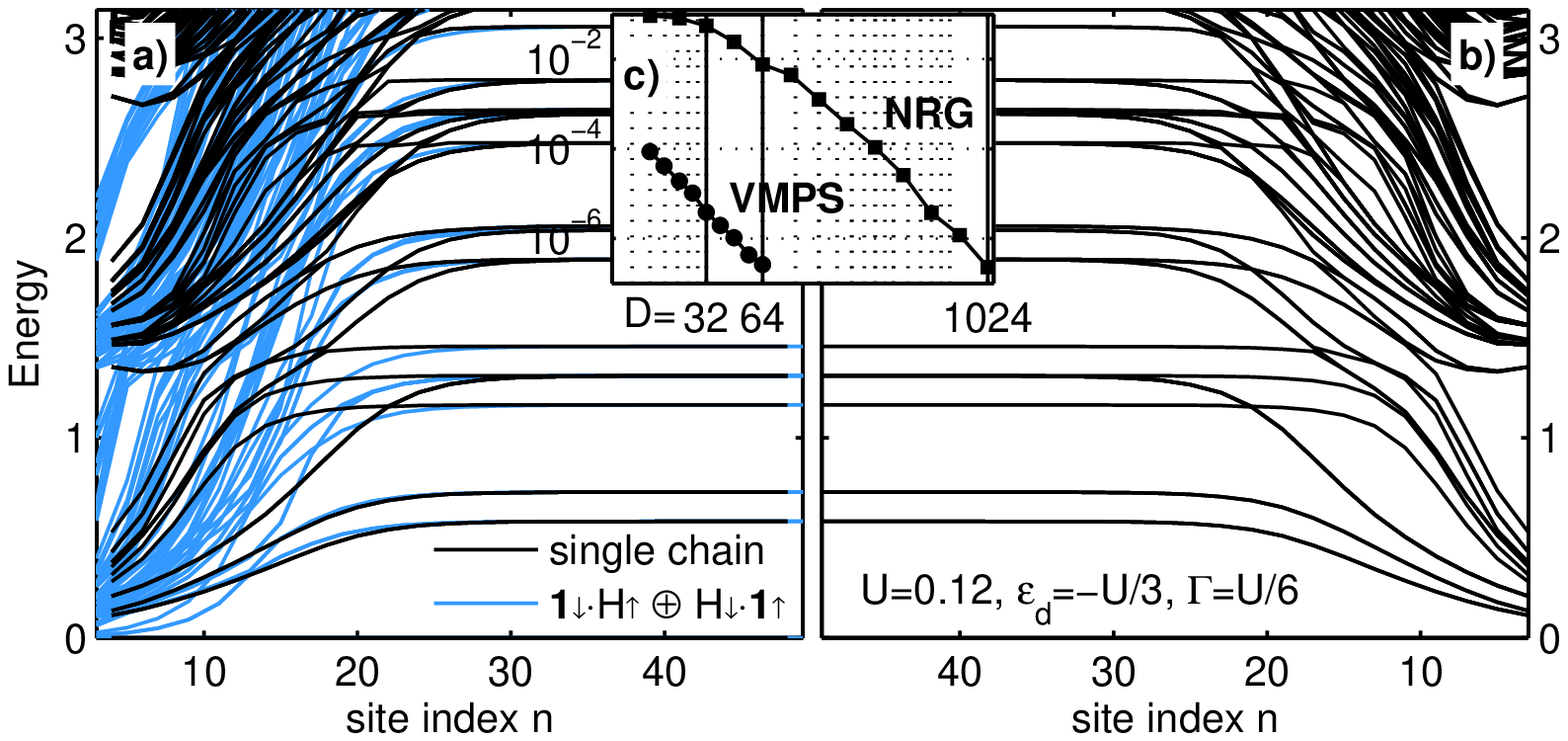}
\caption{(taken from \cite{verstraeteweichselbaum05})
Energy level flow of the SIAM as a
function of the site index $n$ calculated with (a) VMPS using
%$D_\MPS=32$ calculated after sweeping \cite{singlechain} and (b)
$D_\MPS=32$ calculated after sweeping and (b) NRG using $D_\NRG =
32^2=1024$. The blue lines in (a) denote the spectrum of ${\cal H}^{\rm
eff}_\uparrow \otimes 1 \hspace{-1.2mm} 1_\downarrow + 1
\hspace{-1.2mm} 1_\uparrow \otimes {\cal H}^{\rm eff} _\downarrow$
and demonstrate the decoupling of the $\uparrow$- and $\downarrow$
spin chains for large $n$.
% For comparison, NRG data for $D_\NRG=32$ is also shown in blue,
% illustrating that $D_\NRG$ is still too small.
The inset compares the ground state energy as function of $D$ for
VMPS (circles) and NRG (squares). The subtracted extrapolated
ground state energy is $E^* \simeq -3.0714$.}
\label{fig_Eflow}%
\end{figure}

Let us now come back to the variational method explained above for doing NRG. The effective Hamiltonian at chain length $n$, the central object in NRG, is given by $\tilde{\mathcal{H}}^n_{\alpha \beta} =\langle\tilde \psi^n_\alpha | \mathcal{H}^n | \tilde \psi^n_\beta\rangle$, and can hence also easily be recovered. Let us now illustrate the above by applying them to the SIAM given in equation (\ref{SIAM}). The following example is taken out of the paper~\cite{verstraeteweichselbaum05}. Since the Hamiltonian couples $\uparrow$ and $\downarrow$ band electrons only via the impurity, it is possible (see also \cite{raas04}) to ``unfold'' the semi-infinite Wilson chain into an infinite one, with $\uparrow$ band states to the left of the impurity and $\downarrow$ states to the right, and hopping amplitudes decreasing in both directions as $\Gamma^{-|n|/2}$.  Since the left and right end regions of the chain, which describe the model's low-energy properties, are far apart and hence interact only weakly with each other, the
effective Hamiltonian for these low energies will be of the form ${\cal H}^{\rm eff}_\uparrow \otimes 1 \hspace{-1.2mm} 1_\downarrow + 1 \hspace{-1.2mm} 1_\uparrow \otimes {\cal H}^{\rm
eff} _\downarrow$. This is illustrated by the black and blue lines in Fig.~\ref{fig_Eflow}(a).  Since ${\cal H}^{\rm eff}_\uparrow $ and ${\cal H}^{ \rm eff}_\downarrow$ can be calculated separately, instead of simultaneously as in typical NRG programs, the dimensions of the effective Hilbert spaces needed in the VMPS approach and NRG approaches to capture the low energy properties with the same precision are related by $D_{\rm MPS} \simeq \sqrt{ D_{\rm NRG}}$, implying significant computational gain with VMPS (i.e. a square root speed-up).

Figure \ref{fig_Eflow} compares the energy level flows calculated using NRG and VMPS. They agree remarkably well, even though we used $D_{\rm MPS}=32 = \sqrt {D_{\rm NRG}}$. The accuracy can still be improved significantly by using larger $D$ and smaller $\Lambda$ (and exploiting symmetries, not done here for VMPS).

\subsection{Density Matrix Renormalization Group} \label{sec:mps:dmrg}

As discussed in the previous section, DMRG can effectively be understood as the more advanced version of NRG in the sense that one also sweeps back and forth. Historically, the discovery of the DMRG algorithm had an enormous impact on the field of strongly correlated quantum spin systems as it lead to the first algorithm for finding the low energy spectrum of generic spin chains with local or quasi-local interactions. One of the first checks of the accuracy of the method was the calculation of the ground state energy of the critical spin $1/2$ Heisenberg chain, and a dazzling precision was obtained when  $D$ was taken in the order of a few hundreds. The DMRG method also allows one to calculate gaps and gave very strong evidence for the existence of the so-called Haldane-gap in the case of the spin $1$ Heisenberg spin chain.

Still, the variational matrix product state approach discussed in the previous section and the traditional DMRG differ in crucial details, and when looking back at the way DMRG has been understood and derived, it is not completely obvious to see the parallels with the VMPS approach. We will not explain here how DMRG works, but we refer to the many good review papers on the subject~\cite{schollwoeck04,peschel}. One important difference is e.g. that the VMPS approach works as well in the case of open as of periodic boundary conditions, whereas DMRG uses an ansatz that is specifically tailored for systems with open boundary conditions: the basic reason for this is that DMRG implicitly uses the orthonormalized normal form of MPS discussed in section~\ref{calculus}, but such a normal form does not exist in the case of periodic boundary conditions~\cite{verstraeteporras04}. In practice, simulations of spin chains with periodic boundary conditions have been done using DMRG, but this is precisely done in the way that was discussed in section 3.1.4 by folding the MPS with PBC to one with OBC by squaring the dimension of the virtual spins $D\rightarrow D^2$; the cost to pay is an algorithm whose computational cost scales as $D^6$ and, more importantly, for which it is not possible to formulate an ansatz of excited states with a definite momentum (see section 3.4).

Another difference has to do with the fact that the standard DMRG algorithm is set up in such a way that one performs a variational minimization over two sites together instead of over one site described in the VMPS. Such as step is then followed by performing a singular value decomposition and throwing away the least relevant modes; that step is strictly speaking not variational anymore, but there is evidence that the conditioning and the numerical convergence is better when doing it like this. Another reason  why this is done is that the so-called \emph{truncation error} gives some idea about the quality of the energy estimates obtained in a particular DMRG run. But from the point of view of VMPS, the one-site optimization procedure should perform better for the same computational cost, as the computational cost in the two-site setup goes up with a factor of $d^2$ and one could better use that time to increase $D$. This was already observed in the DMRG community \cite{Takasaki99,dukelsky98,WhiteScalapino93,schollwoeck04}, as DMRG can also readily be reformulated with one-site optimizations, but it took until 2005 and a nice paper S. White \cite{White2005} before the numerical conditioning problems were solved (those techniques of course also apply to VMPS) and the one-site DMRG became the standard DMRG method.

But instead of pointing out the differences, let us discuss a few technical improvements of VMPS that parallel the nice features of DMRG. Following Feynman's credo that you should never underestimate the pleasure of readers to read what they already know, let't repeat again how the whole VMPS procedure works. The goal is to minimize
\begin{eqnarray}
\min_{|\psi^N\rangle\in\{ {\rm MPS_D} \}} \left[ \langle \psi^N |
\mathcal{H}^N | \psi^N \rangle -\lambda\langle\psi^N|\psi^N
\rangle \right];\label{costDMRG}
\end{eqnarray}
which is a multiquadratic optimization problem and can  be solved using the alternating least squares method (ALS). This leads to the problem of solving  the generalized eigenvalue problem $H_{eff}x=\lambda N_{eff}x$ which is a standard optimization problem itself that is efficiently solvable if the condition number (i.e. smallest singular value) of $N_{eff}$ is not too small. It happens now that in the case of open boundary conditions, one can always make appropriate gauge transformations (see section \ref{calculus}) such as to assure that $N_{eff}$ is equal to the identity \footnote{Note that in the case of periodic boundary conditions, no such gauge transformation exists that makes $N_{eff}$ equal to the identity, and hence more heuristic methods have to be used. We will discuss this in more detail in the section dealing with periodic boundary conditions; that makes the VMPS algorithms with periodic boundary conditions a bit more tricky to implement, but the basic idea and resulting numerical accuracy is very similar.}. Those gauge transformations are always implicitly done in the case of DMRG, and should also be done in the VMPS approach as this obviously leads to the best conditioning (the computational cost for doing so is small; to see this in practice and also to see how easy this VMPS approach is to implement, we refer to appendix \ref{sec:matlab} for the explicit matlab code for doing so.).

Let us now see whether there is any good justification for using this ALS-method in the context of minimizing the ground state energy of local spin Hamiltonians, both used in the VMPS and in the DMRG approach. Suppose for example that we are optimizing over some site in the VMPS method and the corresponding $D=128$ while the physical spin dimension is $d=2$; then we are effectively doing an optimization over $m=\log_d(dD^2)+1=15$ sites, as the number of degrees of freedom we are varying over is equal to the number of degrees of freedom in $15$ qubits. Ground states of spin models are typically such that there is a finite correlations length (for critical systems, correlations are decaying pretty fast too), such that "boundary effects" of more than 7 sites away will not play a big role. Colloquially, the bulk convergence to a local optimum cannot occur if the gap in this system of $15$ qubits is larger than the effect from that boundary, and this gives a  handwaving argument why DMRG almost always converges to  the global minimum. In other words, the success of the MPS approach is related to its inherent capability of patching together solutions of local (e.g 15 sites) optimization problems, together with the fact that MPS are of course rich enough to approximate ground states of arbitrary ground states that obey an area law (see appendix B for a proof).

Before proceeding, let us look at a completely different way of doing variational optimizations using MPS when the Hamiltonian under interest is translational invariant, and let us consider the thermodynamic limit (i.e. infinitely many sites). We know that the ground state will be translational invariant, so we might well use an ansatz that reflects this by choosing all $A^i$ equal to each other. This approach has been originally proposed by Rommer and Ostlund \cite{rommer97} and later studied by several other authors (see \cite{dukelsky98} and references therein). The optimization problem is now reduced to minimizing the expectation value of one term in the Hamiltonian (i.e. energy per site), and this can be calculated as follows: first consider the "transfer matrix" $E=\sum_\alpha A_\alpha\otimes \bar{A}_\alpha$ and its eigenvalue decomposition $E=\sum_i \lambda_i|r_i\rangle\langle l_i|$ with the $\lambda_i$ in decreasing order. The energy can now be expressed as
\begin{displaymath}
E=\frac{1}{\lambda_0^2}\sum_{\alpha_1\alpha_2\alpha'_1\alpha'_2} H_{\alpha_1\alpha_2;\alpha'_1\alpha'_2}\langle l_0|\left(A_{\alpha_1}\otimes \bar{A}_{\alpha'_1}\right)\left(A_{\alpha_2}\otimes\bar{A}_{\alpha'_2}\right)|r_0\rangle
\end{displaymath}
where we assumed that the Hamiltonian is only acting on nearest neighbors. This cost function is clearly a very nonlinear function of the variables $A^i$, and standard techniques such as conjugate gradient methods can be used to minimize that expression. However, it happens that this optimization procedure may get stuck in local minima, and the situation only gets worse when increasing $D$. In comparison with the DMRG or VMPS approach, this method does not seem to work very well for several problems. At first sight, this seems to be strange as in those latter approaches one has much more variables (i.e. one tensor per site). However, this can be understood from the point of view of optimization theory, where it is standard practice to introduce more variables than needed such as to assure that the problem becomes better behaved. This is precisely what is happening here. However, as we will see later, the idea of using translational invariant MPS can turned into a successful algorithm by combining it with the concept of imaginary time evolution of MPS. It is a bit of a mystery why that approach works better than e.g. conjugate gradient methods, but it is probably related to the inherent robustness of algorithms evolving in imaginary time.

As a last remark, we note that it can be very useful from a computational point of view to exploit symmetries in the system. This can be done both in the DMRG and in the MPS-approach, and we refer to \cite{mcculloch-2002-57,singh-2007,white92b} for more details.

\subsubsection{Excitations and spectral functions}

An issue where the usefulness of VMPS really reveals itself is in the context of the study of excitations. In the standard DMRG approach, the idea is basically to store all the information about the ground state and excited states into one big MPS: during the iteration step at e.g. site $k$, one keeps all tensors $A^1,A^2,...A^{k-1},A^{k+1},...,A^N$ fixed and within the $dD^2$-dimensional subspace one identifies a number of lowest lying states. One then moves to the next site, and continues until convergence (again, for a more complete understanding of how DMRG works, we refer to the many review papers on the subject~\cite{peschel,schollwoeck04}). That procedure is clearly suboptimal and not variational anymore, as there is no reason why the same tensors should be used for the ground and excited states. In the worst case scenario, the tensors $A^i$ are block-diagonal, each block containing the information of a different excited state, and hence the computational cost scales badly with the number of excited states encoded like that (i.e. the cost of doing a simulation with $k$ excited states is $k^3$ times the cost of doing it for the ground state; also, the memory requirements scale as $k^2$).

Looking back at the VMPS approach described above however, this is a more natural way to deal with excitations at a lower cost, both with respect to memory and computational cost. To repeat again what was explained earlier, one can build up the spectrum in a sequential way: first look for the ground state, after this start over again and find the MPS with minimal energy that is furthermore orthogonal to the ground state, and so further~\cite{porrasverstraete06}. This procedure does not have any of the drawbacks mentioned above in the case of DMRG, and is fast and reliable, although it requires to run the whole variational procedure for each required excited state.

If the Hamiltonian under consideration has some symmetries, there might of course alternative ways of finding excited states. Consider e.g. the spin $1$ Heisenberg chain. It is well known that the ground state lives in the sector with total spin $0$, and the really interesting quantity in this case is to find the gap between this ground state and the state with minimal energy out of the spin $1$ sector. As the total spin commutes with the Hamiltonian, we can add a small magnetic field to the Hamiltonian that plays the role of a chemical potential, and within some parameter range, which can easily found by doing some numerics, we are guaranteed that the ground state of the complete Hamiltonian will have spin 1. It will however still be necessary to implement the procedure mentioned above if more excitation energies are to be found.

Let us continue now and find out whether the formulation of DMRG in terms of MPS also allows one to obtain lower bounds to the energies. More precisely, suppose the variational method converges to a MPS $|\psi\rangle$ with associated energy $E=\langle\psi|\mathcal{H}|\psi\rangle$. Let us next define the quantity
\begin{displaymath}
\epsilon=\sqrt{\langle\psi|\left(\mathcal{H}-E\right)^2|\psi\rangle}.
\end{displaymath}
It is a simple exercise to prove that there exists an exact eigenvalue $E_{ex}$ of
$\mathcal{H}$ such that $E\geq E_{ex}\geq E-\epsilon$. We will show that, in the
case of MPS, one can calculate the quantity $\epsilon$  at essentially the same
computational cost as $E$. This implies that the VMPS approach outlined allows to
get both upper and lower bounds to eigenvalues of a given Hamiltonian
$\mathcal{H}$; to our knowledge, this is a truly unique feature for a variational
method. In typical applications like the calculation of the ground state energy of
Heisenberg antiferromagnets, $\epsilon\simeq 10^{-8}$.

Let us now sketch how $\epsilon$ can be calculated efficiently~\cite{verstraeteweichselbaum05}. First of all, we note that expectation values of tensor products of local observables can be calculated by multiplying vectors by matrices:
\begin{eqnarray*}
\langle\psi|\hat{O}_1\otimes\hat{O}_2\cdots
\hat{O}_N|\psi\rangle&=&A^{[1]}_{\hat{O}_1}A^{[2]}_{\hat{O}_2}\cdots
A^{[N-1]}_{\hat{O}_{N-1}}A^{[N]}_{\hat{O}_N}\\
\left(A^{[n]}\right)_{\alpha\beta,\alpha'\beta'}&=&\sum_{ij}P^{i[n]}_{\alpha\alpha'}\bar{P}^{j[n]}_{\beta\beta'}\langle
i|\hat{O}_n|j\rangle
\end{eqnarray*}
Let us now try to evaluate an expression like $\langle\psi|\left(\mathcal{H}-E\right)^2|\psi\rangle$. For simplicity, let us assume that $\mathcal{H}$ represents a spin chains with only nearest neighbor couplings. Naively, one expects that one will have to evaluate in the order of $N^2$ expectation values of local observables. There is however a much more efficient method: going recursively from the first site to the last one, one keeps three $D^2$ dimensional vectors $v_0,v_1,v_2$ containing terms with respectively $0,1,2$ interaction terms (note that $\mathcal{H}^2$ contains at most two interaction terms). At each recursion step, one can easily update $v_0$ as a function of $v_0$ and the local terms of the MPS, and equivalently $v_1$ as a function of $v_1$ and $v_0$ plus the local Hamiltonian; $v_2$ can be updated as a function of the local MPS terms, the local Hamiltonian and $v_0,v_1,v_2$. Therefore the computational complexity  of calculating $\epsilon$ scales as $ND^3$, just as for the case of evaluating the energy $\langle\psi|\mathcal{H}|\psi\rangle$.  Another and perhaps more direct way to see this is do make use of the formalism of matrix product operators (see section 5): there is will be shown that every Hamiltonian with only nearest neighbour interaction has a very simple parametrization as a matrix product operator, and hence also the square of it.

As a side product, this insight allows one to devise efficient algorithms for calculating highly excited eigenstates and eigenenergies of Hamiltonians. Suppose for example that one would like to know the closest eigenstate and -energy to a certain prespecified energy $E_{sp}$. This could be interesting when one wants to calculate gaps in the spectrum. The variational problem that has to be solved in that case is the following:
\begin{displaymath}
\min_{|\psi\rangle\in\{MPS\}}\langle\psi|\left(\mathcal{H}-E_{sp}\right)^2|\psi\rangle-\lambda\langle\psi|\psi\rangle
\end{displaymath}
This is indeed again a multiquadratic problem that can be solved using the same
variational techniques as outlined before, yielding an algorithm with the same
computational complexity as the original one.

Similar techniques also allow us to calculate Green's functions in a variational way~\cite{verstraeteweichselbaum05}. Green's function have been calculated using NRG methods~\cite{costi96,costi98}, but because of the strong interplay between the different energy levels in that context, the MPS-approach should be able to give more precise results. The DMRG techniques that have been developed for calculating Green's functions of spin systems~\cite{hallberg95,kuehner99,jeckelmann02} are related to our approach but differ in several aspects.

\begin{figure}[t]
%\centering
\includegraphics[width=.95\linewidth]{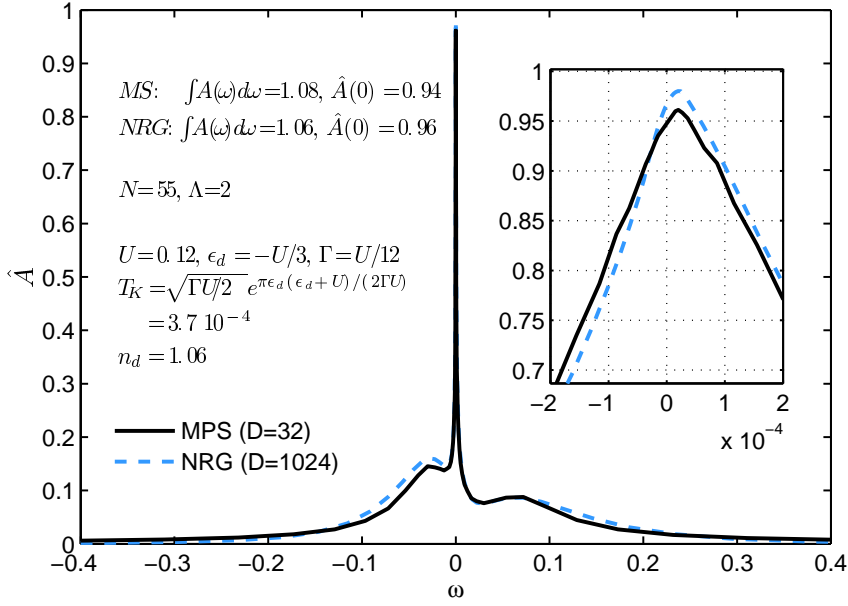}
\caption{(taken from \cite{verstraeteweichselbaum05})
Impurity \protect \rule[4mm]{0mm}{0mm}
spectral function for the SIAM, $\hat {\cal A}_\uparrow \left(\omega\right)
\equiv {\cal A}_\uparrow (\omega) \pi \Gamma / \sin^2(n_d \pi / 2)$,
normalized such that
$\hat {\cal A}_\uparrow (0)$ should be 1,
% with length $N=55$ and other parameters as specified,
calculated  with VMPS (solid)
and NRG (dashed line). } %Inset: blow-up of peak.}
\label{Fig_eDOS}
\end{figure}

The typical Green's functions of interest are of the form
\begin{displaymath}
G(\omega+i\eta)=\langle\psi|f \frac{1}{\mathcal{H}-\omega-i\eta} f^\dagger|\psi\rangle
\end{displaymath}
where $|\psi\rangle$ represents the ground state and $f^\dagger$ is a creation operator. Here we will assume that $|\psi\rangle$ has been calculated using the MPS-approach, and we showed that this can be done to a very good and controlled precision. The basic idea is now to  calculate the Green's function variationally within the set of MPS by finding the unnormalized MPS $|\chi\rangle$, commonly called a correction vector~\cite{verstraeteweichselbaum05}, that minimizes the weighted norm
\begin{displaymath}
\mathcal{N}=\left\||\chi\rangle-\frac{1}{\mathcal{H}-\omega-i\eta} f^\dagger|\psi\rangle\right\|_{W=\left(\mathcal{H}-\omega\right)^2+\eta^2}.
\end{displaymath}
Here we used the notation $\||\xi\rangle\|_W^2=\langle\xi|W|\xi\rangle$, and the weight $W>0$ was introduced to make the problem tractable (it should not really affect the precision as all positive definite norms are essentially equivalent). Writing $|\chi\rangle$ in its real and imaginary part
$|\chi\rangle=|\chi_r\rangle+i|\chi_i\rangle$ and assuming $\mathcal{H},|\psi\rangle$ real, this norm can be written as
\begin{eqnarray} \label{eqn:nrgnorm}
\mathcal{N}&=&\langle\chi_r|\left(\mathcal{H}-\omega\right)^2+\eta^2|\chi_r\rangle-2\langle\chi_r|\left(\mathcal{H}-\omega\right)f^\dagger|\psi\rangle\nonumber\\
&&+\langle\chi_i|\left(\mathcal{H}-\omega\right)^2+\eta^2|\chi_i\rangle-2\langle\chi_i|f^\dagger|\psi\rangle+\langle\psi|\psi\rangle
\end{eqnarray}
Minimizing $\mathcal{N}$ clearly involves two independent optimizations over $|\chi_r\rangle,|\chi_i\rangle$, which we will both parameterize as MPS. Both of the optimizations involve minimizing the sum of a multiquadratic and a multilinear term; as terms like $\langle\chi|\left(\mathcal{H}-\omega\right)^2|\chi\rangle$ can be calculated efficiently, we can again use the same tricks and keep all but one projectors $\{P^{[i]}\}$ fixed and optimize over the remaining one. As a multilinear term instead of a quadratic constraint is present, each iteration step can be done efficiently by solving a sparse linear set of equations \cite{verstraeteweichselbaum05}. Iterating this procedure until convergence, one can evaluate both $|\chi_r\rangle,|\chi_i\rangle$, and exactly determine the precision of the result by calculating the norm (\ref{eqn:nrgnorm}); if this norm is too large, one can always increase the control parameter $D$ associated to the MPS $|\chi_r\rangle,|\chi_i\rangle$. Finally, one can easily evaluate $G(\omega+i\eta)$ in function of $|\chi\rangle$. The precision of this evaluation can again be bounded. To illustrate this with an example, we again consider the SIAM model (see figure \ref{Fig_eDOS}).

\subsection{DMRG and periodic boundary conditions}

The variational matrix product state approach discussed in the previous sections is both applicable to the case of open (OBC) and periodic boundary conditions (PBC). The main difference between both methods is the computational cost of the contraction of the associated tensor network ($ND^3$ versus $ND^5$) and the fact that a useful orthonormalization can be used in the case of OBC. On the other hand, there are several reasons why simulations with PBC are preferred: 1. the boundary effects are much smaller than in the case with OBC. This has been identified as a serious problem for standard DMRG since its conception, as even for very long chains the boundary effects can be seen in the bulk.  2. Due to the fact that any nontrivial system with OBC cannot be translational invariant, it is not possible to study excitations with a definite momentum. Contrasting this to the case of PBC, it is possible there to construct a variational class of MPS with a definite momentum, allowing e.g. to obtain a convenient picture of the energy-momentum relation. The overall price to be paid however is that the computational cost for working with MPS with PBC scales as $D^5$ as opposed to $D^3$ for OBC.  Note that the traditional DMRG algorithm has also been used to treat systems with periodic boundary conditions. It is clear, however, from the structure of MPS that the only way to represent a MPS with PBC by one with OBC is to use the virtual bonds to "teleport" a maximally entangled state from the first to the last site. This leads to an effective $D$ that is equal to the original one squared (this is the only possibly way to accomodate the extra degrees of freedom to take into account this teleportation), and hence the computational cost for the same accuracy would scale as $D^6$. But even in doing so, it would be very hard to construct states with a definite momentum.

\begin{figure}[t]
\centering
\includegraphics[width=\linewidth]{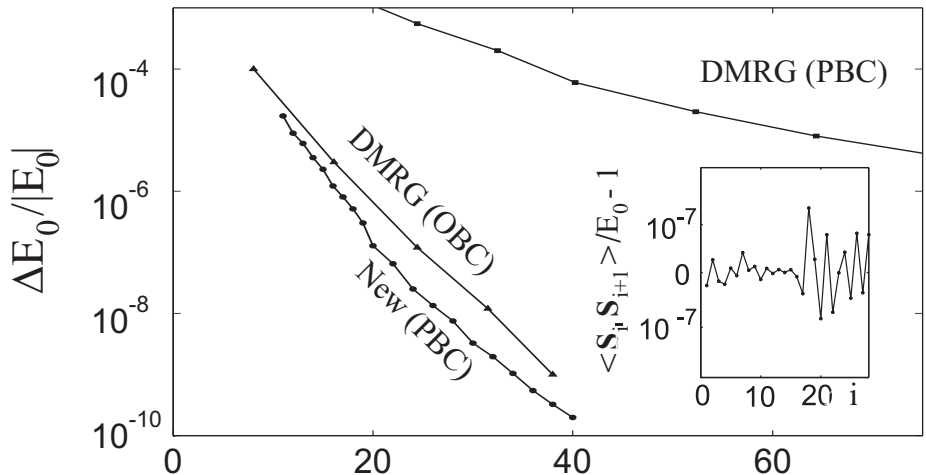}
\label{dmrgperiodic:fig1}
\caption{(taken from \cite{verstraeteporras04}) Comparison between DMRG (squares) \cite{white92b} and the (new) VMPS method (circles) for PBC, and $N=28$. For reference the DMRG results \cite{white92b} for the Heisenberg chain with OBC (triangles) are also shown. Inset: variation of the local bond strength from the average along the chain, calculated with the new method and $D=40$.}
\label{Diego1}
\end{figure}

When implementing the VMPS approach to minimize the energy of a given Hamiltonian with periodic boundary conditions~\cite{verstraeteporras04}, special precautions have to be taken such that the problem remains well conditioned. In practice, this means that we have to make sure that the matrix $N_{eff}$ in the generalized eigenvalue problem $H_{eff}x=\lambda N_{eff}x$ is positive definite and that its smallest eigenvalue is as big as possible. There are several options to achieve this. First of all, we can still play with the gauge conditions which would correspond to a transformation
\begin{displaymath}
N_{eff}\rightarrow \left(X\otimes Y\otimes I_d\right)N_{eff}\left(X^\dagger\otimes Y^\dagger\otimes I_d\right).
\end{displaymath}
Looking back at the case with OBC, the reason why we could always make $N_{eff}=I$ is that the left and right side of the chain factorize such that $N_{eff}$ was always a tensor product to start with. Here, the PBC enforce the existence of correlations between those sides, hence enforcing $N_{eff}$ to be "correlated". However, in practice the amount of correlations will be small (as these are finite size effects), and hence $N_{eff}$ will be close to a tensor product. A sensible way of choosing the gauge transformation therefore consists out of two steps: 1. find the tensor product $N_1\otimes N_2\otimes I_d$ that approximates best $N_{eff}$ (this can easily be achieved by doing a singular value decomposition in the space of Hermitean operators); 2. choose the gauge transformations $X,Y$ such that $XN_1X^\dagger=I=YN_2Y^\dagger$. During the first sweeps, when absolute accuracy is not relevant yet and $N_{eff}$ could still be far from a tensor product, one could e.g. add some identity to $N_{eff}$ such as to make it better conditioned ($N_{eff}\rightarrow N_{eff}+\epsilon I$). The procedure outlined here is only one of many possible ones, and it might well be that another choice of gauge conditions leads to better performance, but the one described here  seems to give good results in practice~\cite{verstraeteporras04}. An example is provided in figure \ref{Diego1} where we looked at the Heisenberg spin $1/2$ chain with $N=28$ (we choose this example with a small number of sites as we can still compare this with exact results, and for reference also the results of a DMRG calculation for the same Hamiltonian is provided). As seen in the figure, the scaling of the error for MPS with PBC nicely follows the scaling of the error obtained in a DMRG-calculation for a Hamiltonian with OBC.

\begin{figure}
  \center
  \includegraphics[width=5.5cm]{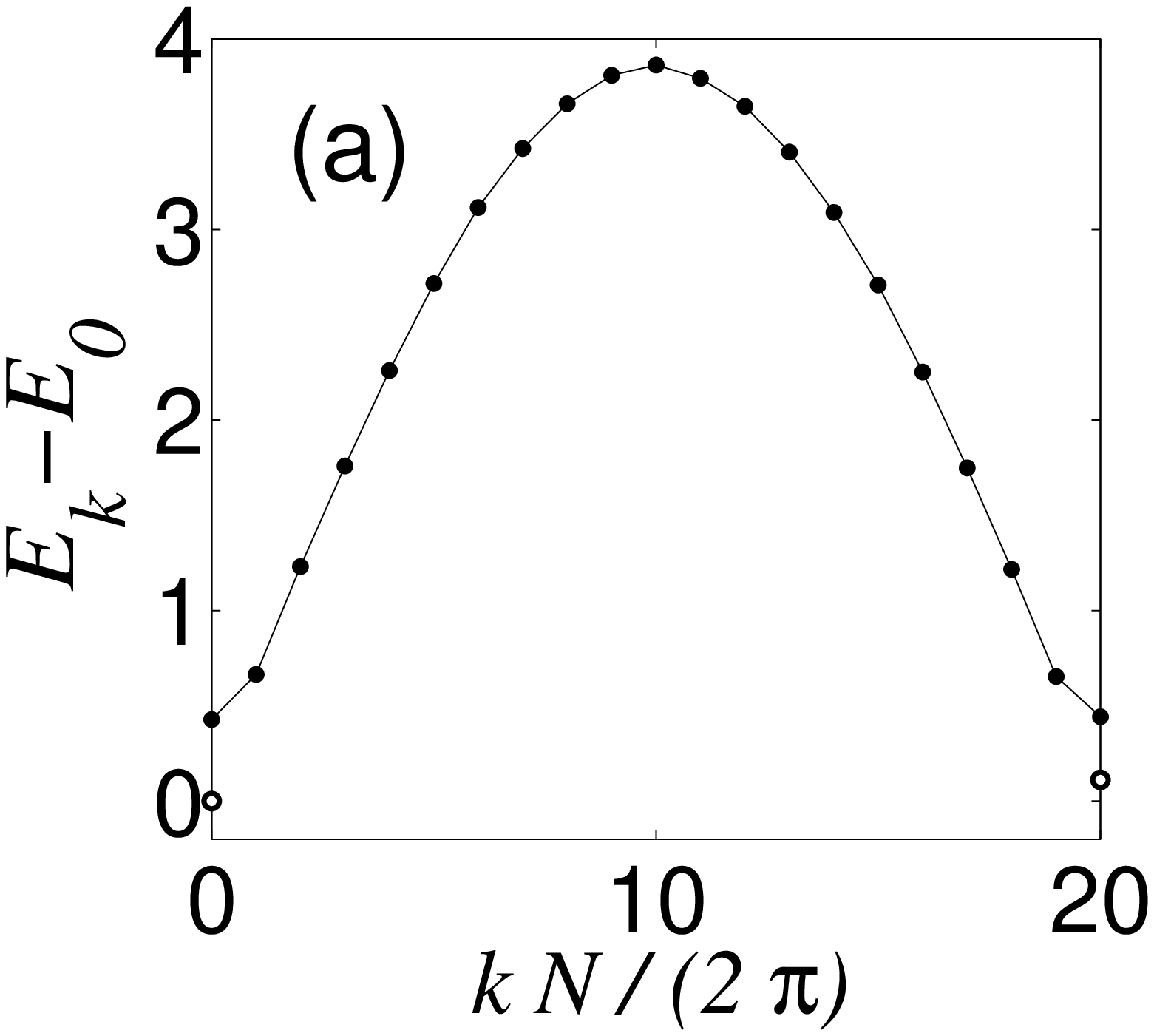}
  \hspace{0.1cm}
  \includegraphics[width=5cm]{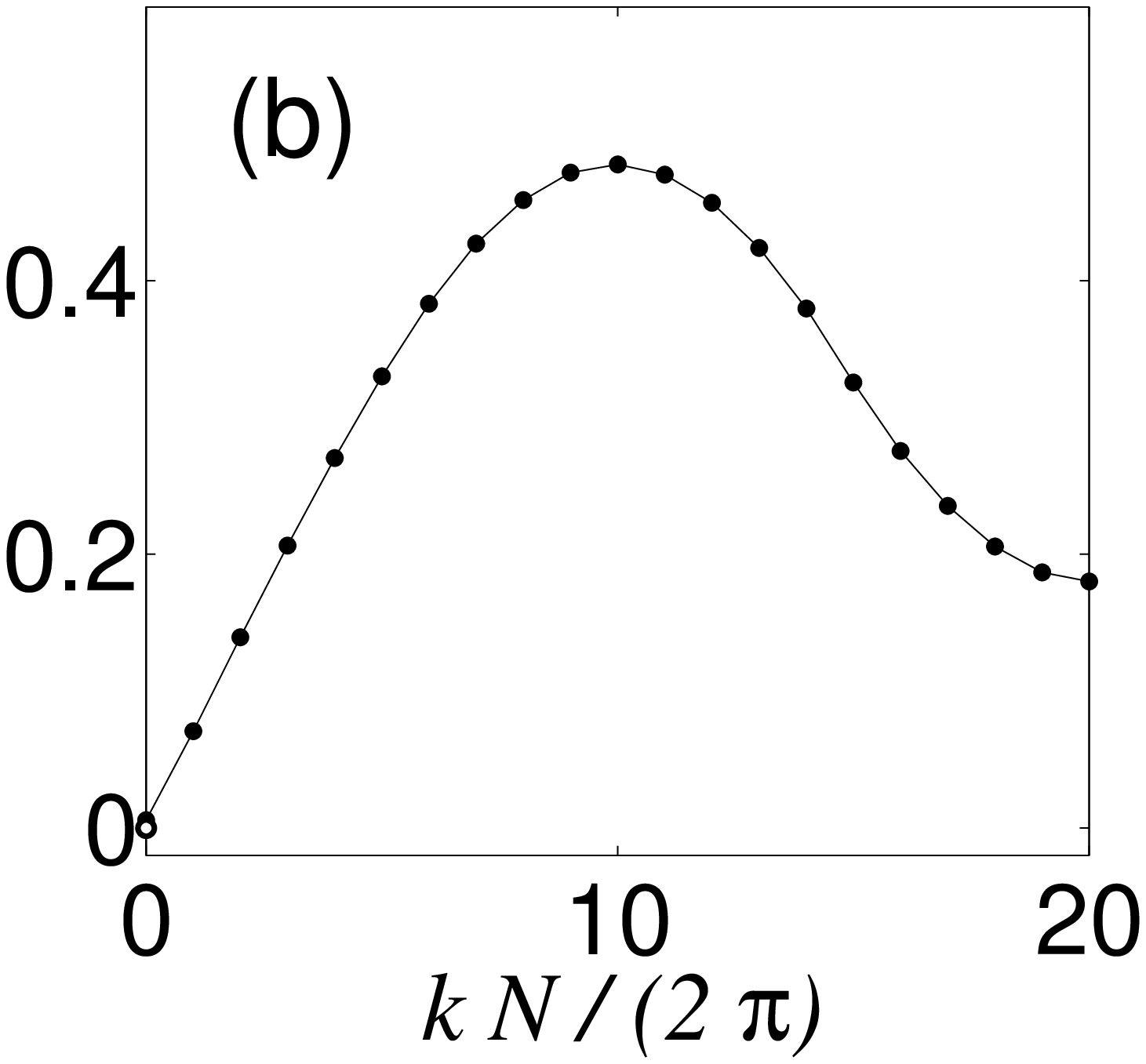}
  \caption{(taken from \cite{porrasverstraete06}) Lowest states of a bilinear--biquadratic $S=1$ chain, $N=40$ sites, $D=10$. (a) $\theta = -\pi/2$, $E_0 = -2.7976 N$, (b) $\theta = - 0.74 \pi$, $E_0 = -1.4673 N$. Empty circles: lowest energy states. Filled circles: first branch of excitations. We estimate an absolute error $\Delta E_k \approx \ 10^{-3}$, by comparison with calculations with larger $D$.}
\label{figDiego2}
\end{figure}

As already explained in the previous section, it is easy to look for excited states using an extra Lagrange constraint. However, the VMPS approach can also easily be generalized such as to become a varational method over states with a definite momentum. The basic idea is very simple. Consider the state~\cite{porrasverstraete06}
\begin{eqnarray*}
|\psi_k\rangle&=&\sum_{n=0}^{N-1}\frac{e^{ikn}}{\sqrt{N}} \hat{T}_n |\chi\rangle\\
|\chi\rangle&=&\sum_{\alpha_1,...}{\rm Tr}\left(A^1_{\alpha_1}A^2_{\alpha_2}...A^N_{\alpha_N}\right)|\alpha_1\rangle|\alpha_2\rangle...|\alpha_N\rangle
\end{eqnarray*}
where the operator $\hat{T}_k$ is the shift operator implementing a translation over $k$ sites. The state $|\psi_k\rangle$ has the momentum $k$ by construction and is obviously a superposition of $N$ MPS (note that this implies that the block entropy of the state $|\psi_k\rangle$ can scale as $\log(ND^2)$ as opposed to $\log(D^2)$ for a normal MPS). What is clear, however, is that the whole VMPS procedure outlined above can now be repeated for this extended class of MPS: the energy is still a multiquadratic function of all tensors $A^i$ involved, and expectation values of local observables can still be calculated efficiently by contracting the corresponding tensor network. The price to pay is an extra factor of $N$ (the number of sites) in all the calculations, due to the fact that cross terms between the different superpositions enter the optimization. To achieve this rather mild slowdown, a rather involved scheme of bookkeeping has to be maintained, and this is explained in great detail in the paper~\cite{porrasverstraete06}. As an illustration of the power of this technique, figure \ref{figDiego2} represents the calculation of the energy-momentum relation for two values of the bilinear-biquadratic $S=1$ spin chain parameterized by the Hamiltonian
\begin{displaymath}
\mathcal{H}=\sum_i \cos(\theta)\vec{S}_i\vec{S}_{i+1}+\sin(\theta)\left(\vec{S}_i\vec{S}_{i+1}\right)^2.
\end{displaymath}

\newpage
\section{Time evolution using MPS} \label{sec:timeevol}

One of the big advantages of the formulation of renormalization group methods in terms of matrix product states is that it also allows to describe time evolution. This opens up the possibility of simulating the non-equilibrium properties of spin chains, a topic that is currently very relevant given the recent breakthroughs of creating strongly correlated quantum spin systems in e.g. optical lattices~\cite{ciraczoller04}. Since the development of DMRG, several methods have been proposed \cite{cazalilla02,luo03,cazalilla03,whitefeiguin04,vidal04,daley04,verstraeteripoll04,Feiguin2005,ripoll06,gobert05}. In the spirit of this paper, we will only review the variational approach, and refer to the review of Schollwock and White for DMRG-related approaches \cite{schollwoeck-2006}.

\subsection{Variational formulation of time evolution with MPS} \label{timeevol}

Mathematically, the problem is to evolve an initial MPS in real time by updating the tensors in the MPS-description under Hamiltonian evolution. In practice, this could be used to investigate how excitations travel through the spin chain or to get spectral information about a Hamiltonian of interest. In a similar spirit as the previous sections, we would like to do this in a variational way: given a Hamiltonian and an initial MPS, evolve that state within the manifold of MPS in such a way that the error in approximating the exact evolution is minimized at every infinitesimal step~\cite{verstraeteripoll04}.

\begin{figure}[t]
  \centering
  \includegraphics[width=\linewidth]{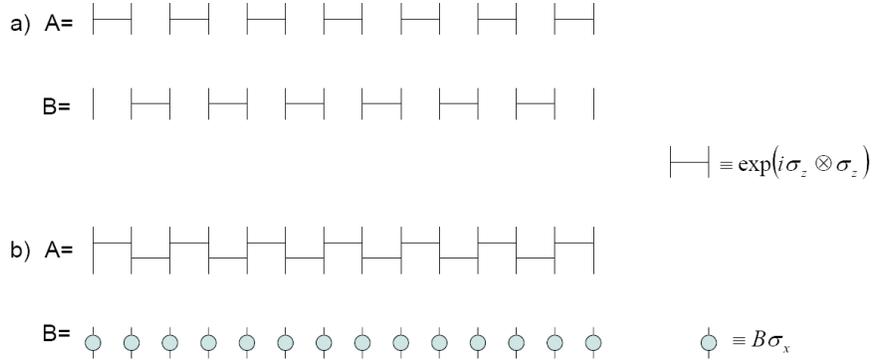}
  \label{trotter:fig1}
\caption{Spin network representation of the different operators obtained in the case of a Trotter expansion of the Ising Hamiltonian in transverse field with a) nearest neighbour decomposition and b) Ising versus magnetic field decomposition.}
  \label{Trotter}
\end{figure}

We are particularly interested in the case where the Hamiltonian, which can be time-dependent, is a sum of local terms of the form
\begin{displaymath}
\mathcal{H}(t)=\sum_{<ij>}f_{ij}(t)\hat{O}_i\otimes \hat{O}_j
\end{displaymath}
and where $<ij>$ means that the sum has to be taken over all pairs of nearest neighbours (note that the situation with long-range interactions can be treated in a very similar way).  There are several tools to discretize the corresponding evolution. This is not completely trivial because generically the different terms in the Hamiltonian don't commute. A standard tool is to use the Trotter decomposition \cite{suzuki90,suzuki91}
\begin{displaymath}
e^{A+B}=\lim_{n\rightarrow\infty}\left(e^{\frac{A}{n}}e^{\frac{B}{n}}\right)^n.
\end{displaymath}
Suppose e.g. that the Hamiltonian can be split into two parts $A$ and $B$ such that all terms within $A$ and within $B$ are commuting: $\mathcal{H}=A+B; A=\sum_iA_i; B=\sum_i B_i$; $[A_i,A_j]_-=0=[B_i,B_j]_-$. This ensures that one can efficiently represent $e^{iA}$ as a product of terms. The evolution can then be approximated by evolving first under the operator $e^{i\delta tA}$, then under $e^{i\delta tB}$, again under $e^{i\delta tA}$ and so further. The time step can be choosen such as to ensure that the error made due to this discretization is smaller than a prespecified error, and there is an extensive literature of how to improve on this by using e.g. the higher order Trotter decompositions~\cite{sornborger99,omelyan02}. In the case of nearest neighbour Hamiltonians, a convenient choice for $A$ is to take all terms that couple the even sites with the odd ones to the right of it and for $B$ the ones to the left of it. In that case, $e^{iA}$ and $e^{iB}$ are tensor products of nearest-neigbour 2-body operators; see figure \ref{Trotter}a for a pictorial representation of this in terms of spin networks. However, it is important to note that different choices of $A$ and $B$ might work better practice. Consider e.g. the Ising Hamiltonian in an transversal field and its decomposition into two different terms $A$ and $B$ (see also Fig \ref{Trotter}b):
\begin{equation}
\mathcal{H}_{Ising}=\underbrace{\sum_k\sigma^k_z\otimes \sigma^{k+1}_z}_{A}+\underbrace{H\sum_{k}\sigma^k_x}_{B}. \label{Ising}
\end{equation}
Obviously, a similar kind of decomposition is possible in the case of the Heisenberg model, but there three terms are needed instead. The advantage of evolution with such a decomposition is that it does not break translational invariance as in the even-odd case.

Let us next investigate how to treat this time-evolution in a variational way within the class of MPS. A sensible cost function  to minimize is given by
\begin{equation}
\|A|\psi(k)\rangle-|\psi(k+1)\rangle\|_2^2
\label{costevol}
\end{equation}
where $|\psi(k)\rangle$ is the initial MPS, $|\psi(k+1)\rangle$ is the one to be found, and $A$ is the operator arising out of the Trotter expansion \footnote{Remark that as alternative  cost-function we could have chosen $\frac{|\langle\psi(k+1)|A|\psi(k)\rangle|^2}{\langle\psi(k+1)|\psi(k+1)\rangle}$
which has the aim at finding the normalized MPS $|\psi(k+1)\rangle$ that has maximal overlap with the evolved version of the original one. It is however an easy exercise to find out that this optimization  problem leads exactly to the same optimal MPS, up to a normalization factor.}. First of all, it is important to note that this cost function can indeed be evaluated efficiently when $|\psi(k)\rangle$ and $|\psi(k+1)\rangle$ are MPS and $A$ is of the form discussed above. In principle, this cost function can be made equal to zero by increasing the bond dimension $D$ by a factor of at most $d^2$ (this is indeed the largest possible Schmidt number of an operator acting on two sites). However, the whole point of using MPS is that MPS with low bond dimension are able to capture the physics needed to describe the low-energy sector of the Hilbert space. So if we stay within that sector, the hope is that the bond dimension will not have to be multiplied by a constant factor at each step (which would lead to an exponential computational cost as a function of time), but will hopefully saturate. This is certainly the case if we evolve using e.g. imaginary time evolution of a constant local Hamiltonian, as we know that in that case the ground state is indeed well represented with a MPS with not too large bond dimension. It is however important to keep in mind that an exponential explosion is in principle possible for other kinds of evolution: the worst-case computational complexity for time evolution using MPS is exponential as a function of time.

Taking this into account, a justified way of dealing with time evolution is to prespecify an error $\epsilon$ that can be tolerated, and then look for the minimal $D$ for which there exists a MPS $|\psi(k+1)\rangle$ that yields an error smaller than $\epsilon$. Looking back at the cost function (\ref{costevol}), it looks pretty familiar how to minimize it:  the cost function has only quadratic and linear terms, and we will hence be able to minimize it by a method very similar to the  alternating least squares method discussed in the previous section. More specifically, the cost function has only multiquadratic and multilinear terms in the variables of the MPS, and we will solve this in a recursive way where at each recursion step an optimization of the form
\begin{displaymath}
x^\dagger A_{eff}x-2x^\dagger y_{eff}
\end{displaymath}
has to be solved. The solution to this problem is the simple solution of the linear set of equations
\begin{displaymath}
A_{eff}x=y_{eff},
\end{displaymath}
and this hence  leads to a very efficient way of minimizing the cost function for time evolution in a recursive way: sweeping back and forth, we solve the above optimization subproblem at each step, and we are guaranteed that the total cost function goes down at every step and hence will converge. After convergence, we can check how big the error has been for the particular value of $D$ that we choose, and if this error is too big, we can increase $D$ and repeat the optimization. Note that the only requirement for the complete Trotter step evolution to be successful is that the tensor network when sandwiching the operators A and B between two MPS can be contracted efficiently. It is clear that this method both works in real and imaginary time evolution, and both for time-independent and time-dependent Hamiltonians. As a note, and this will be crucial in the applications of this variational time evolution of MPS to higher dimensional spin systems, we also want to point out that nowhere we used the fact that the operator $A$ was close to the identity; i.e. this method is applicable in a more general way than only for small Trotter steps. Note also that this method is applicable to systems with both open and periodic boundary conditions, and also with long-range interactions.

\subsubsection{time-evolving block-decimation}

%\begin{figure}[t]
%  \centering
%  \includegraphics[width=8cm]{WHITEfig3.eps}
%\caption{(taken from \cite{whitefeiguin04})
%The single magnon line of the spin-1 Heisenberg antiferromagnetic chain.
%The entire spectrum is obtained from one DMRG run, by Fourier transforming
%the time and position dependent correlation function $\langle S^-_l(t) S^+_0(0)\rangle$.
%The broad solid curve shows the location of the maximum in the spectra for
%a particular $q$, in units of the Haldane gap, 0.41050(2),
%for a system of $L=600$ sites, using a time step $\tau=0.02$, running for $T=27.3$,
%and keeping $m=200$ states. For comparison, results from two other runs are shown:
%$L=400$, $\tau=0.1$, $T=60$, and $m=150$ (dashed curve); and
%$L=400$, $\tau=0.4$, $T=72$, and $m=200$ (dotted curve).
%The solid curve peaked at $q=\pi$, shown only for the first run,
%is the weight $A_0$ in this quasiparticle peak, i.e.
%$S(\omega) \simeq A_0 \delta(\omega-\omega_0)$.}
%  \label{figthree}
%\end{figure}

There has recently been a lot of attention on studying time evolution using MPS, but instead of using the optimal variational way described above, the vast majority of the works has been using the so-called time-evolving block-decimation (TEBD) procedure introduced by Vidal \cite{vidal04}. The reason for this is that this method can readily be implemented with existing DMRG code~\cite{whitefeiguin04,daley04}. It explicitely uses the Schmidt normal form described in section \ref{calculus}, and is hence more suitable for MPS with open boundary conditions. Also, one has to use the Trotter expansion with even-odd sites decomposition. It can also be understood as a variational method, but in the more restricted sense in which we consider the following set-up: given a MPS and one (possibly non-unitary) gate acting on two nearest-neighbors, find the new MPS with given dimension $D$ that approximates this optimally. This can be done using the singular value decompisition: using the normal form for MPS with OBC, we know that the original MPS is of the form
\begin{displaymath}
|\psi\rangle=\sum_{n=1}^D |\psi_n^L\rangle|\chi_n^R\rangle
\end{displaymath}
with $\{|\psi_n^L\}$ and $\{|\psi_n^R\}$ orthogonal sets. The gate which acts on the nearest neighbours locally increased the dimension of the bond with a factor of at most $d^2$, and we would like to reduce the dimension of that bond to~$D$ again. One can orthonormalize everything again and obtain the Schmidt decomposition over that particular bond, and then the reduction can trivially be done by discarding (i.e. projecting out) the smallest Schmidt coefficients\footnote{This is indeed the virtue of the singular value decomposition (SVD): if we want to approximate a given matrix with one of lower rank in an optimal way (in the sense as to minimize the Frobenius norm of the difference), then the solution is given by taking the SVD and put the smallest singular values equal to zero.}. In the next step, evolution between the next nearest neighbours is considered, and so further. Note that the variational method discussed above could deal with all those gates at once; nevertheless, the computational cost of both methods is very similar, and in practice both methods seem to achieve a similar accuracy if the time steps are small. The first one, however, can be improved by choosing longer time steps and higher Trotter orders~\cite{ripoll06}.

It is also possible to devise more sophisticated methods that somehow make use of the fact that superpositions of MPS can still be dealt with in an efficient way. For a nice review that compares all such methods and the ones described before, see~\cite{ripoll06}. For nice examples of the power of real-time evolution, we refer to the review article \cite{schollwoeck-2006}.

\subsection{Finding ground states by imaginary time evolution}

The tools discussed in the previous section apply both to real and imaginary time evolution. This provides a completely different way of finding ground states of local quantum spin Hamiltonians: if we start with an arbitrary state $|\psi_0\rangle$, evolution in imaginary time leads to
\begin{displaymath}
|\psi(t)\rangle=e^{-Ht}|\psi_0\rangle=\sum_{k=1}^n e^{-\lambda_k t}|\chi_k\rangle\langle\chi_k|\psi_0\rangle
\end{displaymath}
with $H=\sum_{k=1}^n \lambda_k|\chi_k\rangle\langle\chi_k|$ the eigenvalue decomposition of $H$. Hence, as long as the ground state is not degenerate and for times longer than the inverse gap, the state $|\psi(t)\rangle$ will converge exponentially fast to the ground state, and the speed of convergence is exactly quantified by the gap. This is indeed something that is a recurring theme: the smaller the gap, the slower all variational methods seem to converge. This is of course not unexpected because it is difficult to discriminate excited states with small energy from the ground state. However, it is interesting to note that the closer a system is to criticality (i.e. gap goes to zero), the bigger $D$ has to be for a MPS to approximate the ground state for a fixed error (see appendix B to see that in critical systems, this scaling is polynomial in the number of spins in the worst case scenario). A very interesting question would be to relate the density of states above the gap to the decay of the Schmidt coefficients that one gets by dividing the ground state into two pieces.

In practice, finding a ground state using imaginary time evolution is pretty reliable. Of course, the time steps have to be adjusted such that they become smaller and smaller, but one of the great features of imaginary time evolution is its inherent robustness: it does not really matter if one makes mistakes in the beginning, as there is anyway an exponential convergence to the ground state afterwards. This is in contrast to the evolution in real time.

\subsubsection{Infinite spin chains}

Another advantage of the imaginary time evolution approach is that one can easily treat systems with periodic boundary conditions or treat the thermodynamic limit (number of sites $\rightarrow\infty$) by making use of the inherent translational invariance of MPS with all tensors $A^i$ equal to each other.

To illustrate how this can be done, let us consider the specific case treated in \cite{inpreparation}  and  assume that we want to model the ground state of the Ising Hamiltonian in a transverse field defined on a spin chain. As discussed in section \ref{timeevol} following equation \ref{Ising}, one can choose the decomposition in the Trotter step in such a way that both the operators $A$ and $B$
\begin{eqnarray*}
A&=&\sum_k \sigma_z^k\otimes \sigma_z^{k+1}\\
B&=&h\sum_k \sigma_x^k
\end{eqnarray*}
are completely translational invariant, and hence we can stay within the manifold of translational invariant states to describe its evolution. The next step is to see how $\exp(\delta t A)$ and  $\exp(\delta t B)$ look like. The latter is simple, as it is just equal to
\begin{displaymath}
\exp(\delta t B)=\otimes_k \exp(\delta t h \sigma_x).
\end{displaymath}
The former expression is obviously a product of commuting operators, and a simple exercise allows one to see that there is a simple matrix product description for it:
\begin{eqnarray}
\exp(\delta t A)&=&\sum_{i_1i_2...} {\rm Tr}\left(C_{i_1}C_{i_2}...\right)X_{i_1}\otimes X_{i_2}\otimes ...\label{jdsf}\\
     C_0&=&\left(\begin{array}{cc} \alpha&0\\ 0 & \beta\end{array}\right)\hspace{1cm}
     C_1=\left(\begin{array}{cc} 0&\sqrt{\sinh(\delta t)\cosh(\delta t)}\\ \sqrt{\sinh(\delta t)\cosh(\delta t)} & 0\end{array}\right)\nonumber\\
     X_0&=&\left(\begin{array}{cc} 1&0\\ 0 & 1\end{array}\right)\hspace{1cm}
     X_1=\left(\begin{array}{cc} 1&0\\ 0 & -1\end{array}\right)\nonumber
\end{eqnarray}
This can justifiably be called a matrix product operator (MPO) [see next section]. The associated bond dimension is $2$, and this means that when acting on a MPS of dimension $D$ with this MPO, the exact representation of the new MPS will at most be $2D$. In this particular example of the Ising Hamiltonian, imaginary time evolution would now amount to act subsequently with $\exp(\delta t A)$ and  $\exp(\delta t B)$ on an initial state. Combined together, it happens that their product $Q=\exp(\delta t B/2)\exp(\delta t A)\exp(\delta t B/2)$ is again exactly of the form~(\ref{jdsf}) with unchanged tensor $C_i$ and $X_i$ but where we have to replace $X_0=I$ with $X_0=\exp(\delta t \sigma_x)$. So $Q$ is a very simple MPO with bond dimension $2$, and the goal is to evolve a translational invariant MPS using this MPO. This can be done is a very simple way: given a MPS with tensor $A_i$ of dimension $D$, the action of $Q$ is such that we have to replace
\begin{displaymath}
A_i\rightarrow \sum_{k,l} A_k\otimes C_l \langle i|X_l|k\rangle.
\end{displaymath}
The new MPS corresponding to this $A_i$ has bond dimension $2D$, and this has to be reduced as otherwise its size would increase exponentially. The optimal choice is again simple if we consider a system with open boundary conditions: we consider its normal form  (section \ref{calculus}), and cut it in the appropriate way without losing translational invariance
\footnote{More specifically, a possible procedure is as follows \cite{inpreparation}: calculate the leading left $|v_l\rangle$ and right eigenvector $|v_r\rangle$ of $\sum_k C_k\otimes \bar{C}_k$ (note that this can be done in a sparse way). Reshape those eigenvectors in the square matrices $X_l$ and $X_r$, and as those matrices are the fixed points of the CP-maps $\sum_k C_k . C_k^\dagger$ and  $\sum_k C_k^\dagger . C_k$, $X_l$ and $X_r$ are guaranteed to be positive. Next take the singular value decomposition of $\sqrt{X_r}\sqrt{X_l}=U\Sigma V^\dagger$, and reduce the number of columns of $U$ and $V$ to $D$ and discard the $D$ lowest singular values in $\Sigma$ ($U$ and $V$ hence become $2D\times D$ matrices and $\Sigma$ a $D\times D$ matrix). Define $G_r=\sqrt{X_r^{-1}}U\sqrt{\Sigma}$ and $G_l=\sqrt{\Sigma}V^\dagger\sqrt{X_l^{-1}}$, and make the transformation $A_i\rightarrow G_l .A_i G_r$ such that it corresponds to a MPS with $D$-dimensional bonds instead of $2D$.}.
That last step is not so easy to justify rigorously, but seems to work very well in practice: what we do is calculating the Schmidt normal form with respect to the $2D$ dimensional bonds, and then cut all bonds together. The tricky thing is that cutting the bond somewhere changes the Schmidt coefficients somewhere else, but the point is that these changes are only of the order of the Schmidt coefficients that are cut and those are very small anyway. Amazingly, this procedure works very well, and even a small bond dimension of $D=32$ already reproduces the ground state energy-density $E$ of the critical Ising model ($h=1$) up to a precision better than $E-4/\pi<10^{-7}$. Much better conditioning can also be obtained \cite{inpreparation} by working with hermitean matrices $\{A^i\}$.

Clearly, this procedure can be repeated for any translational invariant Hamiltonian like the Heisenberg model (note that the bond dimension will be bigger there), and  the big advantage is that it allows to treat infinite systems. The finite system with periodic boundary conditions can be dealt with in a similar way, although the cutting procedure is more involved there.

A variant of this procedure can be obtained by using the even-odd  Trotter decomposition. In that case, exact translational invariance is broken, but it is still perfectly translational invariant with period 2:
\begin{displaymath}
|\psi\rangle=\sum_{i_1i_2...}...A_{i_1}B_{i_2}A_{i_3}B_{i_4}...|i_1\rangle|i_2\rangle|i_3\rangle|i_4\rangle
\end{displaymath}
This type of imaginary time evolution has been studied in detail by G. Vidal \cite{vidal07}, and convergence seems to be fast. The updating and cutting works in a very similar way as discussed in the example discussed above. Note that from the point of view of variational states,  it could be advantagous to work with such an $ABAB...$ scheme when an antiferromagnetic ordering is expected.

\newpage
\section{Matrix Product Operators} \label{sec:mpo}

Instead of restricting our attention to pure  matrix product states, we can readily generalize the approach and deal with matrix product operators (MPO). In its most general case, a MPO is defined as~\cite{verstraeteripoll04,zwolak04}
\begin{equation}
\hat{O}=\sum_{i_1i_2...}{\rm Tr}\left(A^1_{i_1}A^2_{i_2}...\right)\sigma_{i_1}\otimes \sigma_{i_2}\otimes ... \label{MPO}
\end{equation}
with $\sigma_i$ a complete single particle basis (e.g. the Pauli matrices for a qubit). We already encountered an example of such a MPO in the previous section, where the evolution following a Trotter step was expressed by a MPO;  that MPO essentially  played the role of a transfer matrix during the evolution. As we will show later, any transfer matrix arising in the context of classical partition function will have an exact representation in terms of such a MPO. This is the reason why all the renormalization methods described above can also be used in the context of 2-D classical spin systems. As a side remark, it is also true that any translational invariant spin Hamiltonian with nearest neighbour interactions has an exact MPO representation\footnote{For this, let us consider the spin 1/2 case. We first need the fact that there always exists a basis such that the Hamiltonian is of the form
\begin{displaymath}
\mathcal{H}=\sum_{\alpha,i} \lambda_\alpha\sigma_\alpha^i\otimes \sigma_\alpha^{i+1} +\sum_j \hat{O}^j
\end{displaymath}
where $\hat{O}$ can be any one-qubit operator. It is now a small exercise to prove that

\begin{eqnarray*}\mathcal{H}&=&\sum_{i_1i_2...}\left(v_lA_{i_1}A_{i_2}...v_r^T\right)X_{i_1}\otimes X_{i_2}\otimes ...\\
X_0&=&I\hspace{1cm} X_1=\sigma_x \hspace{1cm} X_2=\sigma_y \hspace{1cm} X_3=\sigma_z \hspace{1cm} X_4=\hat{O}\\
v_l&=&|0\rangle\hspace{1cm} v_r=|4\rangle\\
A_0&=&|0\rangle\langle 0|+|4\rangle\langle 4|\hspace{1cm} A_1=|0\rangle\langle 1|+|1\rangle\langle 4|\hspace{1cm}A_2=|0\rangle\langle 2|+|2\rangle\langle 4|\\
A_3&=&|0\rangle\langle 3|+|3\rangle\langle 4|\hspace{1cm} A_4=|0\rangle\langle 4|
\end{eqnarray*}
Actually, one can easily prove that $D=5$ is optimal because this is the operator Schmidt number of the Hamiltonian when splitting it into two pieces.}. Matrix product operators have been shown to be very useful to obtain spectral information about a given Hamiltonian. Examples are the parametrization of Gibbs states (section \ref{Gibbs}), the simulation of random quantum spin systems (section \ref{randomQS}), the calculation of classical partition functions (section \ref{class}),  and the determination of the density of states (section \ref{dos}).

If a MPO is a positive operator and has trace one, then it becomes a matrix product density operator (MPDO). A simple systematic way of constructing MPDO~\cite{verstraeteripoll04} is by making use of the fact that every mixed state can be seen as a part of a bigger pure system, namely its purification. If we model this purification as a MPS, then the MPDO is obtained by tracing over the purifying degrees of freedom. This picture is especially useful if the purification is such that to every original spin, a locally accompanying purifying spin is present, and that the pure state is such that these pairs of spins correspond to one bigger site:
\begin{eqnarray}
|\psi\rangle&=&\sum_{i_1i_2...}{\rm Tr}\left(A^1_{i_1j_1}A^2_{i_2j_2}...\right)|i_1\rangle|j_1\rangle\otimes |i_2\rangle|j_2\rangle\otimes ...\label{purif}\\
\rho&=&{\rm Tr}_{j_1j_2...}\left(|\psi\rangle\langle\psi|\right)\nonumber\\
&=&\sum_{i_1i_2...i_1'i_2'...}{\rm Tr}\left(\underbrace{\sum_{j_1j'_1}A^1_{i_1j_1}\otimes \bar{A}^1_{i_1'j_1'}}_{E^1_{i_1i_1'}}\underbrace{\sum_{j_2j'_2}A^2_{i_2j_2}\otimes \bar{A}^2_{i_2'j_2'}}_{E^2_{i_2i_2'}}...\right)|i_1\rangle\langle i_1'|\otimes |i_2\rangle\langle i_2'|\otimes ...\nonumber\\
&=&\sum_{i_1i_2...i_1'i_2'...}{\rm Tr}\left(E^1_{i_1i_1'}E^2_{i_2i_2'}...\right)|i_1\rangle\langle i_1'|\otimes |i_2\rangle\langle i_2'|\otimes ...\nonumber
\end{eqnarray}
This representation of matrix product density operators in terms of purifications will turn out to be very useful to describe Gibbs states in thermal equilibrium. Note that not all MPO can arise from locally tracing out an ancilla: the tensors $E^i$ arising in the above description are very special in the sense that they correspond to completely positive maps and hence live in a convex positive cone.

\subsection{Finite temperature systems\label{Gibbs}}

Matrix Product Operators are very convenient to approximate Gibbs states of local Hamiltonians. More specifically, we would like to approximate the operator
\begin{displaymath}
e^{-\beta \mathcal{H}}=e^{-\frac{\beta}{2} \mathcal{H}}I e^{-\frac{\beta}{2} \mathcal{H}}
\end{displaymath}
where $I$ is the identity operator. In the spirit of last sections, this amounts to evolving the maximally mixed state in imaginary time for a period $\beta/2$~\cite{verstraeteripoll04,zwolak04}. However, instead of implementing this evolution on the MPO directly, a square root speed up in computational complexity can be gained\footnote{Note that such a square root speed-up is guaranteed when the temperature is very low as there as the number of {\em classical} correlations is zero for pure states.} by considering the evolution on its purification (see previous section). More specifically, the initial state $I$ can be seen as a collection of halve of maximally entangled pairs. In the notation of equation (\ref{purif}), the inititial state is a MPS with $D=1$ and where $A_{ij}=\delta_{ij}$. We can now use the variational algorithm discussed above to evolve this initial state over a time $\beta/2$, and once we have that state it can be used to calculate e.g. the free energy and any correlation function.

Note that without time evolution, it would be very hard to do a variational calculation over all possible matrix product operators to approximate the Gibbs state, as that state is variationally characterized by the condition that it minimizes the free energy ${\rm Tr}\left(\rho\mathcal{H}\right)-TS(\rho)$; the problem here is the calculation of the von-Neumann entropy $S(\rho)$, which we do not know how to calculate for matrix product density operators in general. But as a by-product of the imaginary time approach, we can readily calculate the entropy of the Gibbs state $\rho$

\[S\left(\rho\right)= \log{\rm Tr}\left(e^{-\beta H}\right)+\beta\frac{{\rm Tr}\left(He^{-\beta H}\right)}{{\rm Tr}\left(e^{-\beta H}\right)},\]
as all the quantities on the right hand side of this equation can efficiently be calculated and approximated using the MPS/MPO representation obtained via evolution in imaginary time.

Let us now see whether it is possible to calculate non-equilibrium properties of such MPO by evolving them. The most general type of evolution allowed by quantum mechanics is described by the Lindblad equation, which takes into account noise and can therefore increase the entropy. In general, this means that the system is interacting with an extra auxiliary one, and this poses somehow problems for treating the evolution via the purification described above: the dimension of the ancilla has to grow continuously. The straightforward remedy is to work with the full matrix product operator picture, and evolution can again be treated in a straightforward variational way~\cite{verstraeteripoll04,zwolak04}.

\subsection{Random quantum spin systems} \label{randomQS}
Another nontrivial application of matrix product operators and more specifically of the corresponding purifications is the fact that those allow to simulate random quantum spin systems in a highly efficient way. By making effective use of the entanglement present between system and an appropriately chosen ancilla, it happens that one can simulate exponentially many realizations of the random system in one run; this is quantum parallelism in the strongest possible sense.  To explain this, we closely follow the exposition given in the paper~\cite{paredesverstraete05}.

Let us consider a quantum system with Hilbert space $\mathcal{H}$ that evolves accordingly to a
Hamiltonian $H(r_1, \ldots, r_n)$ where $r_1, \ldots, r_n$ are random variables that take values within a finite discrete set, $r_\ell \in \Gamma_{\ell}=\{\lambda_1^{\ell}, \ldots \ \lambda^{\ell}_{m_{\ell}}\}$, with a probability distribution given by $p(r_1, \ldots, r_n)$. In order to simulate exactly the dynamics of such a system one would need to perform $\prod_{\ell=1}^n m_{\ell}$ simulations, one per each possible realization of the set of random variables $\mathbf{r}={(r_1, \ldots r_n)}$. For each realization the system evolves to a different state $\vert \psi_{\mathbf{r}}(t) \rangle=e^{-iH(\mathbf{r})t /\hbar} \vert \psi_0 \rangle$, where $\vert \psi_0 \rangle$ is the initial state. Given this set of evolved states and a physical observable $\hat{O}$, one is typically interested in the average of the expectation values of that observable in the different evolved states, that is, in quantities of the form:
\begin{equation}
{\big \langle} \langle \hat{O}(t) \rangle {\big \rangle} :=
\sum_{\mathbf{r}} p(\mathbf{r}) \langle \psi_{\mathbf{r}}(t)\vert
\hat{O} \vert \psi_{\mathbf{r}}(t) \rangle.
\label{valores_esperados_dobles}
\end{equation}
On the following we describe an algorithm that allows to simulate in parallel all possible time evolutions of the random system described above. We consider an auxiliary system with Hilbert space $\mathcal{H}_a$ and a Hamiltonian acting on $\mathcal{H}\otimes \mathcal{H}_a$ of the form $\widetilde{H}=H(\hat{R_1}, \ldots, \hat{R_n})$, where $\hat{R_1}, \ldots, \hat{R_n}$ are operators that act in $\mathcal{H}_a$, commute with each other and have spectra $\Gamma_1, \ldots ,\Gamma_n$. Note that we have replaced the set of random variables $\mathbf{r}$ by a set of quantum operators $\mathbf{\hat{R}}$ with the same spectra. The algorithm works as follows. 1) {\em Initialization}. Let us prepare the auxiliary system in an initial superposition state of the form:
\begin{equation}
\vert \psi_a \rangle= \sum_{\mathbf{r}}
\sqrt{p(\mathbf{r})}\,\,\vert \mathbf{r}\rangle,
\label{estado_ancila}
\end{equation}
where the states $\vert \mathbf{r}\rangle$ are simultaneous eigenstates of the set of operators $\mathbf{\hat{R}}$, with $\hat{R_{\ell}}\vert \mathbf{r}\rangle=r_\ell \vert \mathbf{r}\rangle$. Each state $\vert \mathbf{r} \rangle$ is therefore in one to one correspondence with one realization of the the set of random variables $\mathbf{r}$, its weight in the superposition state (\ref{estado_ancila}) being equal to the probability with which the corresponding realization occurs for the random system. 2) {\em Evolution}. We evolve the initial state of the composite system $\vert \psi_0 \rangle \otimes \vert \psi_a \rangle$ under the Hamiltonian $\widetilde{H}$. The evolved state is
%\begin{equation}
%$\vert \Psi(t)\rangle = e^{-i\widetilde{H}t / \hbar}\! \left(\vert
%\psi_0 \rangle \otimes \vert \psi_a \rangle
%\right)=\sum_{\mathbf{r}}\!
%\sqrt{p(\mathbf{r})}\,e^{-iH(\mathbf{\hat{R}})t  / \hbar}\!
%\left(\vert \psi_0 \rangle \otimes \vert
%\mathbf{r}\rangle\right).$
%\nonumber
%\end{equation}
%Since $\vert \hat{\mathbf{r}} \rangle$ is an eigenstate of all
%operators in $\mathbf{\hat{R}}$ it follows that
%$e^{-iH(\mathbf{\hat{R}})t / \hbar}\! \left(\vert \psi_0 \rangle
%\otimes \vert \mathbf{r}\rangle\right)= \left(e^{-iH(\mathbf{r})t
%/ \hbar}\! \vert \psi_0 \rangle \right) \otimes \vert
%\mathbf{r}\rangle$, so that we have
\begin{equation}
\vert \Psi(t)\rangle = \sum_{\mathbf{r}}\!
\sqrt{p(\mathbf{r})}\,\vert \psi_{\mathbf{r}}(t) \rangle \otimes
\vert \mathbf{r}\rangle.
\label{estado_evolucionado_sistema_mas_ancila}
\end{equation}
This superposition state contains the complete set of evolved states we are interested in. 3) {\em Read-out}. In order to obtain the quantities (\ref{valores_esperados_dobles}) we just need to measure the observable $\hat{O} \otimes 1$,
% Since we have that
%$\langle \mathbf{r} \vert \mathbf{r^{\prime}} \rangle
%=\delta_{\mathbf{r}, \mathbf{r^{\prime}}}$, we obtain
\begin{eqnarray}
\langle \Psi (t)\vert \hat{O} \otimes 1 \vert \Psi (t)
\rangle=\big{\langle} \langle \hat{O}(t) \rangle \big{\rangle}.
\end{eqnarray}
The algorithm above allows us, in particular, to obtain the averaged properties of a random system over the collection of all possible {\em ground states}. Let us assume that the interaction between the system and the ancilla is introduced adiabatically, so that the Hamiltonian is now $\widetilde{H}(t)= H (\beta(t)\mathbf{\hat{R}})$, where $\beta (t)$ is a slowly varying function of time with $\beta(0)=0$, $\beta(T)=1$, $T$ being the time duration of the evolution. If the system is prepared in the ground state of the Hamiltonian $H(\mathbf{0})$, the algorithm above will simulate in parallel all possible adiabatic paths, so that the composite superposition state (\ref{estado_evolucionado_sistema_mas_ancila}) will contain all possible ground states of the random system~\cite{paredesverstraete05}.

\begin{figure}
\begin{center}
\includegraphics[width=0.7\linewidth]{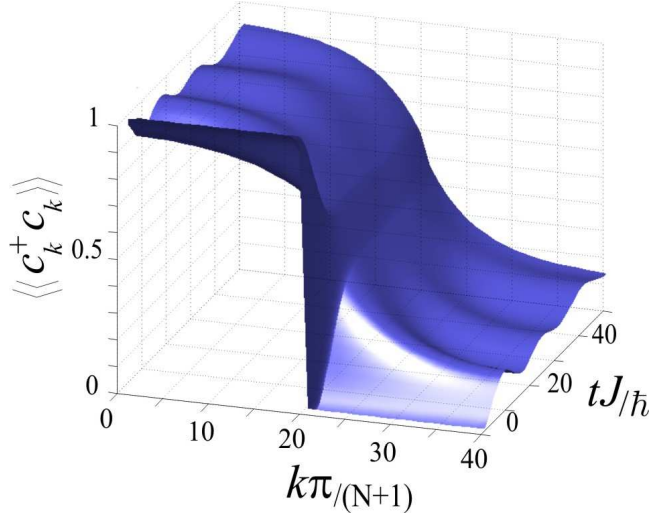}
\caption{Numerical simulation of the time evolution of a random
field XY spin chain with $N=40$. We show the correlation function
${\big \langle} \langle c^{\dagger}_k c_k \rangle {\big \rangle}$
as a function of time and momentum $k$. Here $c_k \propto
\sum_\ell sin(k \ell) \widetilde{c}_\ell$, $k=\frac{\pi}{N+1},
\ldots, \frac{\pi N} {N+1}$ and $\widetilde{c}_\ell=\prod_{\ell <
\ell'} S^z_{\ell'} (S^x_{\ell} + i S^y_{\ell})$ are the fermionic
operators given by the Jordan-Wigner transformation. The evolution Hamiltonian is
$H(b_1, \ldots, b_N)=H_0+B\sum_{\ell=1}^N b_\ell S_\ell^{z}$ with $H_0$ being the XY Hamiltonian
with $B_0=0$, $B/J=-4$ and $p({\bf b})=1/2^N$. The initial state
$\vert \psi_0 \rangle$ is the ground state of $H_0$. As the system
evolves in time the initially sharp Fermi sea disappears.}
\end{center}
\end{figure}

Additionally, the scheme above can be easily extended for the computation of other moments of the distribution of physical observables (higher than (\ref{valores_esperados_dobles})), which are sometimes important in the understanding of QRS \cite{paredesverstraete05}. For example, quantities like $\big{\langle} \langle \hat{O}^2 \rangle- \langle \hat{O} \rangle ^2 \big{\rangle} $, can be computed by using an additional copy of the system.

The algorithm described above reduces the simulation of a quantum random system to the simulation of an equivalent non-random interacting problem. This exact mapping allows us to integrate the simulation of randomness in quantum systems within the framework of numerical methods that are able to efficiently simulate the corresponding interacting problem. As an illustrative example we consider the case of a 1D spin $s=1/2$ system with random local magnetic field. The Hamiltonian of the system is:
\begin{equation}
H(b_1, \ldots, b_N)=H_0+B\sum_{\ell=1}^N b_\ell S_\ell^{z},
\label{ham_campo_mag_random}
\end{equation}
where $H_0$ is a short range interaction Hamiltonian, $\mathbf{b}=(b_1, \ldots, b_N)$ is a set of classical random variables that take values $\{1/2,-1/2\}$ with probability distribution $p(\mathbf{b})$. Following the algorithm above the $2^N$ simulations required for the exact simulation of the dynamics (or the ground-state properties) of this random problem can be simulated in parallel as follows. We consider an auxiliary 1D spin $\sigma=1/2$ system. We prepare this ancilla in the initial state $\vert \psi_a \rangle= \sum_{\mathbf{b}} \alpha_{\mathbf{b}}\vert \mathbf{b}\rangle$, where the states $\vert \mathbf{b}\rangle$ have all $z$ components of the $N$ spins well defined, $\widehat{\sigma}^z_\ell \vert \mathbf{b}\rangle=b_{\ell}\vert \mathbf{b}\rangle$, and $\alpha_{\mathbf{b}}=\sqrt{p(\mathbf{b})}$. The entangled properties of the state of the ancilla reflect the classical correlations among the random variables. For example, for a
uniform distribution of the random field, $p(\mathbf{b})=1/2^N$, the state of the ancilla is just a product state, $\vert \psi_a \rangle \propto \left( \vert \uparrow \rangle + \vert \downarrow \rangle \right)^{\otimes N}$. We evolve the system and the ancilla under the interaction Hamiltonian
\begin{equation}
\widetilde{H}=H(\widehat{\sigma}^z_1, \ldots,
\widehat{\sigma}^z_n)=H_0+\beta \sum_\ell \widehat{\sigma}^z_\ell
\widehat{S}_\ell^{z}. \label{ham_dos_cadenas}
\end{equation}

Here, $\beta=B$ if we want to simulate dynamics under Hamiltonian (\ref{ham_campo_mag_random}), and $\beta$ is a slowly varying function of time with $\beta(0)=0$ and $\beta(T)=B$ for the simulation of the ground state properties. We have then reduced the simulation of the random problem to that of the time evolution of two coupled spin $1/2$ chains with Hamiltonian (\ref{ham_dos_cadenas}). This problem is equivalent to a 1D lattice problem of $N$ sites with physical dimension $d=2 \times 2$, which can be easily incorporated to the framework of the variational numerical methods introduced in the previous sections.

\subsection{Classical partition functions and thermal quantum states} \label{class}

In this section, we will show how the concept of MPO can be used to calculate the free energy of a classical 2-dimensional spin system. This also leads to an alternative way of treating thermal states of 1-D quantum spin systems, as those can be mapped to classical 2-D spin systems by the standard classical-quantum mapping \footnote{In the context of MPS and especially PEPS, there also exist a different mapping between classical and quantum spin models in the same dimension \cite{verstraetewolf06}. There, the thermal classical fluctuations map onto quantum ground state fluctuations, and this leads to  a lot of insights in the nature of quantum spin systems.}. Historically, Nishino was the first one to pioneer the use of DMRG-techniques in the context of calculating partition functions of classical spin systems \cite{nishino95}. By making use of the Suzuki--Trotter decomposition, his method has then subsequently been used to calculate the free energy of translational invariant 1-D quantum systems \cite{bursill96,shibata97,wang97}, but the main restriction of those methods is that it cannot be applied in situations in which the number of particles is finite and/or the system is not homogeneous; furthermore, one has to explicitely use a system with periodic boundary condition in the quantum case, a task that is not well suited for standard DMRG. The variational MPS-approach gives an easy solution to those problems.

Our method relies in reexpressing the partition and correlation functions as a contraction of a collection of 4-index tensors, which are disposed according to a 2--D configuration~\cite{murgverstraete05}. We will perform this task for both 2--D classical and 1--D quantum systems.

 Let us consider first the partition function of an inhomogeneous classical 2--D n--level spin system on a $L_1\times L_2$ lattice. For simplicity we will concentrate on a square lattice and nearest--neighbor interactions, although our method can be easily extended to other short--range situations. We have
\begin{displaymath}
    Z= \sum_{x^{1 1},\ldots,x^{L_1 L_2}}\exp\left[-\beta H(x^{1 1},\ldots,x^{L_1 L_2}) \right],
\end{displaymath}
where
\begin{displaymath}
    H\left(x^{1 1},\ldots\right) = \sum_{i j} \left[ H_{\downarrow}^{ij}\left(x^{i j}, x^{i+1,j}\right) +
    H_{\rightarrow}^{i j}\left(x^{i j}, x^{i,j+1}\right) \right]
\end{displaymath}
is the Hamiltonian, $x^{i j}=1,\ldots,n$ and $\beta$ is the inverse temperature. The singular value decomposition allows us to write
\begin{displaymath}
    \exp\left[ -\beta H_{q}^{ij}(x,y)\right]=\sum_{\alpha=1}^{n}
    f_{q\alpha}^{ij}(x)g_{q\alpha}^{ij}(y),
\end{displaymath}
with $q \in \{\downarrow, \rightarrow \}$. Defining the tensors
\begin{displaymath}
    X_{lrud}^{ij}=\sum_{x=1}^n
    f_{\downarrow d}^{ij}(x)g_{\downarrow u}^{ i-1,j}(x)f_{\rightarrow r}^{ij}(x)g_{\rightarrow l}^{i,j-1}(x),
\end{displaymath}
the partition function can now be calculated by contracting all 4-index tensors $X^{ij}$ arranged on a square lattice in such a way that, e.g., the indices $l,r,u,d$ of $X^{ij}$ are contracted with the indices $r,l,d,u$ of the respective tensors $X^{i,j-1},X^{i,j+1},X^{i-1,j},X^{i+1,j}$. In order to determine the expectation value of a general operator of the form $O(\{x^{i j}\})=Z \prod_{i j} O^{i j}(x^{i j})$, one just has to replace each tensor $X^{i j}$ by
\begin{displaymath}
     X_{lrud}^{i j}\left(O^{i j}\right)=\sum_{x=1}^n O^{i j}(x)
     f_{\downarrow d}^{ij}(x) g_{\downarrow u}^{ i-1,j}(x) f_{\rightarrow r}^{ij}(x) g_{\rightarrow l}^{i,j-1}.
\end{displaymath}

% =============================================

The quantum case if very similar. We consider the partition function of an inhomogeneous 1--D quantum system composed of $L$ $n$-level systems,
\begin{displaymath}
    Z=\tr \exp\left(-\beta H\right).
\end{displaymath}
It is always possible to write the Hamiltonian $H$ as a sum $H=\sum_k H_k$ with each part consisting of a sum of commuting terms. Let us, for simplicity, assume that $H=H_1+H_2$ and that only local and 2-body nearest neighbor interactions occur, i.e. $H_k=\sum_i O^{i,i+1}_k$ and $\left[O^{i,i+1}_k,O^{j,j+1}_k\right]=0$, with $i,j=1,\ldots,L$. The more general case can be treated in a similar way. Let us now consider a decomposition
\begin{equation} \label{eqn:decompop}
     \exp\left(-\frac{\beta}{M} O^{i,i+1}_k\right)=\sum_{\alpha=1}^{\kappa}
     \hat{S}^{i}_{k \alpha}\otimes
     \hat{T}^{i+1}_{k \alpha}.
\end{equation}
The singular value decomposition guarantees the existence of such an expression with $\kappa \leq n^2$. As we will see later, a smart choice of $H=\sum_k H_k$ can typically decrease~$\kappa$ drastically. Making use of the Suzuki--Trotter formula~\footnote{Note that in practice, it will be desirable to use the higher order versions of the Trotter decomposition.}
\begin{displaymath}
     Z={\rm Tr} \left(\prod_k\exp\left(-\frac{\beta}{M} H_k\right)\right)^M
     + \Omicron{\frac{1}{M}}
\end{displaymath}
it can be readily seen that the partition function can again be calculated by contracting a collection of 4-index tensors $X^{ij}$ defined as
\begin{displaymath}
    X^{ij}_{(ll')(rr')ud} \equiv
    \left[ \hat{T}^{j}_{1 l} \hat{S}^{j}_{1 r} \hat{T}^{j}_{2 l'} \hat{S}^{j}_{2 r'}
    \right]_{\left[u d\right]},
\end{displaymath}
where the indices $(l,l')$ and $(r,r')$ are combined to yield a single index that may assume values ranging from~$1$ to~$\kappa^2$. Note that now the tensors $X^{ij}$ and $X^{i'j}$ coincide, and that the indices~$u$ of the first and~$d$ of the last row have to be contracted with each other as well, which corresponds to a classical spin system with periodic boundary conditions in the vertical direction. A general expectation value of an operator of the form $O=Z O^1 \otimes \cdots \otimes O^N$ can also be reexpressed as a contraction of tensors with the same structure: it is merely required to replace each tensor~$X^{1 j}$ in the first row by
\begin{displaymath}
    X^{1 j}_{(ll')(rr')ud} \left( O^j \right) =
    \left[ O^j \hat{T}^{j}_{1 l} \hat{S}^{j}_{1 r} \hat{T}^{j}_{2 l'} \hat{S}^{j}_{2 r'}
    \right]_{\left[u d\right]}.
\end{displaymath}

% =============================================

Let us next move on to explain how the tensor contraction can be done. We will use the techniques that were originally developed in the context of PEPS in order to contract the tensors $X^{ij}$ introduced above in a controlled way. The main idea is to express the objects resulting from the contraction of tensors along the first and last column in the 2--D configuration as matrix product states and those obtained along the columns $2,3,\ldots,L-1$ as matrix product operators. More precisely, we define
 \begin{eqnarray*}
 \bra{\XX^1} & := & \sum_{r_1 \ldots r_M=1}^{m}
 \tr \left( \X^{1 1}_{r_1} \ldots \X^{M 1}_{r_M} \right)
 \bra{r_1 \ldots r_M}\\
 \ket{\XX^L} & := & \sum_{l_1 \ldots l_M=1}^{m}
 \tr \left( \X^{1 L}_{l_1} \ldots \X^{M L}_{l_M} \right)
 \ket{l_1 \ldots l_M}\\
 \XX^j & := & \sum_{l_1,r_1,\ldots=1}^{m}
 \tr \left( \X^{1 j}_{l_1 r_1} \ldots \X^{M j}_{l_M r_M} \right)
 \ket{l_1 \ldots} \bra{r_1 \ldots},
 \end{eqnarray*}
 % \begin{displaymath}
 % \XX^j := \sum_{l_1,r_1,\ldots=1}^{m}
 % \tr \left( \X^{1 j}_{l_1 r_1} \ldots \X^{M j}_{l_M r_M} \right)
 % \ket{l_1 \ldots} \bra{r_1 \ldots},
 % \end{displaymath}
where $m=n$ for 2--D~classical systems and $m=\kappa^2$ for 1--D~quantum systems. These MPS and MPOs are associated to a chain of $M$ $m$--dimensional systems and their virtual dimension amounts to $D=n$. Note that for 2--D~classical systems the first and last matrices under the trace in the MPS and MPO reduce to vectors. The partition function (and similarly other correlation functions) reads $Z=\bra{\XX^1} \XX^{2} \cdots \XX^{L-1}
\ket{\XX^{L}}$.
% \begin{equation}
%     \label{eqn:partitionfunction_MPS}
%     Z=\bra{\XX^1}
%     \XX^{2} \cdots
%     \XX^{L-1}
%      \ket{\XX^{L}}.
% \end{equation}
Evaluating this expression iteratively by calculating step by step $\bra{\XX^j}:=\bra{\XX^{j-1}} \XX^{j}$ for $j=2,\ldots,L-1$ fails because the virtual dimension of the MPS $\bra{\XX^{j}}$ increases exponentially with~$j$. A way to circumvent this problem is to replace in each iterative step the MPS $\bra{\XX^{j}{}}$ by a MPS $\bra{\XXt^{j}}$ with a reduced virtual dimension~$\tilde{D}$ that approximates the state $\bra{\XX^{j}}$ best in the sense that the norm $\delta K:=\|\bra{\XX^{j}}-\bra{\XXt^{j}} \|$ is minimized. Due to the fact that this cost function is multiquadratic in the variables of the MPS, this minimization can be carried out very efficiently;
the exponential increase of the virtual dimension can hence be prevented and the iterative evaluation of~$Z$ becomes tractable, such that an approximation to the partition function can be obtained from $Z \simeq \scal{\XXt^{L-1}}{\XX^{L}}$. The accuracy of this approximation depends only on the choice of the reduced dimension~$\tilde{D}$ and the approximation becomes exact for~$\tilde{D} \geq D^L$. As the norm $\delta K$ can be calculated at each step, $\tilde{D}$ can be increased dynamically if the obtained accuracy is not large enough. In the worst case scenario, such as in the NP-complete Ising spin glasses~\cite{barahona82}, $\tilde{D}$ will probably have to grow exponentially in $L$ for a fixed precision of the partition function. But in less pathological cases it seems that $\tilde{D}$ only has to grow polynomially in $L$; indeed, the success of the methods developed by Nishino~\cite{nishino95} in the translational invariant case indicate that even a constant $\tilde{D}$ will produce very reliable results.

% =============================================

We will illustrate this with an example of bosons in optical lattices. A system of trapped bosonic particles in a 1--D optical lattice of~$L$ sites is described by the Bose-Hubbard Hamiltonian~\cite{jaksch98}
\begin{displaymath}
 H = -J \sum_{i=1}^{L-1} ( \adj{a}_i a_{i+1} + h.c.) + \frac{U}{2}
 \sum_{i=1}^L  \hat{n}_i (\hat{n}_i-1) +
 \sum_{i=1}^L V_i \hat{n}_i,
\end{displaymath}
where $\adj{a}_i$ and $a_i$ are the creation and annihilation operators on site~$i$ and $\hat{n}_i=\adj{a}_i a_i$ is the number operator. This Hamiltonian describes the interplay between the kinetic energy due to the next-neighbor hopping with amplitude~$J$ and the repulsive on-site interaction~$U$ of the particles. The last term in the Hamiltonian models the harmonic confinement of magnitude $V_i = V_0 (i-i_0)^2$. The variation of the ratio~$U/J$ drives a phase-transition between the Mott-insulating and the superfluid phase, characterized by localized and delocalized particles respectively~\cite{fisher89}. Experimentally, the variation of~$U/J$ can be realized by tuning the depth of the optical lattice~\cite{jaksch98,buechler03}. On the other hand, one typically measures directly the momentum distribution by letting the atomic gas expand and then measuring the density distribution. Thus, we will be mainly interested here in the (quasi)--momentum
distribution
\begin{displaymath}
 n_k = \frac{1}{L} \sum_{r,s=1}^{L} \expect{\adj{a}_r a_s} e^{i 2 \pi k
 (r-s)/L}.
\end{displaymath}

Our goal is now to study with our numerical method the finite-temperature properties of this system for different ratios~$U/J$. We thereby assume that the system is in a thermal state corresponding to a grand canonical ensemble with chemical potential~$\mu$, such that the partition function is obtained as $Z=\tr e^{-\beta(H - \mu \hat{N})}$. Here, $\hat{N}=\sum_{i=1}^L \hat{n}_i$ represents the total number of particles. For the numerical study, we assume a maximal particle--number~$q$ per lattice site, such that we can project the Hamiltonian~$H$ on the subspace spanned by Fock-states with particle-numbers per site ranging from~$0$ to~$q$. The projected Hamiltonian~$\tilde{H}$ then describes a chain of~$L$ spins, with each spin acting on a Hilbert-space of dimension~$n=q+1$. A Trotter decomposition that turned out to be advantageous for this case is
\begin{equation} \label{eqn:decomp_bosehubbard}
    e^{-\beta (\tilde{H} - \mu \hat{N} )}= \left( \adj{\hat{V}} \hat{V} \right)^M+\Omicron{\frac{1}{M^2}},
\end{equation}
with $\tilde{H}=H_R+H_S+H_T$, $H_R = - \frac{J}{2} \sum_{i=1}^{L-1} R^{(i)} R^{(i+1)}$, $H_S = - \frac{J}{2} \sum_{i=1}^{L-1} S^{(i)} S^{(i+1)}$, $H_T = \sum_{i=1}^L T^{(i)}$, $R^{(i)}=\adj{\tilde{a}}_i+\tilde{a}_i$, $S^{(i)}=-i (\adj{\tilde{a}}_i-\tilde{a}_i)$, $T^{(i)}=\frac{1}{2} \tilde{n}_i (\tilde{n}_i-1)+V_i \tilde{n}_i$ and $\hat{V} = e^{-\frac{\beta}{2 M} H_R} e^{-\frac{\beta}{2M} H_S} e^{-\frac{\beta}{2M} (H_T-\mu \hat{N})}$. $\adj{\tilde{a}}_i$, $\tilde{a}_i$ and $\tilde{n}_i$ thereby denote the projections of the creation, the annihilation and the number operators $\adj{a}_i$, $a_i$ and $n_i$ on the $q$-particle subspace. The decomposition~(\ref{eqn:decompop}) of all two-particle operators in expression~(\ref{eqn:decomp_bosehubbard}) then straightforwardly leads to a set of 4-index tensors $X^{i j}_{l r u d}$, with indices~$l$ and~$r$ ranging from~$1$ to~$(q+1)^3$ and indices~$u$ and~$d$ ranging from~$1$ to~$q+1$. Note that the typical second order Trotter decomposition with $H=H_{\rm even}+H_{\rm odd}$ would make the indices~$l$ and~$r$ range from~$1$ to~$(q+1)^6$.

\begin{figure}[t]
    \begin{center}
        \includegraphics[width=\linewidth]{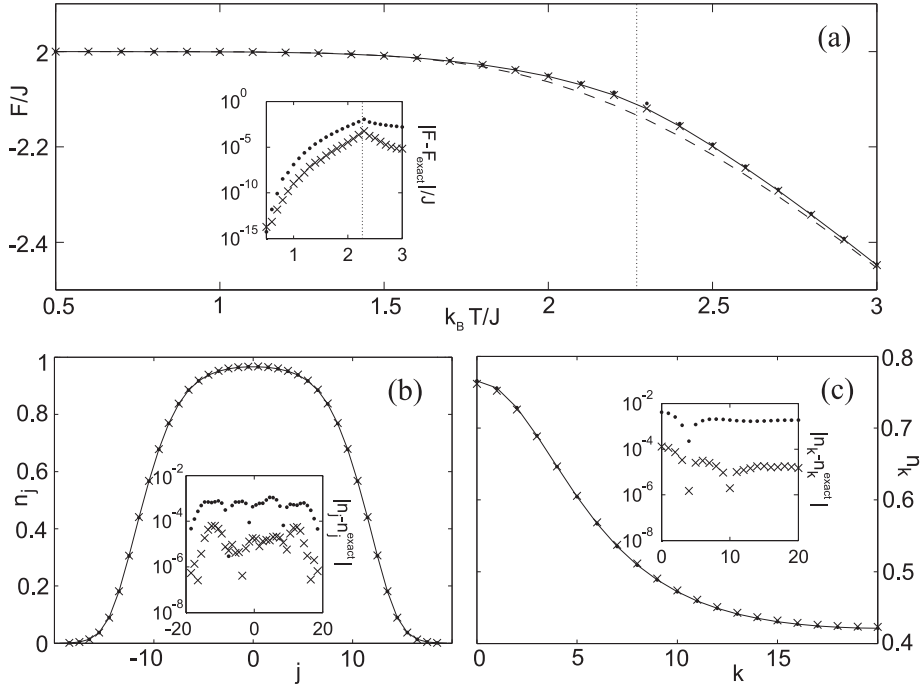}
    \end{center}
    \caption{
        (taken from \cite{murgverstraete05}) Density and (quasi)-momentum distribution in the Tonks-Girardeau gas limit, plotted for $\beta J=1$,
        $L=40$, $N=21$ and $V_0/J=0.034$. The dots (crosses) represent the numerical
        results for $\tilde{D}=2$ ($\tilde{D}=8$) and the solid
        line illustrates the exact results. From the
        insets, the error of the numerical
        results can be gathered.
        }
    \label{fig:tonks}
\end{figure}

Let us start out by considering the limit $U/J \to \infty$ in which double occupation of single lattice sites is prevented and the particles in the lattice form a Tonks--Girardeau gas~\cite{paredes04}. In this limit, the Bose-Hubbard Hamiltonian maps to the Hamiltonian of the exactly solvable (inhomogeneous) XX-model, which allows to benchmark our algorithm. The comparison of our numerical results to the exact results can be gathered from fig.~\ref{fig:tonks}. Here, the density and the (quasi)-momentum distribution are considered for the special case $\beta J=1$, $L=40$, $N=21$ and $V_0/J=0.034$. The numerical results shown have been obtained for Trotter-number~$M=10$ and two different reduced virtual dimensions~$\tilde{D}=2$ and~$\tilde{D}=8$. The norm~$\delta K$ was of order~$10^{-4}$ for $\tilde{D}=2$ and~$10^{-6}$ for~$\tilde{D}=8$~\footnote{We note that we have stopped our iterative algorithm at the point the variation of~$\delta K$ was less than~$10^{-8}$.}. From the insets, it can be gathered that the error of the numerical calculations is already very small for~$\tilde{D}=2$ (of order $10^{-3}$) and decreases significantly for~$\tilde{D}=8$. This error can be decreased further by increasing the Trotter-number~$M$.

\begin{figure}[t]
    \begin{center}
        \includegraphics[width=\linewidth]{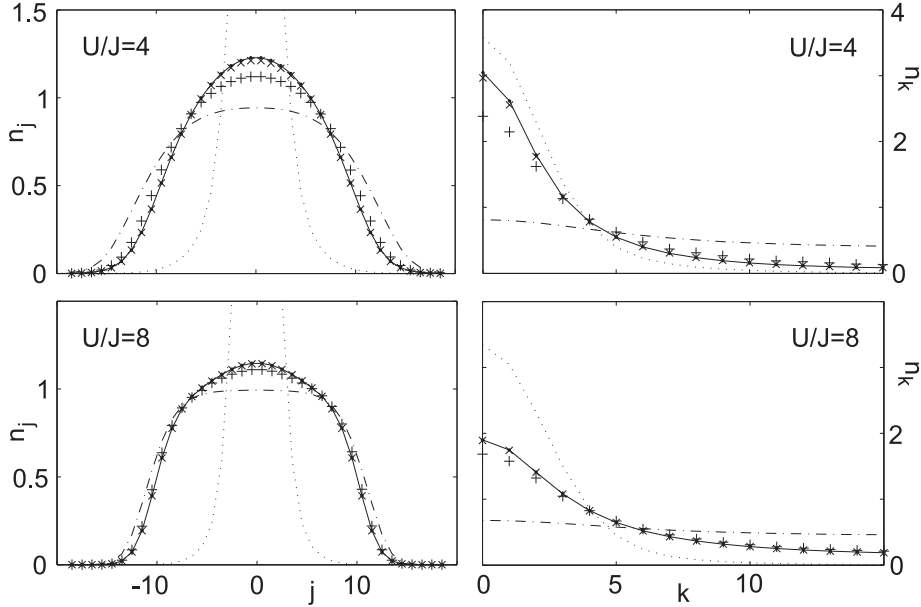}
    \end{center}
    \caption{
        (taken from \cite{murgverstraete05}) Density and (quasi)-momentum distributions for interaction strengths
        $U/J=4$ and~$8$. Here, $\beta J=1$,
        $L=40$, $N=21$ and $M=10$.
        Numerical results were obtained for~$q=2$ (plus-signs), $q=3$ (crosses) and~$q=4$ (solid line).
        For comparison, the distributions for
        $U/J=0$ (dotted lines) and $U/J \to \infty$ (dash-dotted lines) are also
        included.
        }
    \label{fig:bosehubbard_mdist}
\end{figure}

As the ratio~$U/J$ becomes finite, the system becomes physically more interesting, but lacks an exact mathematical solution. In order to judge the reliability of our numerical solutions in this case, we check the convergence with respect to the free parameters of our algorithm ($q$, $\tilde{D}$ and $M$). As an illustration, the convergence with respect to the parameter~$q$ is shown in figure~\ref{fig:bosehubbard_mdist}. In this figure, the density and the (quasi)-momentum distribution are plotted for $q=2,3$ and~$4$. We thereby assume that $\beta J=1$, $L=40$ and $N=21$ and consider interaction strengths~$U/J=4$ and~$8$. The harmonic potential~$V_0$ is chosen in a way to describe Rb-atoms in a harmonic trap of frequency~$\textrm{Hz}$ (along the lines of~\cite{paredes04}). We note that we have taken into account that changes of the ratio~$U/J$ are obtained from changes in both the on-site interaction~$U$ and the hopping amplitude~$J$ due to variations of the depth of the optical lattice. The numerical calculations have been performed with~$M=10$ and $\tilde{D}=q+1$. From the figure it can be gathered that convergence with respect to~$q$ is achieved for $q\geq3$.

\subsection{Density of States} \label{dos}
Information about the density of states can be obtained by studying the quantity
\begin{displaymath}
f(t)={\rm Tr}\left(e^{-iHt}\right)
\end{displaymath}
as a function of time. Indeed, assume that we know the function $f(t)$ exactly from $t=-\infty\rightarrow +\infty$. Its Fourier transform
\begin{displaymath}
F(\omega)=\int_{t=-\infty}^{\infty}e^{i\omega t}{\rm Tr}e^{-iHt}=\sum_{k=0}^\infty \delta\left(\omega-E_k\right)
\end{displaymath}
with $\{E_k\}$ the spectrum of the Hamiltonian $H$. Hence the Fourier transform of $f(t)$ gives all the information about the spectrum.

Of course, in practice we will not be able to determine $f(t)$ for all times, and we will only be able to approximate it within some time window. The effect on its Fourier transform is that it becomes convolved with a sinc-function of some width inversely proportional to the time window
\footnote{Of course, we can also make use of more advanced signal processing tricks to get a better conditioning. We refer to~\cite{osborne06} for more details.}.
This means that the resolution of such simulations will depend on the time frames in which we can calculate $f(t)$. Furthermore, only a discrete number of points will be available, such that we have to do a discrete Fourier transform instead, but this is still sufficient to get the spectral information we're interested in.

As first discussed by T. Osborne, the calculation of $f(t)$ can efficiently be implemented with the techniques discussed before~\cite{osborne06}. Basically, it is the real-time equivalent of calculating $e^{-\beta H}$, and as shown in the previous sections, there are several options for doing this. First of all, we can evolve a collection of maximally entangled states with $I\otimes e^{-iHt}$ using small Trotter steps, and then calculate its overlap with the original maximally entangled states (note again that several options exist for how to choose the Trotter decomposition). Indeed, it happens that
\begin{displaymath}
\langle I|A\otimes I|I\rangle={\rm Tr}A
\end{displaymath}
with $|I\rangle=\sum_k |k\rangle|k\rangle$ a maximally entangled state.

Alternatively, we  can consider the Suzuki-Trotter decomposition for an infinitesimal step, write down the complete tensor network corresponding to $f(t)$, and contract it using the techniques discussed in the previous section \ref{class}. Note that  in this particular case, there is again the possibility of contracting the corresponding tensor network in two directions, namely in the time or space direction. Experience will have to tell us which way is more reliable.

In any case, it is really exciting to see how many new tools and possibilities can be explored by reformulating it in the MPS-language.

\newpage
\section{Projected Entangled Pair States and ground states of 2-D quantum spin systems} \label{sec:peps}

In this section, we present a natural generalization of the 1D MPS to two and higher dimensions and build simulation techniques based on those states which effectively extend DMRG to higher dimensions. We call those states {\em projected entangled--pair states} (PEPS) \cite{verstraetecirac04,murgverstraete07}, since they can be understood in terms of pairs of maximally entangled states of some auxiliary systems that are locally projected in some low--dimensional subspaces. This class of states includes the generalizations of the 2D AKLT-states known as tensor product states \cite{hieida99,okunishi00,nishino00,nishio04,niggeman97,delgado01,verstraetecirac04b} which have been used for 2D problems but is much broader since every state can be represented as a PEPS (as long as the dimension of the entangled pairs is large enough). We also develop an efficient algorithm to calculate correlation functions of these PEPS, and which allows us to extend the 1D algorithms to higher dimensions. This leads to many interesting applications, such as scalable variational methods for finding ground or thermal states of spin systems in higher dimensions as well as to simulate their time-evolution. For the sake of simplicity, we will restrict to a square lattice in 2D. The generalization to higher dimensions and other geometries is straightforward.

We want to emphasize however that the PEPS method is not yet as well established as the 1-D MPS or DMRG methods; this is mainly due to its bigger complexity, but also to a big extent due to the fact that these PEPS methods are relatively unexplored which makes that there is a lot of way for improvement and exciting research.

\subsection{Construction and calculus of PEPS}

\begin{figure}[t]
    \begin{center}
        \includegraphics[width=\linewidth]{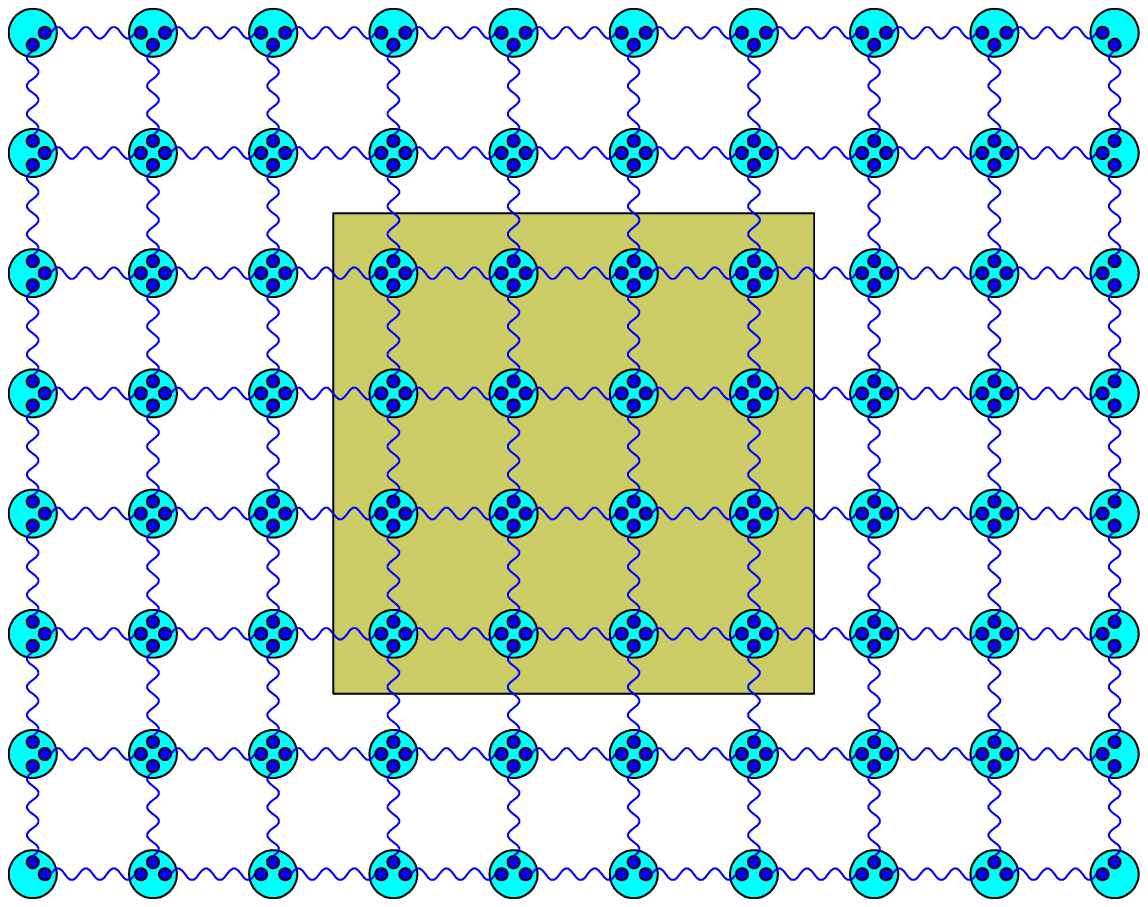}
    \end{center}
    \caption{Representation of a quantum spin system in 2 dimensions using the PEPS representation. If we calculate the entropy of a block of spins, then this quantity will obey an area law and scale as the length of the boundary between the block and the rest. PEPS-states are constructed such as to have this property build in.
    }
    \label{arealawPEPS}
\end{figure}

There have been various attempts at using the ideas developed in the context of the numerical renormalization group and DMRG to simulate 2-D quantum spin systems. However, in hindsight it is clear why those methods were never very successful: they can be reformulated as variational methods within the class of 1-dimensional matrix product states, and the structure of those MPS is certainly not well suited at describing ground states of 2-D quantum spin systems. This can immediately be understood when reconsidering the area law discussed in the first section (see figure \ref{arealawPEPS}): if we look at the number of degrees of freedom needed to describe the relevant modes in a block of spins, this has to scale as the boundary of the block, and hence this increases exponentially with the size of that boundary. This means that it is impossible to use a NRG or DMRG approach\footnote{Clearly, the VMPS/DMRG methods can reveal very valuable information in the case of quasi 2-dimensional systems such as ladders with a few rungs; see e.g. \cite{Whiteladder07} for a nice illustration.}, where the degrees of freedom is bounded to $D$.

However, it is straightforward to generalize the MPS-picture to higher dimensions: the main reason of the success of the MPS approach is that it allows to represent very well local properties that are compatible with e.g. the translational symmetry in the system. These strong local correlations are obtained by sharing maximally entangled states between neighbours, and the longer range correlations are basically mediated by the intermediate particles. This is of course a very physical picture, as the Hamiltonian does not force any long-range correlations to exist a priori, and those only come into existence because of frustration effects. This generalization to higher dimensions can therefore be obtained by distributing virtual maximally entangled states between all neighbouring sites \cite{verstraetecirac04b}, and as such a generalization of the AKLT-picture is obtained.

\begin{figure}[t]
    \begin{center}
        \includegraphics{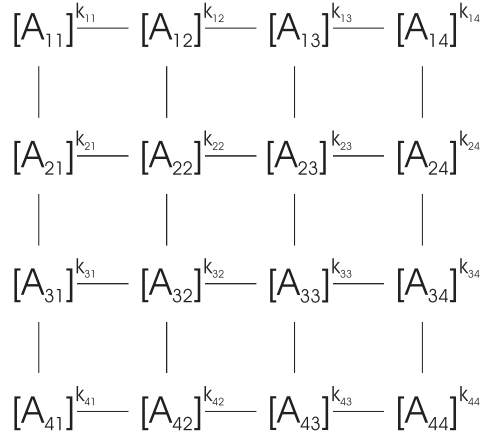}
    \end{center}
    \caption{
        Structure of the coefficient related to the state
        $\ket{k_{11},..,k_{44}}$ in the PEPS $\ket{\Psi_A}$.
       The bonds represent the indices of the tensors $[A_i]^k$
       that are contracted.
    }
    \label{fig:structpeps}
\end{figure}

More specifically, each physical system at site $i$ is represented by four auxiliary systems $a_i$, $b_i$, $c_i$, and $d_i$ of dimension~$D$ (except at the borders of the lattices). Each of those systems is in a maximally entangled state
\begin{displaymath}
\ket{I} = \sum_{i=1}^D \ket{i i}
\end{displaymath}
with one of its neighbors, as shown in the figure. The PEPS $\ket{\Psi}$ is then obtained by applying to each site one operator $Q_i$ that maps the four auxiliary systems onto one physical system of dimension~$d$. This leads to a state with coefficients that are contractions of tensors according to a certain scheme. Each of the tensors is related to one operator~$Q_i$ according to
\begin{displaymath}
\big[ A_{i} \big]^k_{lrud} = \bra{k} Q_i \ket{l,r,u,d}
\end{displaymath}
and thus associated with one lattice site~$i$. All tensors possess one physical index~$k$ of dimension~$d$ and four virtual indices~$l$, $r$, $u$ and~$d$ of dimension~$D$. The scheme according to which these tensors are contracted mimics the underlying lattice structure: the four virtual indices of the tensors are related to the left, right, upper and lower bond emanating from the corresponding lattice site. The coefficients of the PEPS are then formed by joining the tensors in such a way that all virtual indices related to same bonds are contracted. This is illustrated in fig.~\ref{fig:structpeps} for the special case of a $4 \times 4$ square lattice. Assuming this contraction of tensors is performed by the function~$\mathcal{F} (\cdot)$, the resulting PEPS can be written as
\begin{displaymath}
\ket{\Psi} = \sum_{k_1,...,k_M=1}^d \mathcal{F} \big( \big[ A_1
\big]^{k_1},...,\big[ A_M \big]^{k_M} \big) \ket{k_1,...,k_M}.
\end{displaymath}
This construction can be generalized  to any lattice shape and dimension and one can show that any state can be written as a PEPS if we allow the bond dimension to become very large. In this way, we also resolve the problem of the entropy of blocks mentioned above, since now this entropy is proportional to the bonds that connect such block with the rest, and therefore to the area of the block. Note also that, in analogy to the MPS~\cite{fannes92}, the PEPS are guaranteed to be ground states of local Hamiltonians.

There has recently been a  lot of progress in justifying this PEPS picture; M. Hastings has shown \cite{hastings07} that indeed every ground state of a local quantum spin Hamiltonian has an efficient representation in terms of a PEPS, i.e. one whose bond dimension $D$ scales subexponentially with the number of spins under interest. Also, he has shown that all thermal states have an efficient representation in terms of matrix product operators.  This is great news, as it basically shows that we have identified the relevant  manifold describing the low-energy physics of quantum spin systems. This can lead to many applications in theoretical condensed matter physics, as the questions about the possibility of some exotic phase of  matter can now be answered by looking at the set of PEPS hence skipping the bottleneck of simulation of ground states.

The family of PEPS also seems to be very relevant in the field of quantum information theory. For example, all quantum error-correcting codes such as Kitaev's toric code \cite{kitaev03} exhibiting topological quantum order have a very simple and exact description in terms of PEPS \cite{verstraetewolf06}. Furthermore, the PEPS-picture has been used to show the equivalence between different models of quantum computation  \cite{verstraetecirac04b}; more specifically, the so-called cluster states \cite{raussendorf01} have a simple interpretation in terms of PEPS, and this picture demystifies the inner workings of the one-way quantum computer.

\subsection{calculus of PEPS}

\begin{figure}[t]
    \begin{center}
        \includegraphics{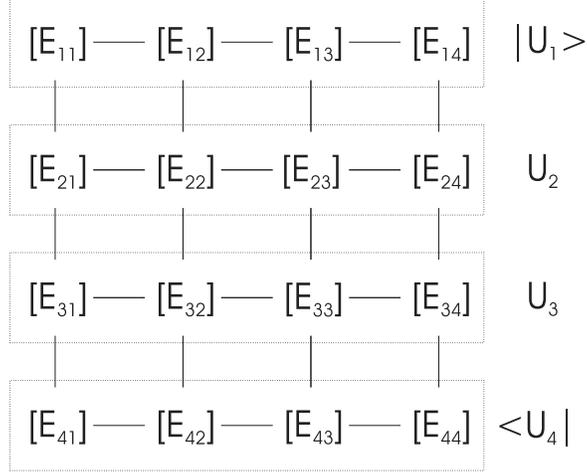}
    \end{center}
    \caption{
        Structure of the contractions in $\scal{\Psi_A}{\Psi_A}$.
    In this scheme, the first and last rows can be interpreted as
    MPS $\ket{U_1}$ and $\bra{U_4}$ and the rows in between
    as MPO $U_2$ and $U_3$. The contraction of all tensors is then
    equal to $\bra{U_4} U_3 U_2 \ket{U_1}$.
        }
    \label{fig:structe}
\end{figure}

We now show how to determine expectation values of operators in the state $\ket{\Psi}$. We consider a general operator $O = \prod_i O_i$ and define the $D^2 \times D^2 \times D^2 \times D^2$--tensors
\begin{displaymath}
\big[ E_j^{O_j} \big]_{(ll')(rr')}^{(uu')(dd')} = \sum_{k,k'=1}^d
\bra{k} O_j \ket{k'}
\big[A_j^* \big]_{lrud}^{k'} \big[ A_j \big]_{l'r'u'd'}^k.
\end{displaymath}
In this definition, the symbols $(ll')$, $(rr')$, $(uu')$ and $(dd')$ indicate composite indices. We may interpret the~$4$ indices of this tensor as being related to the~$4$ bonds emanating from site~$j$ in the lattice. Then, $\bra{\Psi} O \ket{\Psi}$ is formed by joining all tensors $E_j^{O_j}$ in such a way that all indices related to same bonds are contracted -- as in the case of the coefficients of PEPS. These contractions have a rectangular structure, as depicted in fig.~\ref{fig:structe}. In terms of the function $\mathcal{F}(\cdot)$, the expectation value reads
\begin{displaymath}
\bra{\Psi} O \ket{\Psi} = \mathcal{F} \big( E_1^{O_1},...,E_N^{O_N} \big).
\end{displaymath}
The contraction of all tensors $E_j^{O_j}$ according to this scheme requires a number of steps that scales exponentially with~$N$ -- and makes calculations intractable as the system grows larger. Because of this, an approximate method has to be used to calculate expectation values.

The approximate method suggested in~\cite{verstraetecirac04} is based on matrix product states (MPS) and matrix product operators (MPO). The main idea is to interpret the first and last row in the contraction--scheme as MPS and the rows in between as MPO. The horizontal indices thereby form the virtual indices and the vertical indices are the physical indices. Thus, the MPS and MPO have both virtual dimension and physical dimension equal to~$D^2$. Explicitly written, the MPS read
\begin{eqnarray*}
\ket{U_1} & = & \sum_{\tilde{d}_1,...,\tilde{d}_L=1}^{D^2}
\tr \Big( \big[E_{11}^{O_{11}}\big]^{1 \tilde{d}_1} \cdots \big[E_{1L}^{O_{1L}}\big]^{1 \tilde{d}_L} \Big) \ket{\tilde{d}_1,...,\tilde{d}_L}\\
\bra{U_L} & = & \sum_{\tilde{u}_1,...,\tilde{u}_L=1}^{D^2}
\tr \Big( \big[E_{L1}^{O_{L1}}\big]^{\tilde{u}_1 1} \cdots \big[E_{LL}^{O_{LL}}\big]^{\tilde{u}_L 1} \Big) \bra{\tilde{u}_1,...,\tilde{u}_L}
\end{eqnarray*}
and the MPO at row~$r$ is
\begin{displaymath}
% \begin{split}
U_r = \sum_{\begin{subarray}{1} \tilde{u}_1,...,\tilde{u}_L=1\\ \tilde{d}_1,...,\tilde{d}_L=1 \end{subarray}}^{D^2}
% &
\tr \Big( \big[E_{r1}^{O_{r1}}\big]^{\tilde{u}_1 \tilde{d}_1} \cdots \big[E_{rL}^{O_{rL}}\big]^{\tilde{u}_L \tilde{d}_L} \Big)
% \times\\
% &
% \times
\ket{\tilde{u}_1,...,\tilde{u}_L} \bra{\tilde{d}_1,...,\tilde{d}_L}.
% \end{split}
\end{displaymath}
In terms of these MPS and MPO, the expectation value is a product of MPO and MPS:
\begin{displaymath}
\bra{\Psi} O \ket{\Psi} = \bra{U_L} U_{L-1} \cdots U_2 \ket{U_1}
\end{displaymath}
The evaluation of this expression is, of course, intractable. With each multiplication of a MPO with a MPS, the virtual dimension increases by a factor of~$D^2$. Thus, after~$L$ multiplications, the virtual dimension is~$D^{2L}$ -- which is exponential in the number of rows. The expression, however, reminds of the time--evolution of a MPS. There, each multiplication with a MPO corresponds to one evolution step. The problem of the exponential increase of the virtual dimension is circumvented by restricting the evolution to the subspace of MPS with a certain virtual dimension~$\tilde{D}$. This means that after each evolution step the resulting MPS is approximated by the "nearest" MPS with virtual dimension~$\tilde{D}$. This approximation can be done efficiently, as shown in~\cite{verstraeteripoll04}. In this way, also $\bra{\Psi} O \ket{\Psi}$ can be calculated efficiently: first, the MPS $\ket{U_2}$ is formed by multiplying the MPS $\ket{U_1}$ with MPO $U_2$. The MPS $\ket{U_2}$ is then approximated by $\ket{\tilde{U}_2}$ with virtual dimension~$\tilde{D}$. In this fashion the procedure is continued until $\ket{\tilde{U}_{L-1}}$ is obtained. The expectation value $\bra{\Psi} O \ket{\Psi}$ is then simply
\begin{displaymath}
\bra{\Psi} O \ket{\Psi} = \scal{U_L}{\tilde{U}_{L-1}}.
\end{displaymath}

Interestingly enough, this method to calculate expectation values can be adopted to develop very efficient algorithms to determine the ground states of 2D Hamiltonians and the time evolution of PEPS by extenting DMRG and the time evolution schemes to 2D.

\subsection{Variational method with PEPS}

Let us start with an algorithm to determine the ground state of a Hamiltonian with short range interactions on a square $L\times L$ lattice. The goal is to determine the PEPS $\ket{\Psi}$ with a given dimension $D$ which minimizes the energy:
\begin{equation} \label{eqn:hexpect1}
\expect{H} = \frac{\bra{\Psi} H \ket{\Psi}}{\scal{\Psi}{\Psi}}
\end{equation}
Following~\cite{verstraeteporras04}, the idea is to iteratively optimize the tensors $A_i$ one by one while fixing all the other ones until convergence is reached. The crucial observation is the fact that the exact energy of $\ket{\Psi}$ (and also its normalization) is a quadratic function of the components of the tensor $A_i$ associated with \emph{one} lattice site~$i$. Because of this, the optimal parameters $A_i$ can simply be found by solving a generalized eigenvalue problem.

The challenge that remains is to calculate the matrix--pair for which the generalized eigenvalues and eigenvectors shall be obtained. In principle, this is done by contracting all indices in the expressions $\bra{\Psi} H \ket{\Psi}$ and $\scal{\Psi}{\Psi}$ except those connecting to $A_i$. By interpreting the tensor $A_i$ as a $d D^4$--dimensional vector $\boldsymbol{A}_i$, these expressions can be written as
\begin{eqnarray}
\label{eqn:expect}
\bra{\Psi} H \ket{\Psi} & = & \adj{\boldsymbol{A}_i} \mathcal{H}_i \boldsymbol{A}_i \\
\label{eqn:norm}
\scal{\Psi}{\Psi} & = & \adj{\boldsymbol{A}_i}  \mathcal{N}_i \boldsymbol{A}_i.
\end{eqnarray}
Thus, the minimum of the energy is attained by the generalized eigenvector $\boldsymbol{A}_i$ of the matrix--pair $(\mathcal{H}_i,\mathcal{N}_i)$ to the minimal eigenvalue~$\mu$:
\begin{displaymath}
\mathcal{H}_i \boldsymbol{A}_i = \mu  \mathcal{N}_i \boldsymbol{A}_i
\end{displaymath}

It turns out that the matrix--pair $(\mathcal{H}_i,\mathcal{N}_i)$ can be efficiently evaluated by the method developed for the calculation of expectation values: $\mathcal{N}_i$ relies on the contraction of all but one tensors $E_j^{\openone}$ (with $\openone$ denoting the identity) according to the same rectangular scheme as before. The one tensor that has to be omitted is $E_i^{\openone}$ -- the tensor related to site~$i$. Assuming this contraction is performed by the function~$\mathcal{G}_i(\cdot)$, $\mathcal{N}_i$ can be written as
\begin{displaymath}
\big[ \mathcal{N}_i \big]_{lrud}^{k}
{\phantom{\big]}}_{k'}^{l'r'u'd'} = \mathcal{G}_i \big(
E_1^{\openone},...,E_N^{\openone} \big)_{lrud}^{l'r'u'd'} \delta_{k'}^k.
\end{displaymath}
If we join the indices $(klrud)$ and $(k'l'r'u'd')$, we obtain the $d D^4 \times d D^4$--matrix that fulfills equation~(\ref{eqn:norm}). To evaluate $\mathcal{G}_i(\cdot)$ efficiently, we proceed in the same way as before by interpreting the rows in the contraction--structure as MPS and MPO. First, we join all rows that lie above site~$i$ by multiplying the topmost MPS~$\ket{U_1}$ with subjacent MPO and reducing the dimension after each multiplication to~$\tilde{D}$. Then, we join all rows lying below~$i$ by multiplying~$\bra{U_L}$ with adjacent MPO and reducing the dimension as well. We end up with two MPS of virtual dimension~$\tilde{D}$ -- which we can contract efficiently with all but one of the tensors~$E_j^{\openone}$ lying in the row of site~$i$.

The effective Hamiltonian $\mathcal{H}_i$ can be determined in an analogous way, but here the procedure has to be repeated for every term in the Hamiltonian (i.e. in the order of $2N$ times in the case of nearest neighbor interactions). Assuming a single term in the Hamiltonian has the tensor--product structure $H^s \equiv \prod_i h_i^s$, the effective Hamiltonian $\mathcal{H}_i^s$ corresponding to this term is obtained as
\begin{displaymath}
\big[ \mathcal{H}_i^s \big]_{lrud}^{k}
{\phantom{\big]}}_{k'}^{l'r'u'd'} = \mathcal{G}_i \big(
E_1^{h_1^s},...,E_N^{h_N^s} \big)_{lrud}^{l'r'u'd'} \big[ h_i^s \big]_{k'}^k.
\end{displaymath}
The complete effective Hamiltonian $\mathcal{H}_i$ that fulfills equation~(\ref{eqn:expect}) is then produced as
\begin{displaymath}
\mathcal{H}_i = \sum_s \mathcal{H}_i^s.
\end{displaymath}

Thus, both the matrices $\mathcal{N}_i$ and $\mathcal{H}_i$ are directly related to the expressions $\mathcal{G}_i \big(E_1^{\openone},...,E_N^{\openone} \big)$ and $\mathcal{G}_i \big(E_1^{h_1^s},...,E_N^{h_N^s} \big)$. These expressions, however, can be evaluated efficiently using the approximate method introduced before for the calculation of expectation values. Therefore, the optimal $A_i$ can be determined, and one can proceed with the following site, iterating the procedure until convergence.

\subsection{Time evolution with PEPS} \label{sec:peps:timeevol}

Let us next move to describe how a time--evolution can be simulated on a PEPS. We will assume that the Hamiltonian only couples nearest neighbors, although more general settings can be considered. The principle of simulating a time--evolution step is as follows: first, a PEPS $\ket{\Psi_A^0}$ with physical dimension~$d=2$ and virtual dimension~$D$ is chosen as a starting state. This state is evolved by the time--evolution operator~$U=e^{-i H \delta t}$ (we assume $\hbar=1$) to yield another PEPS $\ket{\Psi_B}$ with a virtual dimension~$D_B$ increased by a factor~$\eta$:
\begin{displaymath}
\ket{\Psi_B} = U \ket{\Psi_A^0}
\end{displaymath}
The virtual dimension of this state is then reduced to~$D$ by calculating a new PEPS $\ket{\Psi_A}$ with virtual dimension~$D$ that has minimal distance to $\ket{\Psi_B}$. This new PEPS is the starting state for the next time--evolution step. The crucial point in simulating a time--evolution with PEPS is thus the development of an efficient algorithm for reducing the virtual dimension of a PEPS.

Before formulating this algorithm, let us recite how to express the product $U \ket{\Psi_A^0}$ in terms of a PEPS. This is done by means of a Trotter--approximation: first, the interaction--terms in~$H$ are classified in \emph{horizontal} and \emph{vertical} according to their orientation and in \emph{even} and \emph{odd} depending on whether the interaction is between even--odd or odd--even rows (or columns). The Hamiltonian can then be decomposed into a \emph{horizontal--even}, a \emph{horizontal--odd}, a \emph{vertical--even} and a \emph{vertical--odd} part:
\begin{displaymath}
H = H_{he} + H_{ho} + H_{ve} + H_{vo}
\end{displaymath}
The single--particle operators of the Hamiltonian can simply be incorporated in one of the four parts (note that different Trotter decompositions are again possible, e.g. grouping all Pauli operators of the same kind in 3 different groups as we discussed earlier, and in some cases this leads to a clear computational advantage). Using the Trotter--approximation, the time--evolution operator~$U$ can be written as a product of four evolution--operators:
\begin{equation} \label{eqn:uapprox}
U = e^{-i H \delta t} \approx
e^{-i H_{he} \delta t} e^{-i H_{ho} \delta t} e^{-i H_{ve} \delta t} e^{-i H_{vo} \delta t}
\end{equation}
Since each of the four parts of the Hamiltonian consists of a sum of commuting terms, each evolution--operator equals a product of two--particle operators $w_{ij}$ acting on neighboring sites~$i$ and~$j$. These two--particle operators have a Schmidt--decomposition consisting of, say, $\eta$ terms:
\begin{displaymath}
w_{ij} = \sum_{\rho=1}^{\eta} u_i^{\rho} \otimes v_j^{\rho}
\end{displaymath}
One such two--particle operator~$w_{ij}$ applied to the PEPS~$\ket{\Psi_A^0}$ modifies the tensors~$A_i^0$ and~$A_j^0$ associated with sites~$i$ and~$j$ as follows: assuming the sites~$i$ and~$j$ are horizontal neighbors, $A_i^0$ has to be replaced by
\begin{displaymath}
\big[B_i\big]^k_{l(r\rho)ud} = \sum_{k'=1}^d \big[u_{i}^{\rho}\big]^k_{k'} \big[A_i^0\big]^{k'}_{lrud}
\end{displaymath}
and $A_j^0$ becomes
\begin{displaymath}
\big[B_j\big]^k_{(l\rho)rud} = \sum_{k'=1}^d \big[v_{j}^{\rho}\big]^k_{k'} \big[A_j^0\big]^{k'}_{lrud}.
\end{displaymath}
These new tensors have a joint index related to the bond between sites~$i$ and~$j$. This joint index is composed of the original index of dimension~$D$ and the index~$\rho$ of dimension~$\eta$ that enumerates the terms in the Schmidt--decomposition. Thus, the effect of the two--particle operator~$w_{ij}$ is to increase the virtual dimension of the bond between sites~$i$ and~$j$ by a factor of~$\eta$. Consequently, $e^{-i H_{he} \delta t}$ and $e^{-i H_{ho} \delta t}$ increase the dimension of every second horizontal bond by a factor of~$\eta$; $e^{-i H_{ve} \delta t}$ and $e^{-i H_{vo} \delta t}$ do the same for every second vertical bond. By applying all four evolution--operators consecutively, we have found an approximate form of the time--evolution operator~$U$ that -- when applied to a PEPS $\ket{\Psi_A^0}$ -- yields another PEPS $\ket{\Psi_B}$ with a virtual dimension multiplied by a constant factor~$\eta$.

The aim of the approximate algorithm is now to optimize the tensors $A_i$ related to a PEPS $\ket{\Psi_A}$ with virtual dimension $D$, such that the distance between $\ket{\Psi_A}$ and $\ket{\Psi_B}$ tends to a minimum. The function to be minimized is thus
\begin{displaymath}
K \big( A_1,...,A_M \big) = \big\| \ket{\Psi_A} - \ket{\Psi_B}
\big\|^2.
\end{displaymath}
This function is non--convex with respect to all parameters $\{A_1,...,A_M\}$. However, due to the special structure of PEPS, it is quadratic in the parameters~$A_i$ associated with \emph{one} lattice site~$i$. Because of this, the optimal parameters~$A_i$ can simply be found by solving a system of linear equations. The concept of the algorithm is to do this one--site optimization site-by-site until convergence is reached.

The coefficient matrix and the inhomogeneity of the linear equations system can be calculated efficiently using the method developed for the calculation of expectation values. In principle, they are obtained by contracting all indices in the expressions for the scalar--products $\scal{\Psi_A}{\Psi_A}$ and $\scal{\Psi_A}{\Psi_B}$ except those connecting to ~$A_i$. By interpreting the tensor $A_i$ as a $d D^4$-dimensional vector $\boldsymbol{A}_i$, these scalar--products can be written as
\begin{eqnarray}
\label{eqn:scalaa}
\scal{\Psi_A}{\Psi_A} & = & \adj{\boldsymbol{A}_i} \mathcal{N}_i
\boldsymbol{A}_i \\
\label{eqn:scalab}
\scal{\Psi_A}{\Psi_B} & = & \adj{\boldsymbol{A}_i} \mathcal{W}_i.
\end{eqnarray}
Since
\begin{displaymath}
K = \scal{\Psi_B}{\Psi_B} + \scal{\Psi_A}{\Psi_A} - 2 Re
\scal{\Psi_A}{\Psi_B},
\end{displaymath}
the minimum is attained as
\begin{displaymath}
\mathcal{N}_i \boldsymbol{A}_i = \mathcal{W}_i.
\end{displaymath}

The efficient calculation of $\mathcal{N}_i$ has already been described in the previous section. The scalar product $\scal{\Psi_A}{\Psi_B}$ and the inhomogeneity $\mathcal{W}_i$ are calculated in an efficient way following the same ideas. First, the $D D_B \times D D_B \times D D_B \times D D_B$--tensors
\begin{displaymath}
\big[ F_j \big]_{(ll')(rr')}^{(uu')(dd')} = \sum_{k=1}^d \big[
A_j^* \big]_{lrud}^k \big[ B_j \big]_{l'r'u'd'}^k
\end{displaymath}
are defined. The scalar--product $\scal{\Psi_A}{\Psi_B}$ is then obtained by contracting all tensors~$F_j$ according to the previous scheme -- which is performed by the function~$\mathcal{F}(\cdot)$:
\begin{displaymath}
\scal{\Psi_A}{\Psi_B} = \mathcal{F} \big( F_1,...,F_M \big)
\end{displaymath}
The inhomogenity $\mathcal{W}_i$ relies on the contraction of all but one of the tensors $F_j$, namely the function $\mathcal{G}_i
\big(\cdot)$, in the sense that
\begin{displaymath}
\big[ \mathcal{W}_i \big]_{lrud}^k = \sum_{l'r'u'd'=1}^D
\mathcal{G}_i \big( F_1,...,F_M \big)_{lrud}^{l'r'u'd'}
\big[B_i\big]_{l'r'u'd'}^k.
\end{displaymath}
Joining all indices $(klrud)$ in the resulting tensor leads to the vector of length $d D^4$ that fulfills equation~(\ref{eqn:scalab}). Thus, both the scalar--product $\scal{\Psi_A}{\Psi_B}$ and the inhomogenity $\mathcal{W}_i$ are directly related to the expressions $\mathcal{F}\big( F_1,...,F_M \big)$ and $\mathcal{G}_i \big( F_1,...,F_M \big)$. These expressions, however, can be evaluated efficiently using the approximate method from before.

Even though the principle of simulating a time--evolution step has been recited now, the implementation in this form is numerically expensive.
This is why we append some notes about how to make the simulation more efficient:\\
\emph{1.- Partitioning of the evolution:} The number of required numerical operations decreases significantly as one time--evolution step is partitioned into~$4$ substeps: first the state~$\ket{\Psi_A^0}$ is evolved by $e^{-i H_{vo} \delta t}$ only and the dimension of the increased bonds is reduced back to~$D$. Next, evolutions according to $e^{-i H_{ve} \delta t}$, $e^{-i H_{ho} \delta t}$ and $e^{-i H_{he} \delta t}$ follow. Even though the partitioning increases the number of evolution steps by a factor of~$4$, the number of multiplications in one evolution step decreases by a factor of~$\eta^3$.\\
\emph{2.- Optimization of the contraction order:} Most critical for the efficiency of the numerical simulation is the order in which the contractions are performed. We have optimized the order in such a way that the scaling of the number of multiplications with the virtual dimension~$D$ is minimal. For this, we assume that the dimension~$\tilde{D}$ that tunes the accuracy of the approximate calculation of $\mathcal{N}_i$ and $\mathcal{W}_i$ is proportional to $D^2$, i.e. $\tilde{D}=\kappa D^2$. The number of required multiplications is then of order\footnote{The scaling $D^{10}$ is obtained when at all steps in the algorithm, a sparse matrix algorithm is used. In particular,  we have to use an iterative sparse method for solving the linear set of equations in the approximation step.} $\kappa^2 D^{10} L^2$ and the required memory scales as~$d \eta \kappa^2 D^8$.\\
\emph{3.- Optimization of the starting state:} The number of sweeps required to reach convergence depends on the choice of the starting state for the optimization. The idea for finding a good starting state is to reduce the bonds with increased virtual dimension~$\eta D$ by means of a Schmidt--decomposition. This is done as follows: assuming the bond is between the horizontal neighboring sites~$i$ and~$j$, the contraction of the tensors associated with these sites, $B_i$ and $B_j$, along the bond $i$--$j$ forms the tensor
\begin{displaymath}
\big[\mathcal{M}_{ij}\big]^k_{lud} {\phantom{\big]}}^{k'}_{r'u'd'} =
\sum_{\rho=1}^{D\eta} \big[B_i\big]^k_{l \rho ud} \big[B_j\big]^{k'}_{\rho r'u'd'}.
\end{displaymath}
By joining the indices $(k l u d)$ and $(k' r' u' d')$, this tensor can be interpreted as a $d D^3 \times d D^3$--matrix. The Schmidt--decomposition of this matrix is
\begin{displaymath}
\mathcal{M}_{ij} = \sum_{\rho=1}^{d D^3} c_{\rho} \mathcal{A}_i^{\rho} \otimes \mathcal{A}_j^{\rho}
\end{displaymath}
with the Schmidt--coefficients $c_{\rho}$ ($c_{\rho} \geq 0$) and corresponding matrices $\mathcal{A}_i^{\rho}$ and $\mathcal{A}_j^{\rho}$. We can relate these matrices to a new pair of tensors~$A_i^0$ and~$A_j^0$ associated with sites~$i$ and~$j$:
\begin{eqnarray*}
\big[A_i^0\big]^k_{l \rho ud} & = & \sqrt{c_{\rho}} \big[\mathcal{A}_i^{\rho}\big]_{lud}^k\\
\big[A_j^0\big]^k_{\rho rud} & = & \sqrt{c_{\rho}} \big[\mathcal{A}_j^{\rho}\big]_{rud}^k
\end{eqnarray*}
The virtual dimension of these new tensors related to the bond between sites~$i$ and~$j$ is equal to the number of terms in the Schmidt--decomposition. Since these terms are weighted with the Schmidt--coefficients~$c_{\rho}$, it is justified to keep only the~$D$ terms with coefficients of largest magnitude. Then, the contraction of the tensors~$A_i^0$ and~$A_j^0$ along the bond $i$--$j$ with dimension~$D$ yields a good approximation to the true value $\mathcal{M}_{ij}$:
\begin{displaymath}
\big[\mathcal{M}_{ij}\big]^k_{lud} {\phantom{\big]}}^{k'}_{r'u'd'} \approx
\sum_{\rho=1}^{D} \big[A_i^0\big]^k_{l \rho ud} \big[A_j^0\big]^{k'}_{\rho r'u'd'}.
\end{displaymath}
This method applied to all bonds with increased dimension provides us with the starting state for the optimization.

\subsubsection{Examples}

Let us now illustrate the variational methods with some examples. Models to which the PEPS algorithms have already been applied to include the Heisenberg antiferromagnet~\cite{verstraetecirac04}, the Shastry-Sutherland model~\cite{isacsson06} and the system of hard--core bosons in a 2D optical lattice~\cite{murgverstraete07}. In the following, we recite the results for the latter system -- which include calculations of ground state properties and studies of the time--evolution after sudden changes in the parameters.

The system of bosons in a 2D optical lattice is characterized by the Bose--Hubbard Hamiltonian
\begin{displaymath}
H = -J \sum_{<i,j>} \big( \adj{a_i} a_j +h.c.\big) + \frac{U}{2} \sum_i \hat{n}_i (\hat{n}_i -1)
+ \sum_i V_i \hat{n}_i,
\end{displaymath}
where $\adj{a}_i$ and $a_i$ are the creation and annihilation operators on site~$i$ and $\hat{n}_i=\adj{a}_i a_i$ is the number operator. This Hamiltonian describes the interplay between the kinetic energy due to the next-neighbor hopping with amplitude~$J$ and the repulsive on-site interaction~$U$ of the particles. The last term in the Hamiltonian models the harmonic confinement of magnitude $V_i = V_0 (i-i_0)^2$. Since the total number of particles $\hat{N}=\sum_i \hat{n}_i$ is a symmetry of the Hamiltonian, the ground--state will have a fixed number of particles. This number can be chosen by appending the term $-\mu \hat{N}$ to the Hamiltonian and tuning the chemical potential~$\mu$. In the limit of hard--core interaction, $U/J \to \infty$, two particles are prevented from occupying a single site. This limit is especially interesting in one dimension where the particles form the so--called Tonks-Girardeau gas~\cite{girardeau60,paredes04}. The particles in this gas are strongly correlated -- which leads to algebraically decaying correlation functions. In two dimensions, the model was studied in detail in~\cite{kennedy88}. In the hard--core limit, the Bose--Hubbard model is equivalent to a spin--system with $XX$--interactions described by the Hamiltonian
\begin{displaymath}
H = -\frac{J}{2} \sum_{<i,j>} \big( \sx{i} \sx{j} + \sy{i} \sy{j} \big) +
\frac{1}{2} \sum_i \big( V_i-\mu \big) \sz{i}.
\end{displaymath}
Here, $\sx{i}$, $\sy{i}$ and $\sz{i}$ denote the Pauli-operators acting on site~$i$. This Hamiltonian has the structure that can be simulated with the PEPS algorithm: it describes $L^2$ physical systems of dimension~$d=2$ on a $L \times L$--square lattice.

\begin{figure}[t]
    \begin{center}
        \includegraphics{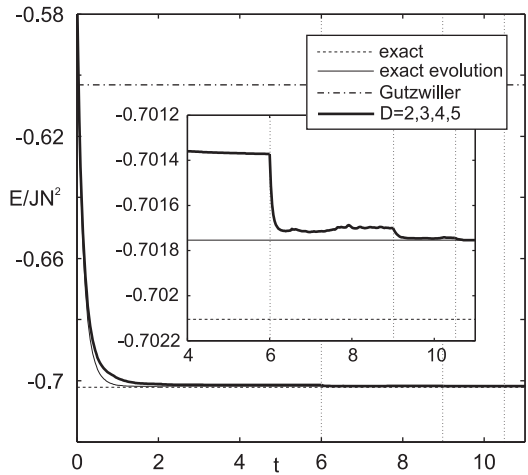}
     \end{center}
    \caption{
    (taken from \cite{murgverstraete07}) Energy as a function of time for the imaginary time--evolution of
    the system of hard--core bosons on a $4 \times 4$--lattice.
    The evolutions are performed sequentially with
    PEPS of virtual dimension $D=2$, $D=3$, $D=4$ and $D=5$.
    The times at which $D$ is increased are indicated by
    vertical lines.
    For comparison, the exact ground state--energy,
    the exact imaginary time--evolution
    and the energy of the optimal Gutzwiller ansatz
    are included.
       }
    \label{fig:gstonks4}
\end{figure}

In fig.~\ref{fig:gstonks4}, the energy in the case of a $4 \times 4$--lattice is plotted as the system undergoes an imaginary time--evolution. Thereby, a time--step $\delta t = - i 0.03$ is assumed and the magnitude of the harmonic confinement (in units of the tunneling--constant) is chosen as $V_0/J=36$. In addition, the chemical potential is tuned to~$\mu/J=3.4$ such that the ground state has particle--number~$\expect{\hat{N}}=4$. With this configuration, the imaginary time--evolution is performed both exactly and variationally with PEPS. As a starting state a product state is used that represents a Mott-like distribution with $4$~particles arranged in the center of the trap and none elsewhere. The variational calculation is performed with~$D=2$ first until convergence is reached; then, evolutions with $D=3$, $D=4$ and $D=5$ follow. At the end, a state is obtained that is very close to the state obtained by exact evolution. The difference in energy is $| E_{D=5}-E_{exact} | \backsimeq 6.4614 \cdot 10^{-5} J$. For comparison, also the exact ground--state energy obtained by an eigenvalue--calculation and the energy of the optimal Gutzwiller ansatz are included in fig.~\ref{fig:gstonks4}. The difference between the exact result and the results of the imaginary time--evolution is due to the Trotter--error and is of order $O(\delta t^2)$. The energy of the optimal Gutzwiller-Ansatz is well seperated from the exact ground--state energy and the results of the imaginary time--evolution.

\begin{figure}[t]
    \begin{center}
        \includegraphics{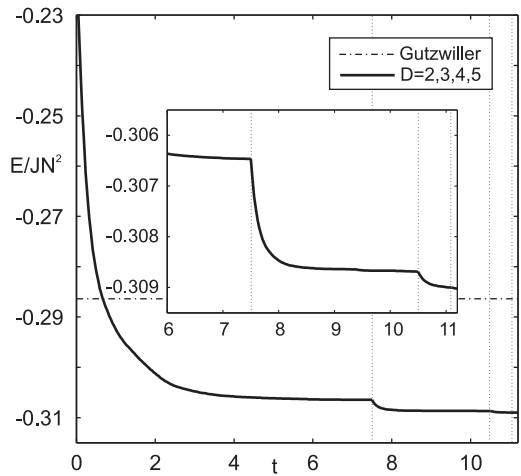}
     \end{center}
    \caption{
    (taken from \cite{murgverstraete07}) Energy as a function of time for the imaginary time--evolution of the system of hard--core bosons on a $11 \times 11$--lattice. The evolutions are performed sequentially with PEPS of virtual dimension $D=2$, $D=3$, $D=4$ and $D=5$. The times at which $D$ is increased are indicated by vertical lines. For comparison, the energy of the optimal Gutzwiller ansatz is included.
    }
    \label{fig:gstonks11}
\end{figure}

In fig.~\ref{fig:gstonks11}, the energy as a function of time is plotted for the imaginary time--evolution on the $11 \times 11$--lattice. Again, a time--step $\delta t = - i 0.03$ is assumed for the evolution. The other parameters are set as follows: the ratio between harmonic confinement and the tunneling constant is chosen as $V_0/J=100$ and the chemical potential is tuned to $\mu/J=3.8$ such that the total number of particles~$\expect{\hat{N}}$ is~$14$. The starting state for the imaginary time--evolution is, similar to before, a Mott-like distribution with $14$ particles arranged in the center of the trap. This state is evolved within the subset of PEPS with $D=2$, $D=3$, $D=4$ and $D=5$. As can be gathered from the plot, this evolution shows a definite convergence. In addition, the energy of the final PEPS lies well below the energy of the optimal Gutzwiller ansatz.

\begin{figure}[t]
    \begin{center}
         \includegraphics{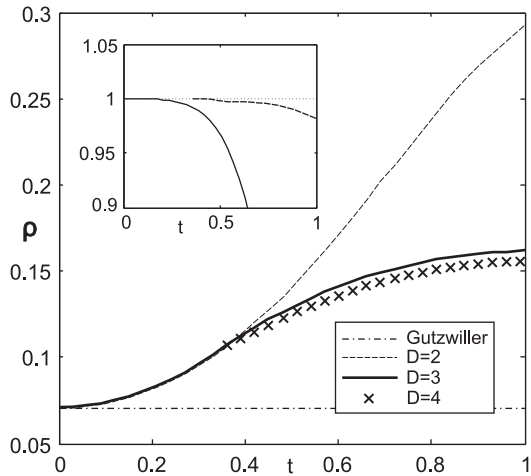}
     \end{center}
    \caption{
    (taken from \cite{murgverstraete07}) Time evolution of the condensate density
    starting from a Mott--distribution
    with $14$--particles arranged in the center of the trap.
    The magnitude of the trapping potential is $V_0/J = 100$.
    For the evolution, the Gutzwiller ansatz and PEPS with $D=2$, $D=3$ and $D=4$ are used.
    The inset shows the overlap between the $D=2$ and $D=3$--PEPS
    (solid line) and the $D=3$ and $D=4$--PEPS (dashed line).
    }
    \label{fig:creation}
\end{figure}

An application of the time--evolution algorithm with PEPS is found in the study of dynamic properties of hard--core bosons on a lattice of size $11 \times 11$. Here, the responses of this system to sudden changes in the parameters are investigated and the numerical results are compared to the results obtained by a Gutzwiller ansatz. An interesting property that is observed is the fraction of particles that are condensed. For interacting and finite systems, this property is measured best by the condensate density~$\rho$ which is defined as largest eigenvalue of the correlation--matrix~$\expect{\adj{a_i} a_j}$.

In fig.~\ref{fig:creation}, the evolution of a Mott-distribution with $14$ particles arranged in the center of the trap is studied. It is assumed that $V_0/J=100$, $\mu/J=3.8$ and $\delta t =0.03$. To assure that the results are accurate, the following procedure was used for simulating the time--evolution: first, the simulation has been performed using PEPS with $D=2$ and $D=3$ until the overlap between these two states fell below a certain value. Then, the simulation has been continued using PEPS with $D=3$ and $D=4$ as long as the overlap between these two states was close to~$1$. The results of this calculation can be gathered from fig.~\ref{fig:creation}. What can be observed is that there is a definite increase in the condensate fraction. The Gutzwiller ansatz is in contrast to this result since it predicts that the condensate density remains constant. The inset in fig.~\ref{fig:creation} shows the overlap of the $D=2$ with the $D=3$--PEPS and the $D=3$ with the $D=4$--PEPS.

\begin{figure}[t]
    \begin{center}
         \includegraphics{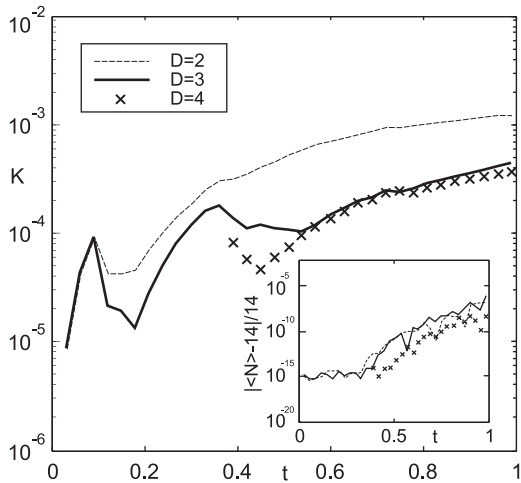}
     \end{center}
    \caption{
    (taken from \cite{murgverstraete07}) Distance $K$ between the time--evolved state and the state with reduced virtual dimension. The virtual dimensions $D=2$, $D=3$ and $D=4$ are included. The distance is plotted for the evolution of a Mott-distribution with $N=14$, as explained in fig.~\ref{fig:creation}. From the inset, the deviation of the particle number from the value~$14$ can be gathered.
    }
    \label{fig:kvalues}
\end{figure}

Finally, we make a few comments about the accuracy of the algorithm. One indicator for the accuracy is the distance between the time--evolved state and the state with reduced virtual dimension. For the time--evolution  of the Mott--distribution that was discussed before, this quantity is plotted in fig.~\ref{fig:kvalues}. We find that the distance is typically of order $10^{-3}$ for $D=2$ and of order $10^{-4}$ for $D=3$ and $D=4$. Another quantity that is monitored is the total number of particles $\expect{\hat{N}}$. Since this quantity is supposed to be conserved during the whole evolution, its fluctuations indicate the reliability of the algorithm. From the inset in fig.~\ref{fig:kvalues}, the fluctuations of the particle number in case of the time--evolution of the Mott--distribution can be gathered. We find that these fluctuations are at most of order~$10^{-5}$.

\subsubsection{PEPS and fermions}

The critical reader should by now have complained that we are only talking about spin systems but not about  fermionic systems. Indeed, one of the long term goals of the numerical approaches discussed here is to be able to simulate e.g. the Fermi-Hubbard model in the relevant parameter regime.

The methods that we discussed in 1 dimension are perfectly applicable to fermionic systems, as the amazing Jordan-Wigner transformation allows to map a local Hamiltonian of fermions to a local Hamiltonian of spins, and we know that the MPS-techniques work provably well on the latter.  The big problem however is that the Jordan-Wigner transformation only works in 1 dimension: if we use it on a 2-D lattice, a local Hamiltonian of fermions is mapped to a highly nonlocal Hamiltonian of spins. PEPS on the other hand are devised to be such that they have extremal local properties; if the Hamiltonian contains a lot of strong nonlocal terms, we can not expect a PEPS to exhibit the corresponding extremal long-range correlations. The natural question to ask is therefore whether there exists a generalization of the Jordan-Wigner transformation to higher dimensions. This was indeed shown to be possible in~\cite{verstraetecirac05b}: given any local Hamiltonian in terms of fermionic operators such as the Hubbard model in 2 or 3 dimensions, then there exists a local spin $3/2$ Hamiltonian whose low-energy sector corresponds exactly to the original fermionic Hamiltonian. The conclusion is that the PEPS methods are equally applicable to quantum spin systems as to fermionic systems.

Another and more efficient approach is to make use of quantum numbers. In general, it is difficult to keep track of quantum numbers on a 2-D lattice. An exception however is given by the parity of the occupation number between two sites: by blocking the PEPS tensors in a specific way, one can invoke the even or odd occupation number between any sites and as such eliminate the fact that sprurious effective long-range interactions arise in fermionic lattice systems \cite{inpreparation}. This is very relevant as it leads to a huge speed-up of the algorithms for fermionic lattice systems.

\subsection{PEPS on infinite lattices}

In parallel with the 1-dimensional MPS, we can also explicitly make use of the translational symmetry in the class of PEPS to simulate low-energy properties of infinite lattices. A naive procedure is to impose complete translational symmetry and do a line-search (such as conjugate gradient) within the parameter space of the tensor; note that this can be done as expectation values of the corresponding PEPS can be calculated using the appropriate 1-D infinite algorithms discussed above. However, the cost function is highly nonlinear and the number of parameters grows very fast with $D$, such that those brute-force methods will typically lead to local minima. A smarter approach is to use imaginary time evolution, but keeping the translational invariance. It was already shown how to do this in the 1-D case, where a particular kind of Trotter expansion lead to a translational invariant matrix product operator, and we can repeat this in 2-D leading to  a translational invariant PEPS-operator. The only nontrivial part is the question of how to reduce the bond dimension after one time evolution step; but here we can again get inspiration of the 1-D case, and ask the question which projectors we can put on the bonds such as to maximize the overlap of the {\emph projected} one with the evolved one.

\begin{figure}
    \begin{center}
\includegraphics[width=8.5cm]{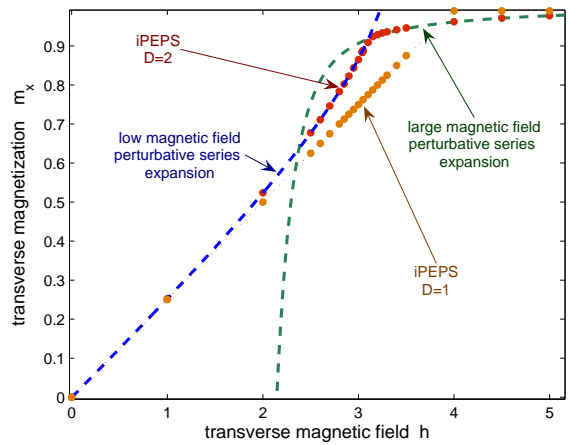}
    \end{center}

\caption{(taken from \cite{jordan07}) Magnetization $m_x(h)$ in the ground state $\ket{\Psi_h}$ of the two-dimensional quantum Ising model with transverse magnetic field. A trivial iPEPS (inner dimension $D=1$) produces a magnetization that grows essentially linearly with the magnetic field until it saturates. Instead, the simplest non-trivial iPEPS (inner dimension $D=2$) produces a magnetization curve that overlaps with the results of series expansions for both small and large magnetic fields. [$D=3$ leads to results that could hardly be distinguished from those for $D=2$ in this figure]. Notice that around $h\approx 3.1$ the derivative of the magnetization $m_x(h)$ changes suddenly.}
\label{fig5ipeps}
\end{figure}

A more straightforward approach is to assume an $..ABAB...$ symmetry where each tensor $A$ has all its neigbouring tensors $B$ and vice-versa (of course this is only possible for bipartite lattices;  a different choice can be made for different types of lattices). Again, imaginary time evolution can be used to find the ground state, and this was first studied in the paper~\cite{jordan07} where the term iPEPS (infinite PEPS) was coined. The idea is as follows: take the even-odd-horizontal-vertical Trotter decomposition as discussed previously, and next evolve with just one operator acting on two nearest neigbour sites. Effectively, this increases the bond dimension between those two sites(A and B). The \emph{environment} of those spins can be readily calculated using the infinite 1-D translational invariant methods discussed above, and then the variational problem becomes the problem of finding new $A'$ and $B'$ that approximate the one with higher bond optimally. Again, this can be done using the alternating least squares method. Subsequently, we replace all tensors $A$ and $B$ with $A'$ and $B'$, and continue until convergence. The last step - replacing all tensors with the optimal local ones - is only justified if the time step in the imaginary time evolution is very small, but in practice, this seems to work very well, as illustrated in figure \ref{fig5ipeps}.

\newpage
\section{Conclusion}

Recent progress in Quantum Information Theory has provided scientists with new mathematical tools to describe and analyze many-body quantum systems. These new tools have given rise to novel descriptions of the quantum states that appear in Nature, which are very efficient both in terms of the number of variables used to parametrize states and the number of operations to determine expectation values of typical observables. In a sense, they allow us to describe the "corner of Hilbert space" where relevant states are located with an effort that only scales polynomially with the number of particles, as opposed to the exponential scaling resulting with other descriptions. These results have automatically led to a diverse set of new powerful algorithms to simulate quantum systems. Those algorithms allow us to describe ground states, thermal equilibrium, low excitations, dynamics, random systems, etc, of many-body quantum systems, and thus to attack new kinds of problems obtaining very precise results. Moreover, the methods work in one and more spatial dimensions. In the first case, the success of some of those methods is directly related to the extraordinary performance of DMRG. In higher dimensions, they also give rise to a better understanding of several many-body systems for which a description has not been possible with the existing techniques.

This paper has reviewed these new methods in a unified form. We have introduced MPS and its extensions to higher dimensions, PEPS, and shown how one can build powerful algorithms that find the best descriptions of states within that family of states. The algorithms are relatively simple to implement, although they require some tricks that have been reported in this paper as well. We have also given simple matlab codes in Appendix~\ref{sec:matlab} to illustrate how one can program some of the methods. Thus, we believe that the present paper may be very useful both to scientist interested in implementing these new algorithms to describe any kind of many-body quantum systems, as well as those interested in creating new algorithms for some specific purposes. We have also provided several appealing evidences of the fact that most interesting states in Nature are well described by MPS and PEPS. This, in fact, indicates that our algorithms as well as future extensions
can provide us with unique tools to explore the fascinating physics of quantum many-body systems.

\vspace{1cm} \emph{Acknowledgements :}
We thank J. Eisert, M. Fannes, J. Garcia-Ripoll, M. Hastings, A. Kitaev, J. Latorre, M. Martin-Delgado, B. Nachtergaele, G. Ortiz, R. Orus, T. Osborne, B. Paredes, D. Perez-Garcia, M. Plenio, M. Popp, D. Porras,  J. Preskill, E. Rico, U. Schollwock, N. Schuch, C. Schoen, E. Solano, G. Vidal, J. von Delft, A. Weichselbaum, R. Werner, S. White, A. Winter, M. Wolf, M. Zwolak for valuable discussions that stimulated the developments of the ideas discussed in this paper.
% ------------------------------------------------------------------------------------

% ------------------------------------------------------------------------------------

\appendix
\section{Local reduced density operators of spin systems} \label{sec:reddensityop}

Due to the variational nature of ground states, there always exists a ground state with the same symmetries as the associated Hamiltonian. If the Hamiltonian has translational symmetry and consists of 2-body nearest neighbor interactions, then it is clear that the energy of a state with the right symmetry is completely determined by its reduced density operator of 2 neighboring spins. The reduced density operators arising from these (eventually mixed) states with a given symmetry form a convex set, and the energy for a given Hamiltonian will be minimized for a state whose reduced density operator is an extreme point in this set. More specifically, the equation $\rm{Tr}(H\rho)=E$ determines a hyperplane in the space of reduced density operators of states with a given symmetry, and the energy will be extremal when the hyperplane is tangent to the convex set ($E=E_{extr}$). The problem of finding the ground state energy of nearest neighbor translational invariant Hamiltonians is therefore equivalent to the determination of the convex set of 2-body reduced density operators arising from states with the right symmetry. Strictly speaking, these two problems are dual to each other. In the case of quadratic Hamiltonians involving continuous variables, the determination of this convex set was solved for fairly general settings in~\cite{wolfverstraete04} by means of Gaussian states. The determination of this convex set in the case of spin systems however turns out to be much more challenging.

\begin{figure}[t]
  %\centering
  \includegraphics[width=\linewidth]{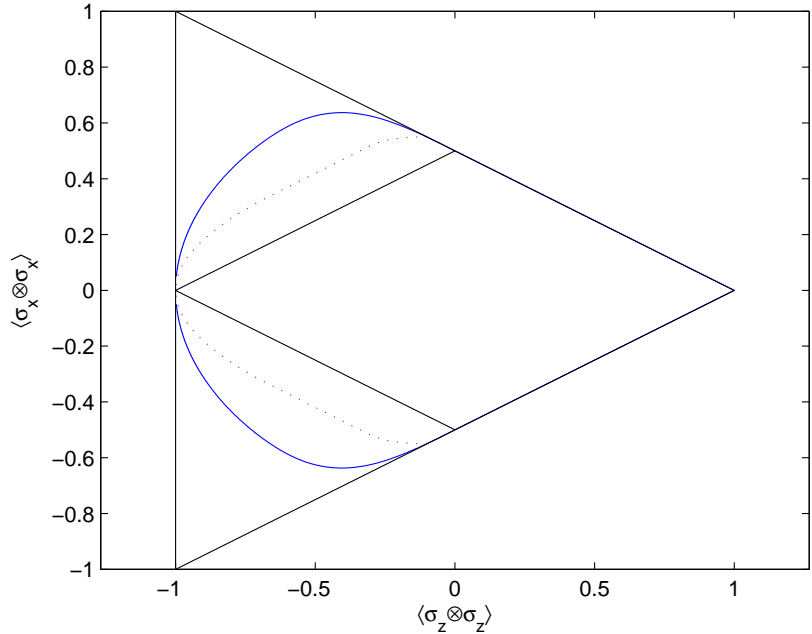}
  \caption{(taken from \cite{verstraetecirac05}) Convex sets of the possible reduced density operators of translational invariant spin $1/2$ states in the XX-ZZ plane.
 The big triangle
  represents all positive density operators; the inner
  parallellogram represents the separable states; the union of the separable cone and the convex hull of the full
  curved line is the complete convex set in the case of a 1-D
  geometry, and the dashed lines represent extreme points in the 2-D case of a square lattice. The singlet corresponds to the point with coordinates
  $(-1,-1)$.}
  \label{convex}
\end{figure}

Let us illustrate this with a simple example.
\begin{displaymath}
\mathcal{H}=-\sum_{<i,j>}S^x_i S^x_j+S^y_iS^y_j+\Delta S^z_iS^z_j
\end{displaymath}
on a lattice of arbitrary geometry and dimension\footnote{We assume that the graph corresponding to the lattice is edge-transitive, meaning that any vertex can be mapped to any other vertex by application of the symmetry group of the graph.}. Due to the symmetries, the reduced density operator of two nearest neighbors can be parameterized by only two parameters
\footnote{This can easily be proven by invoking local twirling operations which leave the Hamiltonian invariant.}:
\begin{displaymath}
\rho= \frac{1}{4}\left(I\otimes I+x (\sigma_x\otimes
\sigma_x+\sigma_y\otimes\sigma_y)+z\sigma_z\otimes\sigma_z\right).
\end{displaymath}
Positivity of $\rho$ enforces $-1\leq z\leq 1-2|x|$, and the state is separable iff $1+z\leq 2|x|$. In the case of an infinite 1-D spin chain, the ground state energy $E(\Delta)$ has been calculated exactly \cite{yang66}, and this determines the tangent hyperplanes \[2x+z\Delta+E(\Delta)=0\] whose envelope makes up the extreme points of the convex set of reduced density operators of translationally invariant 1-D states: the boundary of this convex set is parameterized by
\begin{eqnarray*}
z&=&-\partial E(\Delta)/\partial\Delta\\x&=&-(E(\Delta)+\partial E(\Delta)/\partial\Delta)/2,
\end{eqnarray*}
which we plotted in Figure~\ref{convex}. We also plot the boundary for the 2-dimensional square lattice. These 2-D data were obtained by numerical methods \cite{verstraetecirac04,syljuasen04,lin01}); of course this convex set is contained in the previous one, as all the semidefinite constraints defining the set corresponding to 1-D are strictly included in the set of constraints for the 2-D case. Finally, we plot the set of separable states, which contains the reduced density operators of the allowed states for a lattice with infinite coordination number. The boundary of this separable set is given by the inner diamond; this immediately implies that the difference between the exact energy and the one obtained by mean field theory will be maximized whenever the hyperplane that forms the boundary of the first set will be parallel to this line. This happens when $\Delta=-1$ (independent of the dimension!), which corresponds to the antiferromagnetic case, and this proves that the "entanglement gap" \cite{dowling04} in the XXZ-plane is maximized by the antiferromagnetic ground state for any dimension and geometry. Similarly, it proves that the ground state is separable whenever $\Delta\geq 1$ and $\Delta=-\infty$. Note also that in the 2-D case, part of the boundary of the convex set consists of a plane parameterized by $2x+z+E(1)=0$. This indicates a degeneracy of the ground state around the antiferromagnetic point, and indicates that a phase transition is occurring at that point (more specifically between an Ising and a Berezinskii-Kosterlitz-Thouless phase).

\begin{figure}[t]
  %\centering
  \includegraphics[width=.95\linewidth]{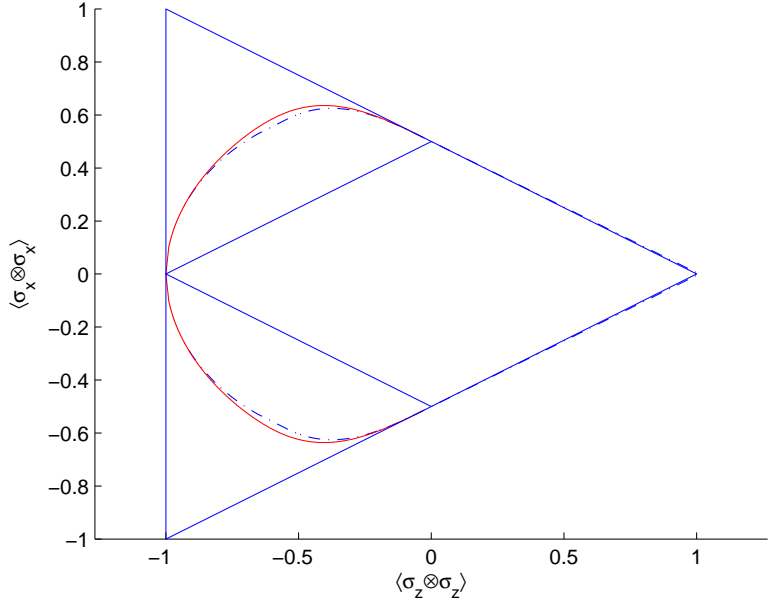}
  \caption{(taken from \cite{verstraetecirac05}) Convex sets in the XXZ-plane: the inner diamond borders the set of separable states (see Fig. 1). Dash-dotted: extreme points of the convex set produced by MPS of $D=2$.}
  \label{error}
\end{figure}

As a good illustration of the actual accuracy obtained with MPS, we calculated the convex set obtained with MPS in the thermodynamic limit for the XXZ-chain with $D=2$, where $D$ is the dimension of the matrices in the MPS (see figure \ref{error}). It is almost unbelievable how good the exact convex set can be approximated. Note that typical DMRG calculations have $D\sim 200$, and that the accuracy grows superpolynomial in $D$. Note also that the $D=1$ case corresponds to mean-field theory, whose corresponding convex set coincides with the set of separable states.

The same argument involving the notion of a correlation length applies in higher dimensions and indicates that PEPS represent ground states of gapped local Hamiltonians well. Note however that the convex set in the 2-D case is much closer to the separable one than in the 1-D case; this gives a hint that PEPS of smaller dimension will suffice to give the same accuracy as in the 1-D case. In the next section we will quantitatively bound how well a translationally invariant state can be represented in terms of a MPS, and will analyze the corresponding implications for the description of ground states of 1D spin chains.

\section{MPS represent ground states faithfully} \label{sec:mpsgs}

We will derive an upper bound to the error made by approximating a general ground state of a 1-D quantum spin system by a MPS. As we will show below, this has very important implications in the performance of the renormalization algorithms to describe ground states of 1D spin chains.

\begin{lemma}
There exists a MPS $|\psi_D\rangle$ of dimension $D$ such that
\begin{displaymath}
\||\psi\rangle-|\psi_D\rangle\|^2\leq
2\sum_{\alpha=1}^{N-1}\epsilon_\alpha(D)
\end{displaymath}
where $\epsilon_\alpha(D)=\sum_{i=D+1}^{N_\alpha}\mu^{[\alpha]i}$.
\end{lemma}

\emph{Proof:} We can always write $\psi$ as a MPS of dimension $D=2^{N/2}$ and fulfilling
\begin{displaymath}
 \sum_i A^{[m]i}A^{[m]i\dagger}=\openone,\quad
 \sum_i A^{[m]i\dagger}\Lambda^{[m+1]}A^{[m]i}=\Lambda^{[m]}.
\end{displaymath}
Let us now consider the $D$-dimensional MPS $|\psi_D\rangle$ which is defined by the $D\times D$ matrices $[A^{[\alpha]i}]_{(1\ldots D,1\ldots D)}$ (i.e. the upper--left block of $A^{[\alpha]i}$). The goal is now to bound $\langle\psi|\psi_D\rangle$. The gauge conditions were chosen such as to make the task simple:
\begin{equation}
 \langle\psi_D|\psi\rangle={\rm Tr}\left[\$_2\left(\cdots
 \$_{N-2}\left(\$_{N-1}\left(\Lambda^{N-1}P\right)P\right)
 P\cdots\right)P\right];\label{HJD}
\end{equation}
here $P=\sum_{k=1}^D|k\rangle\langle k|$ and $\$_m(X)=\sum_i A^{[m]i\dagger}XA^{[m]i}$ represents a trace-preserving completely positive map (TPCP-map) parameterized by the Kraus operators $A^{[m]i}$. Let us now recursively define \[Y^{[k]}=\$_{k}\left(Y^{[k+1]}P\right),\hspace{.5cm} Y^{[N-1]}=\Lambda^{[N-1]}P;\] observe that $\Lambda^{[k]}=\$_{k}\left(\Lambda^{[k+1]}\right)$. We want a bound on $\rm{Tr}|\Lambda^{[1]}-Y_1|$, as equation (\ref{HJD}) is equal to $\rm{Tr}(Y^{[2]})$. The crucial property we need is that TPCP-maps are contractive with relation to the trace-norm \footnote{This can directly be proven by considering the Neumark representation of a TPCP-map as a unitary in a bigger space.}: $\rm{Tr}|\$(X)|\leq \rm{Tr}|X|$. It follows that
\begin{eqnarray*} &&\rm{Tr}|\Lambda^{[k]}-Y^{[k]}|=\rm{Tr}|\$_k\left(\Lambda^{[k+1]}-Y^{[k+1]}P\right)|\leq\\&&\leq
\rm{Tr}|\Lambda^{[k+1]}-Y^{[k+1]}P|\\&&
\leq\rm{Tr}|\Lambda^{[k+1]}-Y^{[k+1]}|+\rm{Tr}|\Lambda^{[k+1]}(\openone-P)|.
\end{eqnarray*}
Note that the last term in the sum is exactly given by $\sum_{\alpha=D+1}^{2^{N/2}}\lambda^{[k+1]\alpha}$. The theorem now follows immediately by recursion and by observing that $\langle\psi_D|\psi_D\rangle\leq 1$ by similar arguments.\qed

The implications of this result are very strong: it shows that for systems for which the $\epsilon_\alpha(D)$ decay fast in $D$, there exist MPS with \emph{small} $D$ which will not only reproduce well the local correlations (such as energy) but also all the nonlocal properties (such as correlation length). The following lemma now relates the derived bound to the Renyi entropies of the reduced density operators, through which one can make the connection to the ground states of 1D Hamiltonians. The Renyi entropies of $\rho$ are defined as
\begin{displaymath}
S^\alpha(\rho)=\frac{1}{1-\alpha}\log\left({\rm Tr}\rho^\alpha\right),
\end{displaymath}
and we will consider  $0<\alpha< 1$. We denote as before $\epsilon(D)=\sum_{i=D+1}^\infty\lambda_i$ with $\lambda_i$ the nonincreasingly ordered eigenvalues of $\rho$. Then we have

\begin{lemma}
Given a density operator $\rho$. If $0<\alpha<1$, then
$\log(\epsilon(D))\leq
\frac{1-\alpha}{\alpha}\left(S^\alpha(\rho)-\log\frac{D}{1-\alpha}\right)$.
\end{lemma}

\emph{Proof:} Let us first characterize the probability distribution that has maximal possible weight in its tail (i.e. $p=\sum_{i=D+1}^\infty p_i$) for a given Renyi-entropy. Introducing a free parameter $0<h\leq (1-p)/D$, such a probability distribution must be of the form
\begin{eqnarray*}
p_1&=&1-p-(D-1)h\\ h&=&p_2=p_3=\cdots p_{D+p/h}\\
p_{D+p/h+1},\cdots p_{\infty}&=&0
\end{eqnarray*}
because this distribution majorizes all other ones with given $p,D,p_D$ (Renyi-entropies are Schur-convex functions). For a given $p,D,h$, it holds that
\begin{eqnarray*}
\sum_i p_i^\alpha&=&(1-p-(D-1)h)^\alpha+(D-1+p/h)h^\alpha\\&\geq& Dh^\alpha+ph^{\alpha-1}.
\end{eqnarray*}
Minimizing this expression with relation to $h$, we get
\begin{displaymath}
\sum_i p_i^\alpha\geq (D^{1-\alpha}p^\alpha)/((1-\alpha)^{1-\alpha}\alpha^\alpha).
\end{displaymath}
Denoting $S^\alpha(p,D)$ the minimal possible entropy for given $p,D$, we get
\begin{displaymath}
S^\alpha(p,D)\geq \frac{1}{1-\alpha}\log\left( \frac{D^{1-\alpha}p^\alpha}{(1-\alpha)^{1-\alpha}\alpha^\alpha}\right)
\end{displaymath}
and hence
\begin{displaymath}
p\leq \exp\left(\frac{1-\alpha}{\alpha}\left(S^\alpha(p,D)-\log\frac{D}{1-\alpha}\right)\right).
\end{displaymath}
The proof now follows by replacing $S^\alpha(p,D)$ by $S^\alpha(\rho)$. \qed

This lemma is very interesting in the light of the fact that in the case of critical systems, arguable the hardest ones to simulate \footnote{For non--critical systems, the renormalization group flow is expected to increase the Renyi entropies in the UV direction. The corresponding fixed point corresponds to a critical system whose entropy thus upper bounds that of the non--critical one.}, the Renyi-entropy of a contiguous block of $L$ spins scales as \cite{peschel99,jin04,calabrese04,vidallatorre03}
\begin{equation}
 S^\alpha(\rho_L)\simeq\frac{c+\bar{c}}{12}
 \left(1+\frac{1}{\alpha}\right)\log(L)\label{infg}
\end{equation}
for all $\alpha>0$; here  $c$ is the central charge. The fact that the eigenvalues of $\rho_L$ decay fast has previously been identified as a indication for the validity of the DMRG-approach \cite{peschel99}. The truncation error \cite{white92,schollwoeck04}, which has been used in the DMRG community as a check for convergence, is essentially given by $\epsilon(D)-\epsilon(2D)$ and therefore indeed gives a good idea of the error in a simulation.

Let us investigate how the computational effort to simulate such critical systems scales as a function of the length $N=2L$ of the chain. Let us therefore consider the Hamiltonian associated to a critical system, but restrict it to $2L$ sites. The entropy of a half chain (we consider the ground state $|\psi_{ex}\rangle$ of the finite system) will typically scale as in eq. (\ref{infg}) but with an extra term that scales like $1/N$. Suppose we want to enforce that $\||\psi_{ex}\rangle-|\psi_D\rangle\|^2\leq\epsilon_0/L$ with $\epsilon_0$ independent of $L$ \footnote{We choose the $1/L$ dependence such as to assure that the absolute error in extensive observables does not grow.}. Denote the minimal $D$ needed to get this precision for a chain of length $2L$ by $D_L$. Following lemma (1) and the fact that the entropy of all possible contiguous blocks reaches its maximum in the middle of the chain (hence $p\leq\epsilon_0/L^2$ is certainly sufficient), lemma (1) and (2) combined yield
\begin{displaymath}
D_L\leq
cst\left(\frac{L^2}{(1-\alpha)\epsilon_0}\right)^{\frac{\alpha}{1-\alpha}}
L^{\frac{c+\bar{c}}{12}\frac{1+\alpha}{\alpha}}.
\end{displaymath}
This shows that $D$ only has to scale polynomially in $L$ to keep the accuracy $\epsilon_0/L$ fixed; in other words, there exists an efficient scalable representation for ground states of critical systems (and hence also of noncritical systems) in terms of MPS! Such a strong result could not have been anticipated from just doing simulations. Furthermore, Hastings has proven that ground states of gapped systems always obey a strict area law \cite{Hastingsarea}. This implies that the ground state of any gapped spin chain is indeed well approximated by a MPS. For a more detailed description of the relations between area laws and approximability, we refer to \cite{Schuchfaith07}.

Now what about the complexity of finding this optimal MPS? It has been observed that DMRG converges exponentially fast to the ground state with a relaxation time proportional to the inverse of the gap $\Delta$ of the system \cite{schollwoeck04}. For translational invariant critical systems, this gap seems to close only polynomially. As we have proven that $D$ only have to scale polynomially too, the complexity of deciding whether the ground state energy of 1-D quantum systems is below a certain value is certainly in NP (as it can be checked efficiently if Merlin gives the MPS-description). The problem would even be in P if the following conditions are met: 1) the $\alpha$-entropy of blocks in the exact ground state grow at most logarithmically with the size of the block for some $\alpha<1$; 2) the gap of the system scales at most polynomially with the system size; 3) given a gap that obeys condition 2, there exists an efficient DMRG-like algorithm that converges to the global minimum. As the variational MPS approach \cite{verstraeteporras04} is essentially an alternating least squares method of solving a non-convex problem which is in worst case NP-hard \cite{bental98,eisert06}, there is a priori no guarantee that it will converge to the global optimum, although the occurrence of local minima seems to be unlikely \cite{schollwoeck04}. Surprisingly, one can indeed construct a family of Hamiltonians with nearest neighbour interactions on a line for which the ground state is an exact MPS with polynomial $D$, but for which it is an NP-complete problem to find it \cite{Schuchinpreparation}. As the corresponding Hamiltonian has a gap that closes polynomially in the size of the system, the only hope that VMPS/DMRG methods to be in P is in the case of gapped systems \footnote{Let us specify an alternative method which should in principle not get trapped in local minima in the case of gapped systems. Like in the adiabatic theorem, we can construct a time dependent Hamiltonian $H(t)$ with $H(0)$ trivial and $H(1)$ the Hamiltonian to simulate; if we discretize this evolution in a number of steps that grows polynomially in the inverse gap, the adiabatic theorem guarantees that we will end up in the ground state of $H(1)$ if we can follow the ground state of $H(t)$ closely. The idea is to make $D$ of $|\psi_D(t)\rangle$ large enough such as to follow the ground state $|\psi(t)\rangle$ close enough in such a way that the optimization is always convex around the global optimum within the domain $\||\chi\rangle-|\psi(t)\rangle\|\leq\epsilon$. As we are simulating this classically, we could even do imaginary time evolution over a longer time during each step.}.

\section{Matlab code} \label{sec:matlab}

As a last part, we would like to give an idea of how to program the variational methods explained in the previous sections. We present two functions, one for the calculation of ground- and first excited states and one for the reduction of the virtual dimension of matrix product states. We demonstrate these functions by means of the antiferromagnetic Heisenberg chain.

\subsection{Minimization of the Energy}

The function \verb"minimizeE" optimizes the parameters of a matrix product state in  such a way that the expectation value with respect to a given Hamiltonian tends to a minimum. The function expects this Hamiltonian to be defined in a $M \times N$ cell \verb"hset", where $N$ denotes the number of sites and $M$ the number of terms in the Hamiltonian. Assuming the Hamiltonian is of the form
\begin{displaymath}
H=\sum_{m=1}^M h_m^{(1)} \otimes \cdots \otimes h_m^{(N)},
\end{displaymath}
the element \verb"hset{m,j}" equals $h_m^{(j)}$. Further arguments are the virtual dimension of the resulting matrix product state, \verb"D", and the expected accuracy of the energy, \verb"precision".

Output arguments are the optimized energy \verb"E" and corresponding matrix product state \verb"mps". The matrix product state is stored as a $1 \times N$ cell, each entry corresponding to one matrix.

Optionally, a matrix product state \verb"mpsB" can be specified as an argument to which the resulting state shall be orthogonal. This is especially useful for calculating the first excited state.

\begin{tiny}
\begin{verbatim}
function [E,mps]=minimizeE(hset,D,precision,mpsB)

[M,N]=size(hset);
d=size(hset{1,1},1);
mps=createrandommps(N,D,d);
mps=prepare(mps);

% storage-initialization
Hstorage=initHstorage(mps,hset,d);
if ~isempty(mpsB), Cstorage=initCstorage(mps,[],mpsB,N); end
P=[];

% optimization sweeps
while 1
    Evalues=[];

    % ****************** cycle 1: j -> j+1 (from 1 to N-1) ****************
    for j=1:(N-1)
        % projector-calculation
        if ~isempty(mpsB)
            B=mpsB{j};
            Cleft=Cstorage{j};
            Cright=Cstorage{j+1};
            P=calcprojector_onesite(B,Cleft,Cright);
        end

        % optimization
        Hleft=Hstorage(:,j);
        Hright=Hstorage(:,j+1);
        hsetj=hset(:,j);
        [A,E]=minimizeE_onesite(hsetj,Hleft,Hright,P);
        [A,U]=prepare_onesite(A,'lr');
        mps{j}=A;
        Evalues=[Evalues,E];

        % storage-update
        for m=1:M
            h=reshape(hset{m,j},[1,1,d,d]);
            Hstorage{m,j+1}=updateCleft(Hleft{m},A,h,A);
        end
        if ~isempty(mpsB)
            Cstorage{j+1}=updateCleft(Cleft,A,[],B);
        end
    end

    % ****************** cycle 2: j -> j-1 (from N to 2) ******************
    for j=N:(-1):2
        % projector-calculation
        if ~isempty(mpsB)
            B=mpsB{j};
            Cleft=Cstorage{j};
            Cright=Cstorage{j+1};
            P=calcprojector_onesite(B,Cleft,Cright);
        end

        % minimization
        Hleft=Hstorage(:,j);
        Hright=Hstorage(:,j+1);
        hsetj=hset(:,j);
        [A,E]=minimizeE_onesite(hsetj,Hleft,Hright,P);
        [A,U]=prepare_onesite(A,'rl');
        mps{j}=A;
        Evalues=[Evalues,E];

        % storage-update
        for m=1:M
            h=reshape(hset{m,j},[1,1,d,d]);
            Hstorage{m,j}=updateCright(Hright{m},A,h,A);
        end
        if ~isempty(mpsB)
            Cstorage{j}=updateCright(Cright,A,[],B);
        end
    end

    if (std(Evalues)/abs(mean(Evalues))<precision)
        mps{1}=contracttensors(mps{1},3,2,U,2,1);
        mps{1}=permute(mps{1},[1,3,2]);
        break;
    end
end

% ************************ one-site optimization **************************

function [A,E]=minimizeE_onesite(hsetj,Hleft,Hright,P)

DAl=size(Hleft{1},1);
DAr=size(Hright{1},1);
d=size(hsetj{1},1);

% calculation of Heff
M=size(hsetj,1);

Heff=0;
for m=1:M
    Heffm=contracttensors(Hleft{m},3,2,Hright{m},3,2);
    Heffm=contracttensors(Heffm,5,5,hsetj{m},3,3);
    Heffm=permute(Heffm,[1,3,5,2,4,6]);
    Heffm=reshape(Heffm,[DAl*DAr*d,DAl*DAr*d]);
    Heff=Heff+Heffm;
end

% projection on orthogonal subspace
if ~isempty(P), Heff=P'*Heff*P; end

% optimization
options.disp=0;
[A,E]=eigs(Heff,1,'sr',options);
if ~isempty(P), A=P*A; end
A=reshape(A,[DAl,DAr,d]);


function [P]=calcprojector_onesite(B,Cleft,Cright)

y=contracttensors(Cleft,3,3,B,3,1);
y=contracttensors(y,4,[2,3],Cright,3,[2,3]);
y=permute(y,[1,3,2]);
y=reshape(y,[prod(size(y)),1]);

Q=orth([y,eye(size(y,1))]);
P=Q(:,2:end);
\end{verbatim}
\end{tiny}

\subsection{Time Evolution}

The function \verb"reduceD" forms the basis for the simulation of a time evolution. It multiplies a given matrix product state with a given matrix product operator and reduces the virtual dimension of the resuling state, i.e. it searches a matrix product state with reduced virtual dimension and minimal distance to the original state. The matrix product state and the matrix product operator are specified in the arguments \verb"mpsA" and \verb"mpoX". As before, they are represented by a cell with entries identifying the matrices. The reduced virtual dimension is specified in the argument \verb"DB". The argument \verb"precision" defines the convergence condition: if fluctutions in the distance are less than \verb"precision", the optimization is assumed to be finished.

The output argument is the optimized matrix product state \verb"mpsB" with virtual dimension \verb"DB".

\begin{tiny}
\begin{verbatim}
function mpsB=reduceD(mpsA,mpoX,DB,precision)

N=length(mpsA);
d=size(mpsA{1},3);
mpsB=createrandommps(N,DB,d);
mpsB=prepare(mpsB);
% initialization of the storage
Cstorage=initCstorage(mpsB,mpoX,mpsA,N);

% optimization sweeps
while 1
    Kvalues=[];

    % ****************** cycle 1: j -> j+1 (from 1 to N-1) ****************
    for j=1:(N-1)
        % optimization
        Cleft=Cstorage{j};
        Cright=Cstorage{j+1};
        A=mpsA{j}; X=mpoX{j};
        [B,K]=reduceD2_onesite(A,X,Cleft,Cright);
        [B,U]=prepare_onesite(B,'lr');
        mpsB{j}=B;
        Kvalues=[Kvalues,K];

        % storage-update
        Cstorage{j+1}=updateCleft(Cleft,B,X,A);
    end

    % ****************** cycle 2: j -> j-1 (from N to 2) ******************
    for j=N:(-1):2
        % optimization
        Cleft=Cstorage{j};
        Cright=Cstorage{j+1};
        A=mpsA{j}; X=mpoX{j};
        [B,K]=reduceD2_onesite(A,X,Cleft,Cright);
        [B,U]=prepare_onesite(B,'rl');
        mpsB{j}=B;
        Kvalues=[Kvalues,K];

        % storage-update
        Cstorage{j}=updateCright(Cright,B,X,A);
    end

    if std(Kvalues)/abs(mean(Kvalues))<precision
        mpsB{1}=contracttensors(mpsB{1},3,2,U,2,1);
        mpsB{1}=permute(mpsB{1},[1,3,2]);
        break;
    end
end



% ************************ one-site optimization **************************

function [B,K]=reduceD2_onesite(A,X,Cleft,Cright)

Cleft=contracttensors(Cleft,3,3,A,3,1);
Cleft=contracttensors(Cleft,4,[2,4],X,4,[1,4]);

B=contracttensors(Cleft,4,[3,2],Cright,3,[2,3]);
B=permute(B,[1,3,2]);

b=reshape(B,[prod(size(B)),1]);
K=-b'*b;
\end{verbatim}
\end{tiny}

\subsection{Auxiliary functions}

The previous two functions depend on several auxiliary functions that are printed in this section.

\begin{itemize}
\item Gauge transformation that prepares the MPS \verb"mpsB" in such a form that $N_{eff}$ is equal to the identity for the first spin (see section \ref{sec:mps:dmrg}):
\begin{tiny}
\begin{verbatim}
function [mps]=prepare(mps)

N=length(mps);

for i=N:-1:2
	[mps{i},U]=prepare_onesite(mps{i},'rl');
    mps{i-1}=contracttensors(mps{i-1},3,2,U,2,1);
    mps{i-1}=permute(mps{i-1},[1,3,2]);
end
\end{verbatim}
\end{tiny}

\begin{tiny}
\begin{verbatim}
function [B,U,DB]=prepare_onesite(A,direction)

[D1,D2,d]=size(A);
switch direction
    case 'lr'
        A=permute(A,[3,1,2]); A=reshape(A,[d*D1,D2]);
        [B,S,U]=svd2(A); DB=size(S,1);
        B=reshape(B,[d,D1,DB]); B=permute(B,[2,3,1]);
        U=S*U;
    case 'rl'
        A=permute(A,[1,3,2]); A=reshape(A,[D1,d*D2]);
        [U,S,B]=svd2(A); DB=size(S,1);
        B=reshape(B,[DB,d,D2]); B=permute(B,[1,3,2]);
        U=U*S;
end
\end{verbatim}
\end{tiny}

\item Initialization of storages:
\begin{tiny}
\begin{verbatim}
function [Hstorage]=initHstorage(mps,hset,d)

[M,N]=size(hset);
Hstorage=cell(M,N+1);
for m=1:M, Hstorage{m,1}=1; Hstorage{m,N+1}=1; end
for j=N:-1:2
    for m=1:M
        h=reshape(hset{m,j},[1,1,d,d]);
        Hstorage{m,j}=updateCright(Hstorage{m,j+1},mps{j},h,mps{j});
    end
end
\end{verbatim}
\end{tiny}

\begin{tiny}
\begin{verbatim}
function [Cstorage]=initCstorage(mpsB,mpoX,mpsA,N)

Cstorage=cell(1,N+1);
Cstorage{1}=1;
Cstorage{N+1}=1;
for i=N:-1:2
    if isempty(mpoX), X=[]; else X=mpoX{i}; end
    Cstorage{i}=updateCright(Cstorage{i+1},mpsB{i},X,mpsA{i});
end
\end{verbatim}
\end{tiny}

\begin{tiny}
\begin{verbatim}
function [Cleft]=updateCleft(Cleft,B,X,A)

if isempty(X), X=reshape(eye(size(B,3)),[1,1,2,2]); end

Cleft=contracttensors(A,3,1,Cleft,3,3);
Cleft=contracttensors(X,4,[1,4],Cleft,4,[4,2]);
Cleft=contracttensors(conj(B),3,[1,3],Cleft,4,[4,2]);
\end{verbatim}
\end{tiny}

\begin{tiny}
\begin{verbatim}
function [Cright]=updateCright(Cright,B,X,A)

if isempty(X), X=reshape(eye(size(B,3)),[1,1,2,2]); end

Cright=contracttensors(A,3,2,Cright,3,3);
Cright=contracttensors(X,4,[2,4],Cright,4,[4,2]);
Cright=contracttensors(conj(B),3,[2,3],Cright,4,[4,2]);
\end{verbatim}
\end{tiny}

\item Creation of a random MPS:
\begin{tiny}
\begin{verbatim}
function [mps]=createrandommps(N,D,d)

mps=cell(1,N);
mps{1}=randn(1,D,d)/sqrt(D);
mps{N}=randn(D,1,d)/sqrt(D);
for i=2:(N-1)
    mps{i}=randn(D,D,d)/sqrt(D);
end
\end{verbatim}
\end{tiny}

\item Expectation value of the MPS \verb"mps" with respect to the operator defined in~\verb"hset":
\begin{tiny}
\begin{verbatim}
function [e,n]=expectationvalue(mps,hset)

[M,N]=size(hset);
d=size(mps{1},3);

% expectation value
e=0;
for m=1:M
    em=1;
    for j=N:-1:1
        h=hset{m,j};
        h=reshape(h,[1,1,d,d]);
        em=updateCright(em,mps{j},h,mps{j});
    end
    e=e+em;
end

% norm
n=1;
X=eye(d); X=reshape(X,[1,1,d,d]);
for j=N:-1:1
    n=updateCright(n,mps{j},X,mps{j});
end

e=e/n;
\end{verbatim}
\end{tiny}

\item Contraction of index \verb"indX" of tensor \verb"X" with index \verb"indY" of tensor \verb"Y" (\verb"X" and~\verb"Y" have a number of indices corresponding to \verb"numindX" and \verb"numindY" respectively):
\begin{tiny}
\begin{verbatim}
function [X,numindX]=contracttensors(X,numindX,indX,Y,numindY,indY)

Xsize=ones(1,numindX); Xsize(1:length(size(X)))=size(X);
Ysize=ones(1,numindY); Ysize(1:length(size(Y)))=size(Y);

indXl=1:numindX; indXl(indX)=[];
indYr=1:numindY; indYr(indY)=[];

sizeXl=Xsize(indXl);
sizeX=Xsize(indX);
sizeYr=Ysize(indYr);
sizeY=Ysize(indY);

if prod(sizeX)~=prod(sizeY)
    error('indX and indY are not of same dimension.');
end

if isempty(indYr)
    if isempty(indXl)
        X=permute(X,[indX]);
        X=reshape(X,[1,prod(sizeX)]);

        Y=permute(Y,[indY]);
        Y=reshape(Y,[prod(sizeY),1]);

        X=X*Y;
        Xsize=1;

        return;

    else
        X=permute(X,[indXl,indX]);
        X=reshape(X,[prod(sizeXl),prod(sizeX)]);

        Y=permute(Y,[indY]);
        Y=reshape(Y,[prod(sizeY),1]);

        X=X*Y;
        Xsize=Xsize(indXl);

        X=reshape(X,[Xsize,1]);

        return
    end
end

X=permute(X,[indXl,indX]);
X=reshape(X,[prod(sizeXl),prod(sizeX)]);

Y=permute(Y,[indY,indYr]);
Y=reshape(Y,[prod(sizeY),prod(sizeYr)]);

X=X*Y;
Xsize=[Xsize(indXl),Ysize(indYr)];
numindX=length(Xsize);
X=reshape(X,[Xsize,1]);
\end{verbatim}
\end{tiny}

\item Economical singular value decomposition:
\begin{tiny}
\begin{verbatim}
function [U,S,V]=svd2(T)

[m,n]=size(T);
if m>=n, [U,S,V]=svd(T,0); else [V,S,U]=svd(T',0); end
V=V';
\end{verbatim}
\end{tiny}

\end{itemize}

\subsection{Examples}

\begin{figure}[t]
  %\centering
  \includegraphics[width=\linewidth]{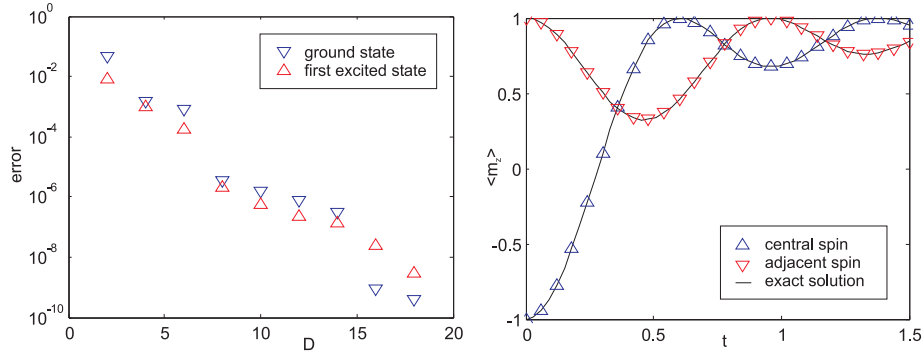}
  \caption{(Left) Error of the variational method as a function of the virtual dimension~$D$ for $N=10$ spins. The blue triangles represent the ground state, the red triangles the first excited state. (Right) Time evolution as described in example 2. The magnetization of the central spin is represented by blue triangles and the magnetization of the spin adjacent to the central spin by red triangles. For comparison, exact results are also included (black lines).}
  \label{fig:matlab}
\end{figure}

As a first example, we show how to calculate the ground-state and the first excited state of the antiferromagnetic Heisenberg chain using the method \verb"minimizeE" from before. In this example, the chain--length~\verb"N" is assumed to be~$10$ and the virtual dimension~\verb"D" is set to~$5$.

\begin{tiny}
\begin{verbatim}
N=10;
D=5;
precision=1e-5;

% Heisenberg Hamiltonian
M=3*(N-1);
hset=cell(M,N);
sx=[0,1;1,0]; sy=[0,-1i;1i,0]; sz=[1,0;0,-1]; id=eye(2);
for m=1:M, for j=1:N, hset{m,j}=id; end; end
for j=1:(N-1)
    hset{3*(j-1)+1,j}=sx; hset{3*(j-1)+1,j+1}=sx;
    hset{3*(j-1)+2,j}=sy; hset{3*(j-1)+2,j+1}=sy;
    hset{3*(j-1)+3,j}=sz; hset{3*(j-1)+3,j+1}=sz;
end

% ground state energy
randn('state',0)
[E0,mps0]=minimizeE(hset,D,precision,[]);
fprintf('E0 = %g\n',E0);

% first excited state
[E1,mps1]=minimizeE(hset,D,precision,mps0);
fprintf('E1 = %g\n',E1);
\end{verbatim}
\end{tiny}

As a second example, we focus on the real time evolution with respect to the Heisenberg antiferromagnet. Starting state is a product state with all spins pointing in $z$-direction except the central spin which is flipped. We evolve the state with the method \verb"reduceD" and calculate at each step the magnetization $m_z$ of the central spin. As before, \verb"N" is equal to $10$ and~\verb"D" is set to~$5$.

\begin{tiny}
\begin{verbatim}
N=10;
D=5;
precision=1e-5;
dt=0.03;
jflipped=5;

% magnetization in z-direction
oset=cell(1,N);
sx=[0,1;1,0]; sy=[0,-1i;1i,0]; sz=[1,0;0,-1]; id=eye(2);
for j=1:N, oset{1,j}=id; end;
oset{1,jflipped}=sz;

% time evolution operator
h=kron(sx,sx)+kron(sy,sy)+kron(sz,sz);
w=expm(-1i*dt*h);
w=reshape(w,[2,2,2,2]); w=permute(w,[1,3,2,4]); w=reshape(w,[4,4]);
[U,S,V]=svd2(w); eta=size(S,1);
U=U*sqrt(S); V=sqrt(S)*V;
U=reshape(U,[2,2,eta]); U=permute(U,[4,3,2,1]);
V=reshape(V,[eta,2,2]); V=permute(V,[1,4,3,2]);
I=reshape(id,[1,1,2,2]);
mpo_even=cell(1,N);
mpo_odd=cell(1,N);
for j=1:N, mpo_even{j}=I; mpo_odd{j}=I; end
for j=1:2:(N-1), mpo_odd{j}=U; mpo_odd{j+1}=V; end
for j=2:2:(N-1), mpo_even{j}=U; mpo_even{j+1}=V; end

% starting state (one spin flipped)
mps0=cell(1,N);
for j=1:N
    if j==jflipped, state=[0; 1]; else state=[1; 0]; end
    mps0{j}=reshape(state,[1,1,2]);
end

% time evolution
mps=mps0;
mzvalues=[];
for step=1:50
    fprintf('Step %2d: ',step);
    [mps,K]=reduceD(mps,mpo_even,D,precision);
    [mps,K]=reduceD(mps,mpo_odd,D,precision);
    mz=expectationvalue(mps,oset);
    mzvalues=[mzvalues,mz];
    fprintf('mz=%g\n',mz);
end
\end{verbatim}
\end{tiny}

A comparison of the results produced by these examples to exact caclulations is shown in figure~\ref{fig:matlab}. It can be seen that already for moderate values of~$D$ the precision is very good.

\end{document}